\documentclass[utf8]{FrontiersinHarvard} 

\usepackage{url,microtype,subcaption}
\usepackage[onehalfspacing]{setspace}
\usepackage[breaklinks, colorlinks, citecolor=blue, urlcolor = blue]{hyperref}
\usepackage{soul}

\newcommand{\msun}{~M$_{\odot}$}

\def \ecut {$E_{\mbox{\scriptsize cut}}$}
\def \kte {$kT_{\rm e}$}

\def \swift {{\em Swift}}
\def \bepposax {{\em BeppoSAX}}
\def \integral {{\em INTEGRAL}}
\def \suzaku {{\em Suzaku}}
\def \nustar {{\em NuSTAR}}

\def \swiftbat {{\em Swift}/BAT}

\def \chandra {{\em Chandra}}

\def \nicer {{\em NICER}}
\def \xmm {{\em XMM-Newton}}
\def \hexp {{\em HEX-P}}
\def \ixpe {{\em IXPE}}
\def \xrism {{\em XRISM}}
\def \athena {{\em Athena}}

\def\keyFont{\fontsize{8}{11}\helveticabold }
\def\firstAuthorLast{Kammoun {et~al.}} 
\def\Authors{E.\,Kammoun\,$^{1,2,3,*}$, 
A.\,M.\,Lohfink\,$^{4}$, 
M.\,Masterson\,$^{5}$, 
D.\,R.\,Wilkins\,$^{6}$, 
X.\,Zhao\,$^{7}$, 
M.\,Balokovi\'{c}\,$^{8,9}$, 
P.\,G.\,Boorman\,$^{10}$, 
R.\,M.\,T.\,Connors\,$^{11}$, 
P.\,Coppi\,$^{12}$, 
A.\,C.\,Fabian\,$^{13}$, 
J.\,A.\,García\,$^{14,10}$, 
K.\,K.\,Madsen\,$^{14}$, 
N.\,Rodriguez Cavero\,$^{15}$, 
N.\,Sridhar\,$^{10, 16, 17}$, 
D.\,Stern\,$^{18}$, 
J.\,Tomsick\,$^{19}$, 
T.\,Wevers\,$^{20}$,
D.\,J.\,Walton\,$^{21}$,
S.\,Bianchi\,$^{3}$,
J.\,Buchner\,$^{22}$,
F.\,Civano\,$^{14}$,
G.\,Lanzuisi\,$^{23}$,
L.\,Mallick\,$^{24,25,10}$,
G.\,Matt\,$^{3}$,
A.\,Merloni\,$^{22}$,
E.\,Nardini\,$^{2}$,
J.\,M.\,Piotrowska\,$^{10}$,
C.\,Ricci\,$^{26, 27}$,
K.-W.\,Wong\,$^{28}$,
A.\,Zoghbi\,$^{29,30,31}$,
and the HEX-P Collaboration
}

\def\Address{\footnotesize
$^{1}$IRAP, Universit\'{e} de Toulouse, CNRS, UPS, CNES, Toulouse, France 

$^{2}$INAF -- Osservatorio Astrofisico di Arcetri, Firenze, Italy 

$^{3}$Dipartimento di Matematica e Fisica, Universit\`{a} Roma Tre,  Rome, Italy

$^4$Department of Physics, Montana State University, , Bozeman, MT , USA

$^5$MIT Kavli Institute for Astrophysics and Space Research, Massachusetts Institute of Technology, Cambridge, MA, USA

$^6$Kavli Institute for Particle Astrophysics and Cosmology, Stanford University,  Stanford, CA, USA

$^7$Center for Astrophysics, Harvard \& Smithsonian, Cambridge, MA, USA

$^8$Yale Center for Astronomy \& Astrophysics, New Haven, CT, USA

$^9$Department of Physics, Yale University, New Haven, CT, USA

$^{10}$Cahill Center for Astronomy and Astrophysics, California Institute of Technology, Pasadena, CA, USA

$^{11}$Department of Physics, Villanova University,  Villanova, PA, USA

$^{12}$Department of Astronomy, Yale University, New Haven, CT, USA

$^{13}$Institute of Astronomy, University of Cambridge, Cambridge, UK

$^{14}$X-ray Astrophysics Laboratory, NASA Goddard Space Flight Center, Greenbelt, MD, USA

$^{15}$Department of Physics \& McDonnell Center for the Space Sciences, Washington University in St. Louis, St. Louis, MO, USA

$^{16}$Department of Astronomy, Columbia University, New York, NY, USA

$^{17}$Theoretical High Energy Astrophysics (THEA) Group, Columbia University, New York, NY, USA

$^{18}$Jet Propulsion Laboratory, California Institute of Technology, Pasadena, CA , USA

$^{19}$Space Sciences Laboratory, University of California, Berkeley, CA, USA

$^{20}$Space Telescope Science Institute, Baltimore, MD, USA

$^{21}$Centre for Astrophysics Research, University of Hertfordshire, College Lane, Hatfield, UK

$^{22}$Max-Planck-Institut f\"{u}r Extraterrestrische Physik (MPE), Garching bei M\"{u}nchen, Germany

$^{23}$INAF-Osservatorio di Astrofisica e Scienza delle Spazio di Bologna, OAS, Bologna, Italy

$^{24}$ CITA National Fellow, University of Manitoba, Department of Physics \& Astronomy, Winnipeg, Canada

$^{25}$ Canadian Institute for Theoretical Astrophysics, University of Toronto, Toronto, Ontario, Canada

$^{26}$Instituto de Estudios Astrof\'{i}sicos, Facultad de Ingenier\'{i}a y Ciencias, Universidad Diego Portales, Santiago, Chile

$^{27}$Kavli Institute for Astronomy and Astrophysics, Peking University, Beijing, China

$^{28}$Department of Physics, SUNY Brockport, Brockport, NY,  USA

$^{29}$Department of Astronomy, University of Maryland, College Park, MD, USA

$^{30}$HEASARC, Code 6601, NASA/GSFC, Greenbelt, MD, USA

$^{31}$CRESST II, NASA Goddard Space Flight Center, Greenbelt, MD, USA
}
\def\corrAuthor{Elias Kammoun}

\def\corrEmail{ekammoun.astro@gmail.com}

\begin{document}
\onecolumn
\firstpage{1}

\title{The High Energy X-ray Probe (HEX-P): Probing the physics of the X-ray corona in active galactic nuclei} 

\author[\firstAuthorLast ]{\Authors} 
\address{} 
\correspondance{} 

\extraAuth{}

\maketitle
\eject
\begin{abstract}

\section{}
The hard X-ray emission in active galactic nuclei (AGN) and black hole X-ray binaries is thought to be produced by a hot cloud of electrons referred to as the corona. This emission, commonly described by a power law with a high-energy cutoff, is suggestive of Comptonization by thermal electrons. While several hypotheses have been proposed to explain the origin, geometry, and composition of the corona, we still lack a clear understanding of this fundamental component. \nustar\ has been playing a key role improving our knowledge of X-ray coron\ae\ thanks to its unprecedented sensitivity above 10 keV. However, these constraints are limited to bright, nearby sources. The {\em High Energy X-ray Probe} (\hexp) is a probe-class mission concept combining high spatial resolution X-ray imaging and broad spectral coverage (0.2-80 keV) with a sensitivity superior to current facilities. In this paper, we highlight the major role that \hexp\ will play in further advancing our insights of X-ray coron\ae\, notably in AGN. We demonstrate how \hexp\ will measure key properties and track the temporal evolution of coron\ae\ in unobscured AGN. This will allow us to determine their electron distribution and test the dominant emission mechanisms. Furthermore, we show how \hexp\ will accurately estimate the coronal properties of obscured AGN in the local Universe, helping address fundamental questions about AGN unification. In addition, \hexp\ will characterize coron\ae\ in a large sample of luminous quasars at cosmological redshifts for the first time and track the evolution of coron\ae\ in transient systems in real time. We also demonstrate how \hexp\ will enable estimating the coronal geometry using spectral-timing techniques. \hexp\ will thus be essential to understand the evolution and growth of black holes over a broad range of mass, distance, and luminosity, and will help uncover the black holes’ role in shaping the Universe.

 \keyFont{ \section{Keywords:} accretion, X-ray astronomy, black holes, active galactic nuclei, quasars, tidal disruption events} 
\end{abstract}

\section{Introduction}
\label{sec:intro}


Active galactic nuclei (AGN) are known to be prolific X-ray emitters. The hard X-rays from AGN are thought to be generated through Compton scattering of thermal UV-soft X-ray seed photons from the accretion disk by a hot electron plasma located in the vicinity of the black hole, known as the corona \citep[e.g.,][]{Vaiana1978, Rybicki1979, Haardt1991, Haardt1993}. 
The resulting X-ray continuum spectrum is usually approximated by a power law of photon index $\Gamma$, with a high-energy cutoff (\ecut) at tens to hundreds of keV \citep[e.g.,][]{Fabian2017, Tortosa2018, Tortosa2023}. Within the Comptonization framework, $\Gamma$ and \ecut\ can be mapped into the electron temperature\footnote{The cutoff energy is commonly estimated to be $\sim 2-3$ times the electron temperature \citep{Petrucci01a}.} (\kte) and the coronal optical depth ($\tau$) plane \citep[see, e.g.,][]{Titarchuk95, Zdziarski96, Beloborodov+1999, Petrucci01a, Petrucci01b, Middei19}. The X-ray corona thus holds key information about the physical processes occurring near the supermassive black hole (SMBH), offering insights into the energetic phenomena associated with AGN. Understanding the physics of the X-ray corona in AGN is crucial for unraveling the accretion process, comprehending the origin of the emission from AGN, studying black hole feedback and its impact on galaxy evolution, probing strong gravity regimes, and investigating magnetic fields and particle acceleration.

\subsection{The geometry of the X-ray corona}

Constraining the geometry (e.g., size and location) of the X-ray corona offers an opportunity to test models for the formation of the corona as each mechanism predicts a distinct geometry. For example, the ``two-phase'' corona model proposed by \citet{Haardt1991} predicts an extended corona. In contrast, within the framework of the failed jet model proposed by \citet{Ghisellini2004}, the corona is expected to be compact. To date, the most promising constraints come from microlensing variability. The microlensing of quasars by stellar components in the lensing galaxy results in a complex magnification pattern \citep[e.g.,][]{Wambsganss2006}. The relative motion of the lensed quasar, the galaxy and its stellar components, and the observer leads to uncorrelated variability. The variability amplitude depends significantly on the size of the emitting region, with a larger amplitude from a smaller emitting region. Thus, the size of the source region can be estimated by modelling the light curve of a lensed quasar \citep[e.g.,][]{Chartas2002, Chartas2009, Chartas2016, Kochanek2004, Kochanek2007, Mosquera2013}. \citet{Reis2013} compiled measurements of the X-ray-emitting region in lensed quasars and found the sizes range from $\sim$ one to tens of gravitational radii ($r_{\rm g} = GM_{\rm BH}/c^2$, where $G$ is the gravitational constant, $M_{\rm BH}$ is the mass of the black hole, and $c$ the speed of light). We highlight that current results are based on only a few strongly lensed quasars.

X-ray spectral and spectral-timing properties of nearby bright Seyfert galaxies, notably the detection of X-ray reverberation lags, led to the similar conclusion that the X-ray source is compact and lies within a few gravitational radii of the black hole \citep{Fabian2009, DeMarco2013, Cackett2014, Emmanoulopoulos2014, Uttley2014, Kara2016}. Modelling the continuum optical-to-X-ray spectral energy distribution (SED) of a few bright Seyfert galaxies also yields similar estimates of the corona size \citep[e.g.,][]{Petrucci2013, Done2013, Porquet2019}. This has been recently applied also to a bright quasar at intermediate redshift \citep{Kammoun2023}. Further evidence of a small physical size of the corona emerges from varying obscuration of the corona by clouds in the broad line region \citep[e.g.,][]{Risaliti2011, Sanfrutos2013, Gallo2021}. These results motivated the adoption of the lamp-post geometry to describe the disk–corona system, where the X-ray corona is assumed to be a point-like source located on the symmetry axis of the disk \citep[e.g.,][]{Matt1991, Martocchia1996, Miniutti2004}. Such a geometry could be physically realized by collisions and shocks within an ejection flow or a failed jet \citep[e.g.,][]{Henri1991, Henri1997, Ghisellini2004}, and has been assumed in detailed models for the calculation of reflection spectra \citep[e.g.,][]{Dauser2013, Garcia2014, Niedzwiecki2016, Ingram2019, Mastroserio2021,Dovciak2022}. We note that recently several works are considering a variety of more realistic 3D coronal geometries \citep[e.g.,][]{Wilkins2016, Gonzalez2017, Zhang2019}.

Several works have adopted an approach similar to the one of \citep{Soltan1982} to constrain the radiative efficiency ($\eta$) of accreting SMBHs \citep[e.g.,][]{Fabian1999, Yu2002, Marconi2004, Shankar2004, Shankar2009, Shankar2020, Raimundo2009, Raimundo2012, Aversa2015, Zhang2017}. This cosmology-independent approach enables constraining the radiative efficiency by comparing the energy density of quasar/AGN radiation with the local mean SMBH mass density. All these works resulted in constraining $\eta$ to be larger than 0.1. More recently, \cite{Shankar2020} favor $\eta \sim 0.12-0.2$ which is in line with spinning SMBH, suggesting an average spin value larger than $\sim 0.5$. For these spin values the half of the accretion disk emission is produced within the inner $20\,r_{\rm g}$ \citep[see e.g.,][]{Agol2000, Dovciak2022}. The inner edge of the accretion disk is thus required to extend close to the black hole. Consequently, the bulk of the emission should be emitted from a compact region, which is in agreement with the results obtained from X-ray observations.

\subsection{The heating and cooling mechanisms of the X-ray corona}

While the heating mechanism of the corona is still uncertain, magnetic reconnection has been suggested to play an important role \citep[e.g.,][]{Galeev1979, DiMatteo1997, Merloni2001a, Merloni2001b, Sironi&Beloborodov20, Sridhar+21, Sridhar+23, Gupta+23}. Within this scenario, the magnetized coron\ae\ could generate cyclo/synchrotron radiation that can be observed in the radio/mm band \citep[e.g.,][]{Laor2008, Inoue2014, Panessa2019}. In fact, low levels of radio emission has been almost ubiquitously detected in radio-quiet AGN. In many cases, the emitting region is unresolved and associated with a compact, sub-kpc nuclear region \citep[see e.g.,][and references therein]{Smith2016, Panessa2019}. Given the expected physical size of the X-ray emitting region, the coronal mm-wave synchrotron emission is expected to be self-absorbed, and it would therefore be more easily detectable in the mm than in the radio. The size ($R$) of the self-absorbed synchrotron source decreases with frequency following $R \propto \nu^{-7/4}$. Thus, the synchrotron emission from an X-ray corona sized source would peak at $\sim 100\,\rm GHz$ \citep[e.g.,][]{Raginski2016, Inoue2018}. Several observational studies hint towards a coronal origin of the mm-wave nuclear emission in non-jetted AGN \citep[e.g.,][]{Behar2015, Behar2018, Doi2016, Inoue2018, Kawamuro2022, Kawamuro2023, Petrucci2023}. In particular, correlations between mm continuum emission and X-ray emission have been found in various samples of AGN, with an average ratio between the $\sim 100-200\,\rm GHz$ and X-ray continuum of $\sim 10^{-5}-10^{-4}$ \citep{Behar2015, Behar2018, Kawamuro2022, Ricci2023}. Interestingly, this relation is consistent with what has been observed in coronally active stars \citep{Guedel1993}, which are magnetically heated, similar to what is expected for AGN coron\ae, further supporting the proposed coronal origin for the $100-200\,\rm GHz$ continuum emission, and in turn the idea that the corona is magnetically heated.

A physically compact X-ray corona is expected to be radiatively compact, meaning that the luminosity to radius ratio ($L/R_{\rm c}$) is large. The compactness is usually described by the dimensionless parameter $\ell = \frac{L}{R_{\rm c}} \frac{\sigma_{\rm T}}{m_{\rm e}c^3}$, where $\sigma_{\rm T}$ is the Thompson scattering cross section and $m_{\rm e}$ is the electron mass \citep{Guilbert1983}. For large values of $\ell$, the number density of the high-energy photons can lead to electron-positron pair production due to photon-photon collisions. In this case, feeding the corona with more energy will lead to the production of more particles to share the energy, which will limit the increase of the temperature of the corona and prevent pair production from becoming a runaway process and exceed annihilation. This can be assimilated to an $\ell$-dependent thermostat \citep{Svensson1984, Zdziarski1985, Stern1995}. Observational constraints on the coronal parameters so far obtained are in agreement with the hypothesis that pair production and annihilation may act as an effective thermostat controlling the coronal temperature \citep{Fabian+15, Fabian2017}. It is common to assume that the coronal plasma is thermal. However, several AGN show electron temperatures that are much lower than expected for a purely thermal pair plasma \citep[see e.g.,][]{Ursini2016, Balokovic2015, Reeves2017, Kara2017, Tortosa2018, Bertola+22}. These low temperatures could be indicative of a hybrid plasma composed of a mixture of thermal and non-thermal particles \citep{Zdziarski1993, Ghisellini1993,Fabian2017}. Only a small fraction of non-thermal electrons with energy above 1\,MeV would be needed to result in runaway pair production. In this case, the cooled pairs share the energy which reduces the temperature of the corona.

It is common to model the Comptonization spectrum using a phenomenological power law with a high energy cutoff. Various physical models are also used and all of them describe the current data equally well. These models differ in their assumptions for the spectrum of the seed photons and the geometry of the Comptonizing region. Various analytic Comptonization models such as {\tt CompTT} \citep{Titarchuk1994} and {\tt CompPS} \citep{Poutanen1996} are used, allowing different simple geometries. Recently, more physically self-consistent calculations such as {\tt EQPAIR} \citep{Coppi1999}, {\tt BELM} \citep{Belmont2008}, {\tt MoCa} \citep{Tamborra2018}, and {\tt MONK} \citep{Zhang2019} include a variety of geometries, hybrid coron\ae, and polarization.

\subsection{Observational constraints}

The unprecedented sensitivity of \nustar\ \citep{Harrison2013} above 10\,keV has been instrumental to study the hard X-ray spectra of AGN. \nustar\ enabled high-energy cut-off measurements from single epoch observations for a relatively large sample of local \citep{Ballantyne2014, Matt2014, Balokovic2015, Balokovic2020, Ricci2017, Tortosa2018, Akylas2021, Kamraj2022, Tortosa2023} and intermediate-redshift AGN for the first time \citep[see e.g.,][and references therein]{Kammoun2017, Kammoun2023,Lanzuisi2019, Bertola+22, Marinucci2022}. It should be mentioned that previous studies (with \textit{BeppoSAX, RXTE, INTEGRAL}, etc.) allowed these measurements for a much smaller sample of the brightest nearby AGN \citep[e.g.,][]{Nicastro2000,Petrucci2004, Malizia2014, Lubinski2016}. \nustar\ observations allowed us not only to constrain the coronal temperature by accurately measuring values of \ecut, but also to probe its variability \citep[e.g.,][]{Ballantyne2014, Zoghbi2017} the origin of which remains unclear.

The \swift/BAT AGN sample comprises all AGN detected by the all-sky $14-195$\,keV BAT survey \citep[][]{Baumgartner2013, Oh2018}. \cite{Ricci2017} have estimated a median high-energy cutoff of local unobscured AGN (i.e., $\log \left(N_{\rm H}/\rm cm^{-2}\right) < 22$) of $E_{\rm cut} = 210 \pm 36\,\rm keV$. \cite{Ricci2018} subsequently found evidence for an inverse correlation between \ecut\ and Eddington ratio ($\lambda_{\rm Edd}$) where the median \ecut\ increases from $ E_{\rm cut} = 160 \pm 41\,\rm keV$ to $ E_{\rm cut} = 370 \pm 51\,\rm keV$ considering AGN with $\lambda_{\rm Edd}$ larger and smaller than 0.1, respectively. This could be suggestive that the properties of the X-ray corona evolve with the accretion rate. More recently, \citet{Kamraj2022} included \nustar\ observations of this sample and estimated the mean electron temperature to be $kT_{\rm e} = 84 \pm 9\,\rm keV$, which is consistent with the values reported previously in the literature. However, the authors were not able to recover the correlation between the electron temperature and any of the accretion parameters. It is worth mentioning that differences in coronal spectra may be expected between obscured and unobscured AGN, within the framework of the orientation-based unified model of AGN \citep{Antonucci1993, Urry1995}, especially if the corona has a net velocity perpendicular to the accretion disk \citep[most likely related to jet formation; see e.g.,][]{Malzac2001, Markoff2005}. However, \cite{Balokovic2020} estimated a median $ E_{\rm cut} = 290 \pm 20\,\rm keV$ for a sample of 130 local obscured AGN, consistent within the current large uncertainties with the values found by \citet{Ricci2017} for unobscured AGN. 

More recently, the launch of the {\em Imaging X-ray Polarimetry Explorer} \citep[\ixpe;][]{Weisskopf2022} has enabled studies of X-ray polarization in AGN for the first time (in the $2-8\,\rm keV$ range). X-ray polarization is a powerful technique to explore the geometry of the X-ray emitting region \citep[see e.g.,][]{Schnittman2010, Beheshtipour2017, Tamborra2018, Zhang2019}. So far, a total of four non-blazar AGN have been observed (NGC\,4151, MCG--05-23-16, the Circinus galaxy, and IC\,4329A). Only upper limits could be determined for MCG--05-23-16 \citep{Marinucci2022, Tagliacozzo2023} and IC\,4329A \citep{Ingram2023}, which are not conclusive on the geometry of the X-ray source. The Circinus galaxy shows a polarization of $28 \pm 7$ per cent (at 68 per cent confidence level), with a polarization angle of $18^\circ \pm 5^\circ$, roughly perpendicular to the radio jet \citep{Ursini2023}. Given the large obscuration in this source, this polarization is most likely due to reflection off cold material in the torus \citep[e.g.,][]{Ghisellini1994, Goosmann2011, Marin2018, Ratheesh2021} and cannot be interpreted as the intrinsic polarization of the X-ray source. As for NGC\,4151, a polarization of $4.9\pm 1.1$ per cent with a polarization angle of $86^\circ \pm 7^\circ$ (at the 68 per cent confidence level) are reported by \cite{Gianolli2023}. This is consistent with a radially extended source of polarization that is perpendicular to the radio jet, i.e., parallel to the accretion disk. Similar results have also been observed in the black hole X-ray binary Cygnus X-1 \citep{Krawczynski2022}.

The {\em High-Energy X-ray Probe} (\hexp; Madsen et al., 2023, in preparation) will address many of the open questions related to the physics of the X-ray corona in AGN. In this paper, we showcase how \hexp\ will play a major role in advancing our understanding of the physics of the X-ray coron\ae\ in AGN. In Section\,\ref{sec:hexp} we briefly present the properties of the mission. Characterisations of the the X-ray coron\ae\ in unobscured and obscured AGN with \hexp\ are discussed in Section\,\ref{sec:unobscured} and Section\,\ref{sec:obscured}, respectively. Section\,\ref{sec:high-z} shows how studying the X-ray coron\ae\ in high-redshift quasars will be possible with \hexp. Section\,\ref{sec:CLAGN} addresses the capabilities of \hexp\ in studying the X-ray coron\ae\ of transient sources. Section\,\ref{sec:geometry} demonstrates how \hexp\ will allow us to better constrain the coronal geometry. Finally, we present a summary and a brief discussion in Section\,\ref{sec:discussion}.

\section{Mission design}
\label{sec:hexp}

\hexp\ is a probe-class mission concept that offers sensitive broad-band coverage ($0.2-80\, \rm keV$) of the X-ray spectrum with exceptional spectral, timing, and angular capabilities. It features two high-energy telescopes (HETs) that focus hard X-rays, and one low-energy telescope (LET) that focuses lower energy X-rays.

The LET consists of a segmented mirror assembly coated with Ir on monocrystalline silicon that achieves a half power diameter of $3.5''$, and a low-energy DEPFET detector, of the same type as the Wide Field Imager \citep[WFI;][]{Meidinger2020} onboard {\it Athena} \citep{Nandra2013}. It has $ 512 \times 512$ pixels that cover a field of view of $11.3' \times 11.3'$. It has an effective passband of $0.2-25$\,keV, and a full frame readout time of 2\,ms, which can be operated in a 128 and 64 channel window mode for higher count-rates to mitigate pile-up and faster readout. Pile-up effects remain below an acceptable limit of $\sim 1\%$ for fluxes up to $\sim 100$\,mCrab in the smallest window configuration. Excising the core of the PSF, a common practice in X-ray astronomy, will allow for observations of brighter sources, with a typical loss of up to $\sim 60\%$ of the total photon counts.

The HET consists of two co-aligned telescopes and detector modules. The optics are made of Ni-electroformed full shell mirror substrates, leveraging the heritage of \xmm\ \citep{Jansen2001}, and coated with Pt/C and W/Si multilayers for an effective passband of $2-80$\,keV. The high-energy detectors are of the same type as flown on \nustar\ \citep{Harrison2013}, and they consist of 16 CZT sensors per focal plane, tiled $4 \times 4$, for a total of $128 \times 128$ pixels spanning a slightly larger field of view than the LET, $13.4' \times 13.4'$.

The broad X-ray band pass and superior sensitivity will provide a unique opportunity to study the X-ray corona physics across a wide range of energies, luminosity, and dynamical regimes.

\subsection{Simulations}
\label{sec:simulations}

All the simulations presented here were produced with a set of response files that represent the observatory performance based on current best estimates (see Madsen et al., 2023, in preparation). The effective area is derived from a ray-trace of the mirror design including obscuration by all known structures. The detector responses are based on simulations performed by the respective hardware groups, with an optical blocking filter for the LET and a Be window and thermal insulation for the HET. The LET background was derived from a GEANT4 simulation \citep{Eraerds2021} of the WFI instrument, and the HET background was derived from a GEANT4 simulation of the \nustar\ instrument; both simulations assume \hexp\ is orbiting at the first Earth-Sun Lagrangian point\,(L1). 

\section{Physics of the corona in unobscured AGN}
\label{sec:unobscured}

Unobscured AGN ($\log \left( N_{\rm H}/{\rm cm^{-2}} \right) < 22$) with their clean view of the central engine have provided most of the knowledge we have today regarding AGN coron\ae, as outlined in Section\,\ref{sec:intro}. The view of the whole Comptonization spectrum, and of the X-ray reflection arising from the interaction with the disk, is what makes X-ray observations of unobscured AGN so valuable. The constraints provided on these sources are key as we move towards refining our picture of AGN coron\ae\ with \hexp. 

\nustar\ observations of unobscured AGN have provided many coronal temperature measurements, which are consistent with the runaway pair production scenario discussed above. Interestingly, in some AGN the coronal temperatures have been found to be much lower than what would be expected for pair-regulated thermal coron\ae. One likely explanation for this is that the energy distribution of the electrons in the corona is not entirely thermal but has a non-thermal contribution. Such a hybrid corona would have a lower observed coronal temperature as the non-thermal high energy electrons create pairs even when the bulk of the electrons have lower temperatures. 

The hybrid nature of the electron distribution can be detected directly from the X-ray spectrum. Comptonization from thermal electrons results in a power-law like X-ray continuum with a fairly abrupt cut-off at high energies. Consequently, an addition of non-thermal electrons, following a power-law like distribution beyond the thermal peak, leads to an increase in hard X-rays, as there are now electrons with even higher energies. This additional hard X-ray emission beyond the thermal cut-off is usually referred to as a ``hard tail". 

Using these differences in the resulting Comptonization spectrum, i.e., the X-ray continuum, some evidence has been found in black hole X-ray binaries that the X-ray continuum is caused by hybrid Comptonization \citep[e.g.,][]{Cadolle2006, Cangemi2021, Zdziarski2021}. The presence of hybrid Comptonization in AGN has been widely hypothesized but a direct detection of the hard tail itself is beyond the capabilities of current and proposed future instruments. \hexp, however, will allow indirect confirmation of the presence of hybrid Comptonization in AGN and determine the strength of its contribution.

\begin{figure}[htbp]
\begin{center}
\includegraphics[width=0.49\textwidth]{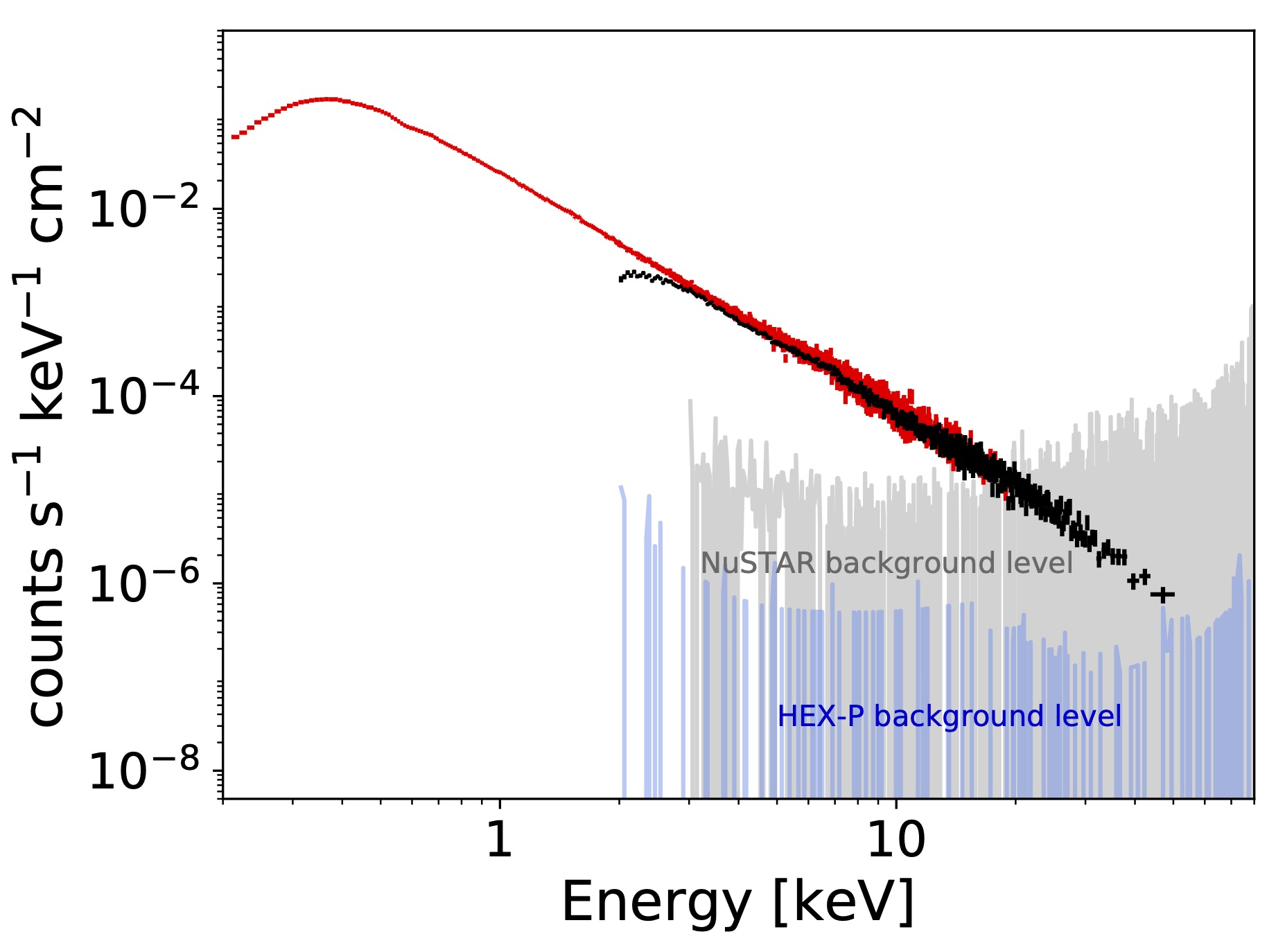}
\includegraphics[width=0.49\textwidth]{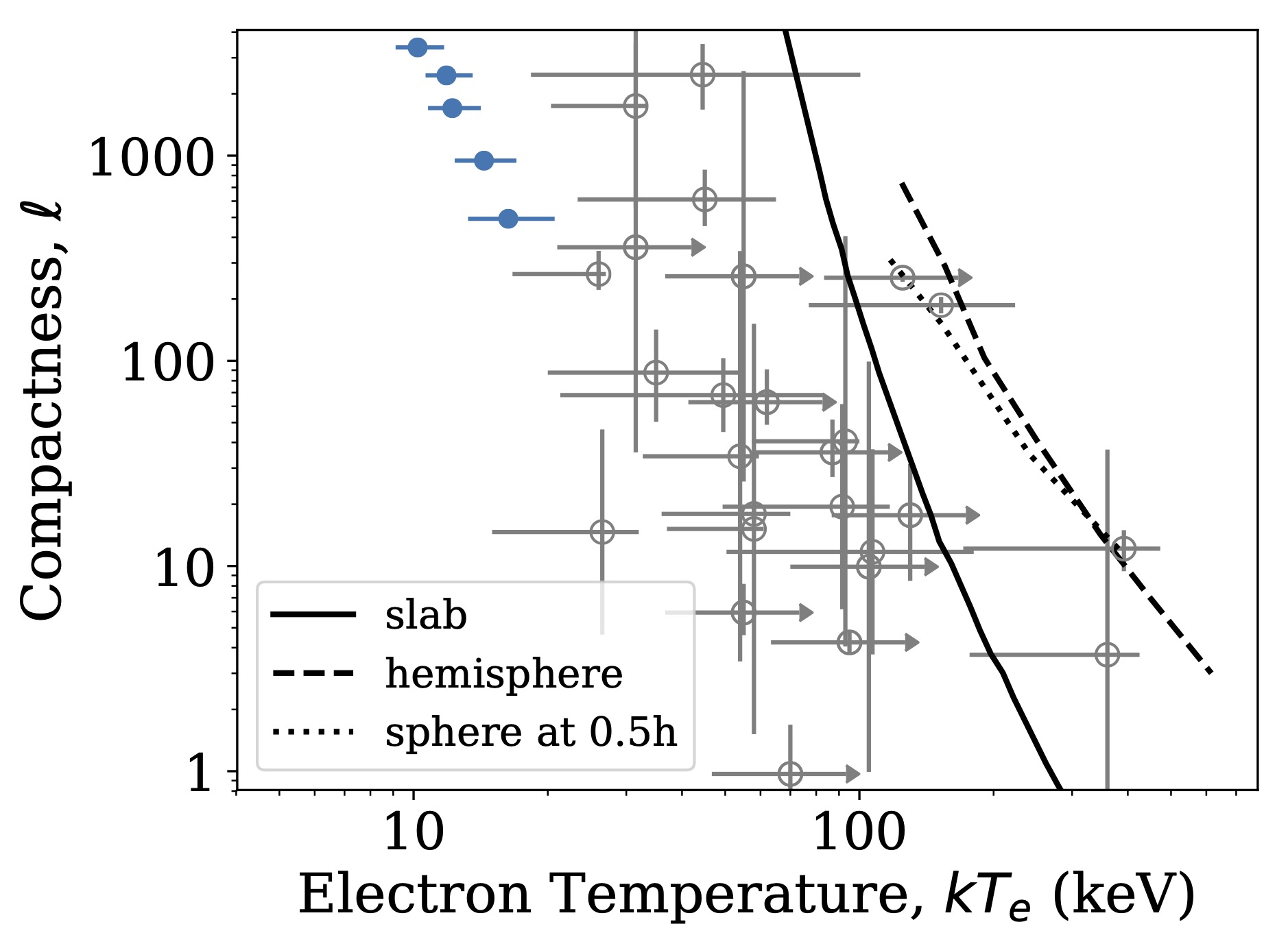}
\end{center}
\caption{{Left:} Simulated LET (red) and HET (black) spectra of Ark\,564, assuming $\Gamma = 2.48$, $kT_{\rm e} = 13\,\rm keV$ and an exposure time of 90\,ks. The figure also shows the background levels of \textit{NuSTAR} (grey) and \textit{HEX-P}-HET (blue):  the background of \textit{HEX-P} is significantly lower than that of \textit{NuSTAR}. {Right:} The blue circles correspond to simulated observations of Ark\,564 with \hexp\ at five different flux levels and five different coronal temperatures from 15 to 10\,keV (see text). For clarity, uncertainities in the the compactness of Ark\,564 are not plotted. The grey empty circles represent typical estimates of $kT_{\rm e}$ from other AGN performed with \nustar.}\label{fig:ltheta}
\end{figure}

There are two primary methods to indirectly detect hybrid Comptonization: one relies on the reflection spectrum of the high energy continuum \citep[][see also Section\,\ref{sec:discussion}]{Garcia2015}, and the other tests temperature regulation by electron-positron pair production. The latter case can be made by studying the evolution of a source with unusually low coronal temperature over a period of time. If the source is variable, we can study the evolution of its compactness and coronal temperature with time. If the coronal temperature is set by pair production and the accretion geometry remains unchanged\footnote{This can be assessed by studying the X-ray reflection spectrum.} then as the source flux increases, the coronal temperature will decrease. If, on the other hand, the coronal temperature is not regulated by pairs then the temperature may not change or change in a different fashion. If we see the expected trend of lower coronal temperature with increased flux, and if the observed coronal temperatures are much lower than expected for thermal Comptonization, that would be strongly indicative of hybrid Comptonization. The findings can then also be compared to theoretical predictions. In rare cases, this is already possible with today's instrumentation for some of the brightest X-ray binaries \citep[see][for an example]{Buisson2019}, but has remained out of reach for AGN. For the observations to yield meaningful constraints our measurements need to be more accurate at low fluxes than is possible with current instruments. However, \hexp\ is able to obtain much tighter constraints on the coronal temperatures due to its low background at high energies (Figure~\ref{fig:ltheta}, left), enabling a high-quality study of coronal temperature variations in AGN for the first time. 

To demonstrate this, we explored the feasibility of a study with \hexp\ that targets the evolution of the coronal temperature in the bright, local narrow-line Seyfert 1 galaxy Ark\,564. Ark\,564 is ideally suited for such a study as it exhibits a low measured coronal temperature \citep{Kara2017,Lewin2022} as well as observed coronal temperature and flux variations \citep{Barua2020}. Our \hexp\ simulations of Ark 564 are based on the absorbed 2-10 keV fluxes observed with \textit{RXTE} over a 4-year period ranging from 1999 to 2003 and our simulations are further assuming a scenario where the coronal size remains unchanged with luminosity$\footnote{Our final conclusions do not depend on this assumption, as the coronal size can also be estimated from the spectral modeling.}$. During this period the observed 2-10\,keV flux varies between $1.0-5.0\times10^{-11}\rm\,erg\,s^{-1}\,cm^{-2}$ \citep{Rivers2013}. We simulate spectra at five evenly spaced levels within the observed flux range assuming a model that consists of a thermal Comptonization continuum and its associated reflection spectrum, both of which are treated with the relativistic reflection model \texttt{relxilllpcP} \citep{Dauser2020}. The second lowest flux level of $2.0\times10^{-11}\rm\,erg\,s^{-1}\,cm^{-2}$ corresponds very well to the 2018 joint \xmm+\nustar\ observations of the source, which were studied in detail in \citet{Lewin2022}. We therefore adopt the spectral parameters from their detailed modeling for this flux level (Table 3 in their paper), in particular they have determined that $\Gamma=2.48$ and $kT_e=14.7$\,keV at this particular flux level. Then as we vary the flux level, we not only adjust the normalization of our spectral model but also evolve the photon index from 2.40 to 2.56 to mimic a softer-when-brighter effect and the coronal temperature from 15.1\,keV to 10.0\,keV to have the coronal evolve mildly with flux, the reflection parameters are kept unchanged. To determine how much exposure time was needed at each flux level, we simulated spectra with increasingly longer exposure times until we were able to recover the coronal temperature with an accuracy of $\pm 2.5\,$keV or better at 90\% confidence. From the simulations, we found that a study of Ark\,564 with \hexp\ with a total exposure time of 445\,ks targeting the source at different flux levels will constrain the coronal temperature to $\leq 2.5$\,keV and could yield important data points in the compactness-temperature plane (blue points, Figure~\ref{fig:ltheta}, right). At higher flux levels it is easy for us to exceed the accuracy requirement of 2.5\,keV and the error bars are therefore smaller. We note here that accuracy of the coronal temperature constraint for a certain simulated exposure time is model-dependent, where more complex models lead to less accurate constraints.

To place the findings from the Ark\,564 study into the context of the wider population of radio-quiet, unobscured AGN, ideally a few similar studies of other sources would also be conducted. Suitable study targets display large variability on timescales of weeks to months and possess a low coronal temperature, several of such sources have already been identified. Together, with the findings from one-time observation of unobscured AGN and high redshift targets, this will highlight whether Ark\,564 is special in any way or whether its coronal variations can be considered representative for the population as a whole.

The much lower background of \hexp\ at higher energies and the therefore much improved constraints on the spectral shape in this energy range with shorter exposure times will likely not only lead to a better understanding of hybrid Comptonization, but also to a more accurate spectral decomposition of unobscured AGN. This aspect is studied in more detail in Piotrowska et al 2023.


\section{Physics of the corona in obscured AGN}
\label{sec:obscured}

\begin{figure}[htbp]
\begin{center}
\includegraphics[width=0.99\linewidth]{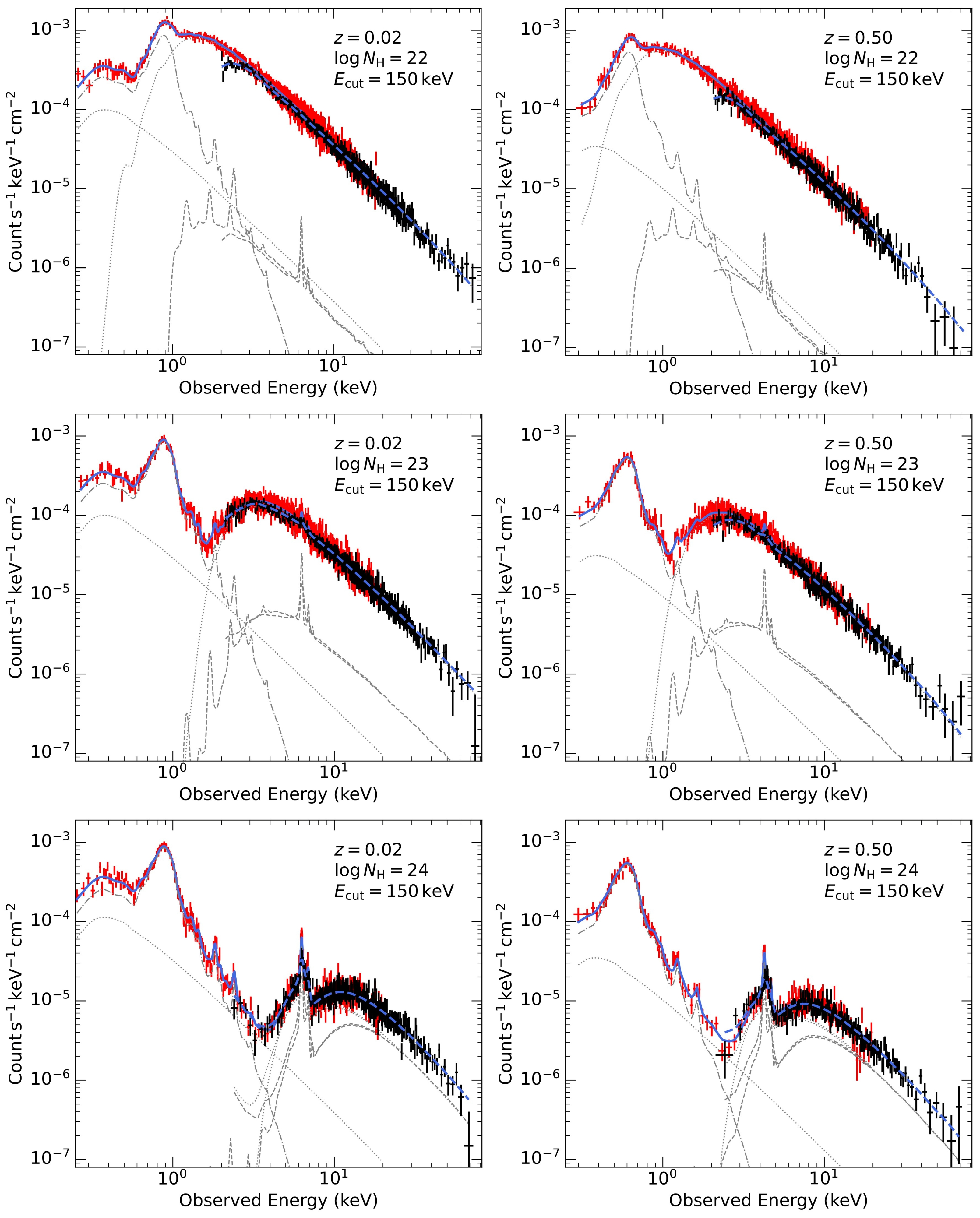}

\end{center}
\caption{Simulated LET (red) and HET (black) spectra of an absorbed AGN at $z=0.02$ (left) and $z=0.5$ (right), assuming $\log \left( N_{\rm H}/\rm cm^{-2} \right) = 22, 23, 24$ (top to bottom), $\Gamma = 1.9$, $E_{\rm cut} = 150\,\rm keV$, and an exposure time of 100\,ks. The grey dotted lines represent the power law components, the grey dashed lines represent the neutral reflection component, and the grey dash-dotted lines represent the thermal component.}\label{fig:obscured_spectra}
\end{figure}

Within the framework of the orientation-based unified model for AGN \citep{Antonucci1993}, differences in coronal spectra between obscured  ($\log \left( N_{\rm H}/{\rm cm^{-2}} \right) \geq 22$) and unobscured AGN could be expected if the corona has significant velocity perpendicular to the accretion disk, possibly related to jet formation \citep{Beloborodov+1999, Markoff2005,liu+2014}. Hard X-rays ($\geq 15\, \rm keV$) are essential to study the intrinsic properties of obscured AGN, as the obscuring material is optically thin to these photons. Constraints on coron\ae\ in obscured AGN remain extremely scarce in the literature despite the prevalence of obscured AGN both in the local universe and, particularly, at high redshift (50–75\% of the population; \citealt{hickox+alexander-2018}). A few constraints based on data from non-focusing hard X-ray instruments are available from, e.g., \bepposax\ \citep{dadina-2007}, \suzaku\ \citep{tazaki+2011}, \integral\ \citep{deRosa+2012}, \swiftbat\ \citep{Ricci2018}, or a combination thereof \citep{molina+2013}. However, the cutoff of the power-law continuum at high energies, \ecut, has been well constrained using \nustar\ data for only about a dozen bright, obscured AGN \citep[e.g.,][]{Balokovic2015,Fabian2017,buisson+2018,Tortosa2018, Ursini2019}, while larger samples representative of the obscured AGN population provide ensemble-level constraints mostly based on short \nustar\ observations with limited photon statistics \citep{Balokovic2020}. In addition to \ecut, we note that constraining the photon index in obscured AGN is crucial to study the intrinsic AGN spectrum and understand the accretion process.

\begin{figure}[htbp]
\begin{center}
\includegraphics[width=0.99\linewidth]{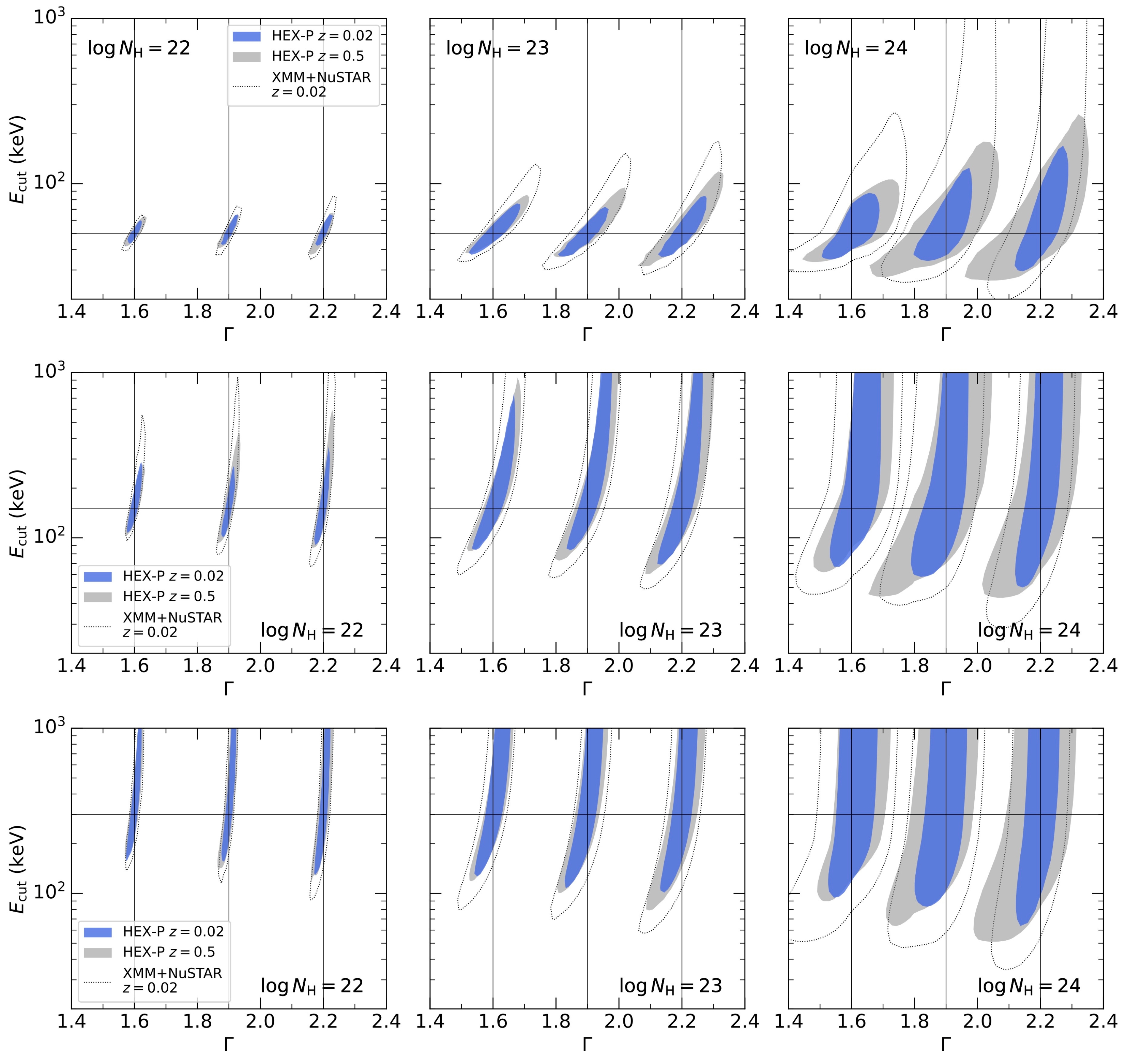}
\end{center}
\caption{$\Gamma-E_{\rm cut}$ contour plots ($3\sigma$ confidence level) for $z = 0.02$ and $0.5$ (blue and grey contours, respectively) assuming $\log \left( N_{\rm H}/\rm cm^{-2} \right) = [22, 23, 24]$ (left to right), $E_{\rm cut} = 50, 150,$ and $300\rm\,keV$ (top to bottom), and simulating a 100-ks \hexp\ observation. The black dotted contours correspond to the $3\sigma$ confidence level constraints for a source at $z=0.02$ with 30-ks exposure by \xmm\ combined with a 100-ks exposure by \nustar\ (including both modules, FPMA and FPMB). Horizontal and vertical lines indicate the input values of $\Gamma$ and \ecut, respectively. }\label{fig:obscured_contours}
\end{figure}


Thanks to its low background, and  broad-band energy coverage, \hexp\ will provide better constraints on the properties of the X-ray corona in obscured AGN. We performed simulations of various configurations of absorbed AGN to test the capabilities of \hexp\ in constraining coronal parameters, assuming uniform exposure times of 100\,ks (typical exposure for the planned \hexp\ surveys; see Civano et al., 2023). We used the {\tt borus02} model \citep{Balokovic2018}, which self-consistently computes the reprocessed emission from  a sphere with conical cutouts at both poles, approximating a torus with variable covering factor. We considered a torus with a half-opening angle of $45^\circ$ and an inclination of $60^\circ$. We also considered a photon index of the incident power law of $\Gamma = [1.6, 1.9,2.2]$, and a high-energy cutoff $E_{\rm cut} = [50, 150,300]\,\rm keV$. The incident luminosity was estimated to match the average values observed by \cite{Ricci2017} for $z = 0.02$ and $z = 0.5$, resulting in unabsorbed $2-10\,\rm keV$ luminosities in the range $10^{42.9-45.5}\,\rm erg\,s^{-1}$. We considered three values for the column densities $\log  \left( N_{\rm H}/\rm cm^{-2} \right) = [22, 23, 24]$. For simplicity, we assume a homogeneous torus with the line-of-sight column density equal to the equatorial value. In addition, the soft X-ray spectrum in obscured AGN is typically characterized by the emission from photoionized gas in the narrow line region. For simplicity, we include a soft {\tt APEC} component, which is commonly used in the literature to model this emission in absorbed AGN. For this component we assume a temperature of $0.9\,\rm keV$ and a normalization that is equal to $10\%$ that of the power law, as typically seen in local AGN \citep[see e.g.,][]{Kammoun2020}. The total model results observed $2-10\,\rm keV$ fluxes in the range $2.5-8.5\times 10^{-12}\,\rm erg\,s^{-1}\,cm^{-2}$. Figure\,\ref{fig:obscured_spectra} shows example \hexp\ spectra for an intermediate configuration for all the simulated $N_{\rm H}$ values, assuming  $\Gamma = 1.9$, and $E_{\rm cut} = 150\,\rm keV$, for $z = 0.02$ and $z=0.5$ (left and right panels, respectively). Figure\,\ref{fig:obscured_contours} shows the \ecut-vs-$\Gamma$ contours at the $3\sigma$ confidence level for all the configurations considered. \hexp\ accurately measures the photon index over a broad range of parameters. The uncertainty on $\Gamma$ is larger for higher column densities, since increasing amounts of absorption limits the access to the power law portion of the continuum to increasingly hard X-rays. This yields larger uncertainties on \ecut\ as well. The uncertainty on \ecut\ also increases with increasing \ecut; only lower limits could be estimated for $E_{\rm cut} = 300\,\rm keV$. We note that the observed flux decreases by a factor of $\sim 3$ between $z=0.02$ and $z=0.5$, which lead to an increase in the uncertainty on the parameters higher redshift. For comparison with current facilities, we simulate a joint 30-ks \xmm\ plus 100-ks \nustar\ (accounting for both modules, FPMA and FPMB) observation, which are typical exposures with these facilities. For all cases, we adopt $z = 0.02$ as it results in tighter constraints than the $z=0.5$ case. The $\Gamma-E_{\rm cut}$ constraints are shown as black dotted contours in Fig.\,\ref{fig:obscured_contours}, demonstrating the improvements enabled by \hexp. We note that, for the flux levels of these sources and the exposure time, many of these sources will be easily detected in the \hexp\ surveys (see Civano et al., 2023).

We would like to note that our {\tt borus02} model is setup in coupled mode in which the line-of-sight and global column densities scale together. The reflection component will be higher at higher line-of-sight column densities, possibly contributing to the increased uncertainties in the higher $N_{\rm H}$ regime that we show in Figure~\ref{fig:obscured_contours}. To definitively test this would require knowledge of the dependence between global column density and a wide range of different physical AGN parameters as well as a wide range of different model prescriptions for the obscurer, which is outside the scope of the current work. It is also worth mentioning that near and above the Compton-thick regime, the measurement of the energy cut-off is degenerate with the assumed obscurer model. Different obscurer geometries have different Compton-scattering behaviour, altering the reprocessed spectrum up to 400\,keV \citep[e.g.,][]{Buchner2019uxclumpy,Buchner2021physicalobscurermodels}. Thus the obscurer geometry and the energy cut-off are degenerate. Merely assuming an obscurer geometry may bias the energy cut-off measurement. Here, for simplicity, a fixed obscurer model ({\tt borus02}) was used for generating and modeling the data. The complex task of inferring the true obscurer geometry and the true energy cut-off thus would have to be achieved simultaneously, but this is  beyond the scope of this paper.  Boorman et al. (in prep.) discusses \hexp's  ability to infer torus geometries in heavily obscured AGN in more detail.


\section{Physics of the corona in intermediate-/high-redshift quasars}
\label{sec:high-z}

\begin{figure} [htbp]
\begin{minipage}[b]{.5\textwidth}
\centering
\includegraphics[width=.98\textwidth]{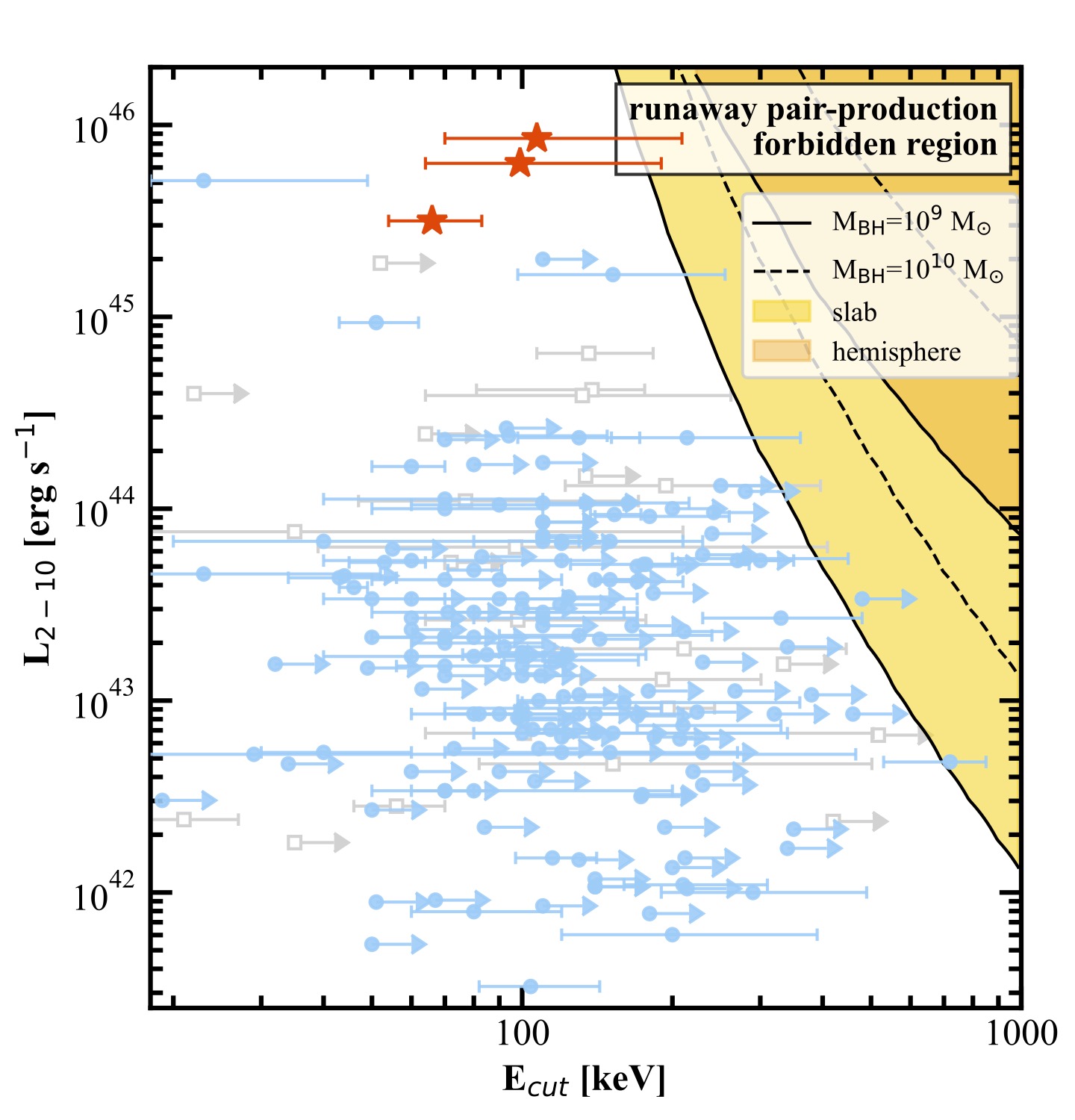}
\end{minipage}
\begin{minipage}[b]{.5\textwidth}
\centering
\includegraphics[width=.98\textwidth]{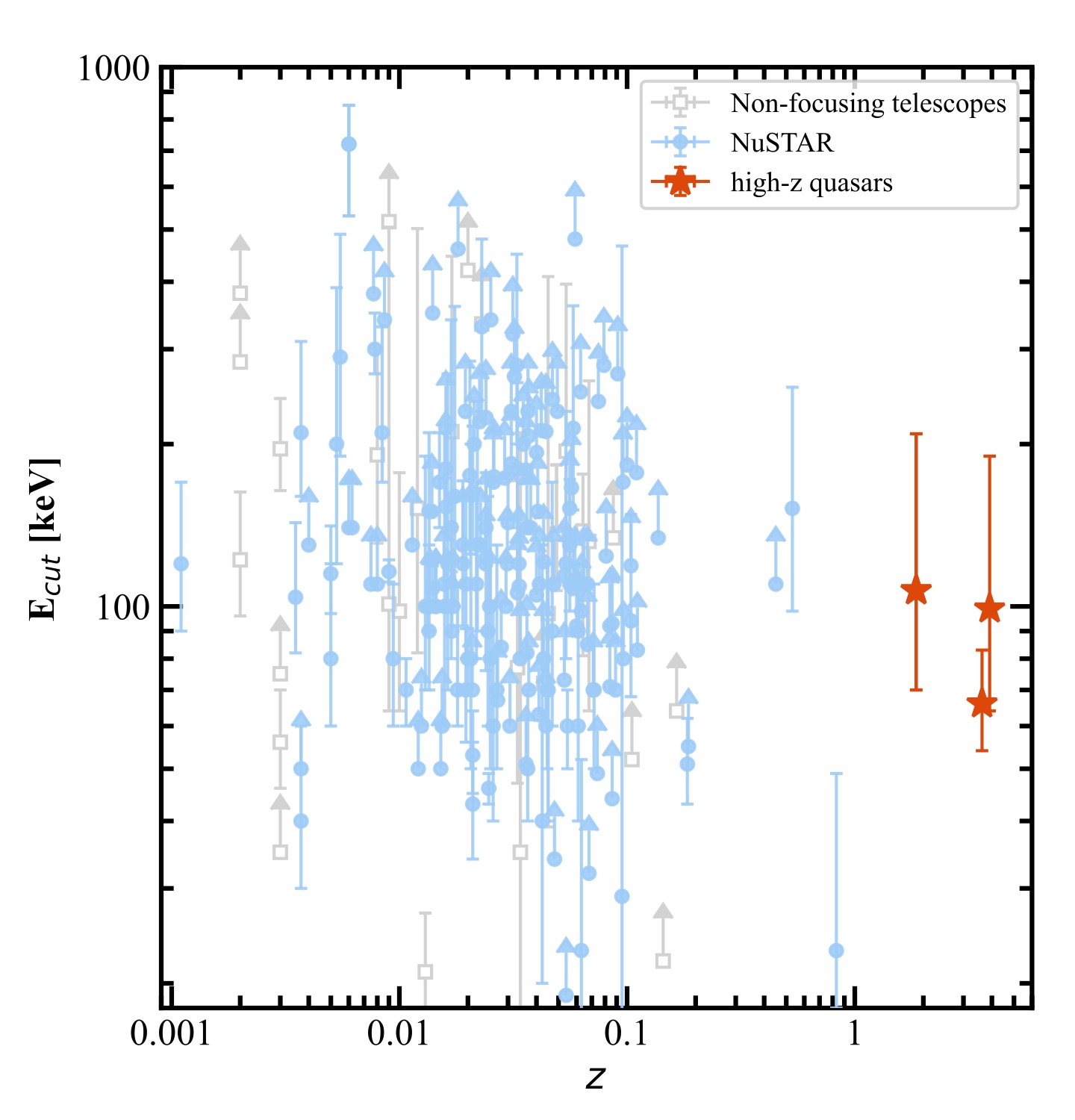}
\end{minipage}
\caption{High-energy cutoff of AGN as a function of 2--10\,keV luminosity (left) and redshift (right) by \nustar\ (light blue circles) and non-focusing telescopes (gray squares), based on \citet{Bertola+22}. The pair-production forbidden regions assuming different coronal geometries are plotted in orange (hemisphere) and yellow (slab). So far, \ecut\ was only measured in three quasars at $z>1$ (red stars). \ecut\ is not constrained for most of the low-redshift and low-luminosity AGN.}\label{fig:high_z_lum}
\end{figure}

The high-energy cutoff in the X-ray spectrum (at tens to hundreds keV) can be directly linked to the properties of the corona. Thanks to \nustar, whose hard X-ray sensitivity is orders of magnitude better than previous missions, a large sample of AGN were observed with unprecedented data quality at hard X-rays. The majority of the observed sources are AGN with low luminosity \citep[$L_{\rm 2-10\,keV}$ = 10$^{42}$--10$^{44}$ erg\,s$^{-1}$; see e.g.,][]{Kamraj2018, Ricci2018,Balokovic2020}, so they are in a region of the $kT_{\rm e}-\ell$ plane where a broad range of $E_{\rm cut}$ is allowed within the framework of the runaway pair production hypothesis. This can be seen in the left panel of Figure~\ref{fig:high_z_lum} (we assume here $E_{\rm cut}$ = 2\,$kT_{\rm e}$, $R_{\rm c}\sim$10\,$r_{\rm g}$, and coronal luminosity $L_{\rm c}$ = $L_{\rm 0.1-200\,keV}$ $\sim$4 $\times$ $L_{\rm 2-10\,keV}$). Furthermore, most of these sources are in the nearby Universe ($z<$0.1, Figure~\ref{fig:high_z_lum}, right). Therefore, only lower limits to their cutoff energy were measured at 90\% confidence level due to the limited bandpass in the rest frame of local sources ($<$80\,keV), covered by \nustar. 

On the contrary, high-redshift ($z\ge1$) radio-quiet quasars (mostly with luminosity $L_{\rm 2-10\,keV}>10^{45}\rm \,erg\,s^{-1}$) are ideal sources to constrain high-energy cutoffs thanks to the limited allowed $E_{\rm cut}$ range on the $kT_{\rm e}-\ell$ plane (assuming $kT_{\rm e}$ is indeed limited by runaway pair production; see Figure\,\ref{fig:high_z_lum}, left) and the cosmological redshifting of $E_{\rm cut}$ to the \nustar\ observable band. Therefore, they constitute a promising sample to constrain the AGN coronal properties at high luminosity and test the runaway pair production theory. So far, only three $z>$1 quasars were observed with \nustar\ and had their high-energy cutoffs constrained \citep[][and Figure~\ref{fig:high_z_lum}, right]{Lanzuisi2019,Bertola+22}. All three quasars (with $L_{\rm 2-10\,keV}>3\,\times\,10^{45}\,\rm erg\,s^{-1}$) were found below the critical line on the $kT_{\rm e}-\ell$ plane as per pair-production models expectations, though note the large uncertainties for two of the sources which leave ambiguous results (Figure~\ref{fig:high_z_lum}, left). The  well constrained cutoff energies of the three high-$z$ quasars ($\lambda_{\rm Edd}>0.1$) were 1.5--2.5 times lower than the 160\,keV measured by \citet{Ricci2018} for sources with high Eddington ratio. Therefore, observing high-redshift quasars, where the coronal properties could be well constrained, is crucial to study the possible evolution of the corona with the accretion parameters such as the accretion rate and the SMBH mass.

Despite the recent progress in constraining the coronal properties of high-redshift, high-luminosity quasars, the observations are extremely time-consuming. A total of $\sim$375~ks \nustar\ and 150~ks XMM observations were spent on the above three sources. The long exposures required in these observations lead to a strong limitation in constraining AGN coronal properties statistically. 

\begin{figure}[htbp]
\begin{minipage}[b]{.5\textwidth}
\centering
\includegraphics[width=.99\textwidth]{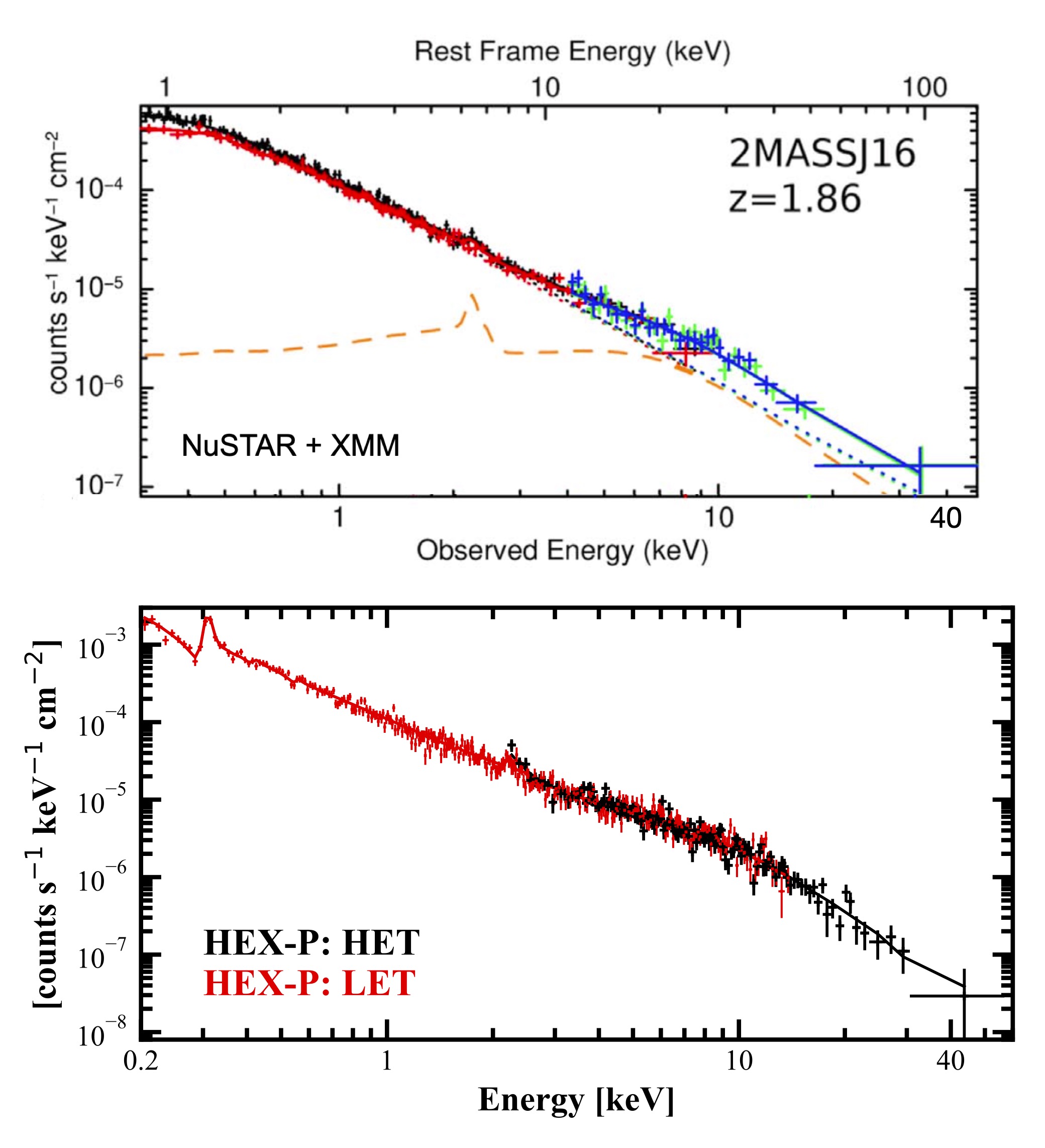}
\end{minipage}
\begin{minipage}[b]{.5\textwidth}
\centering
\includegraphics[width=.96\textwidth]{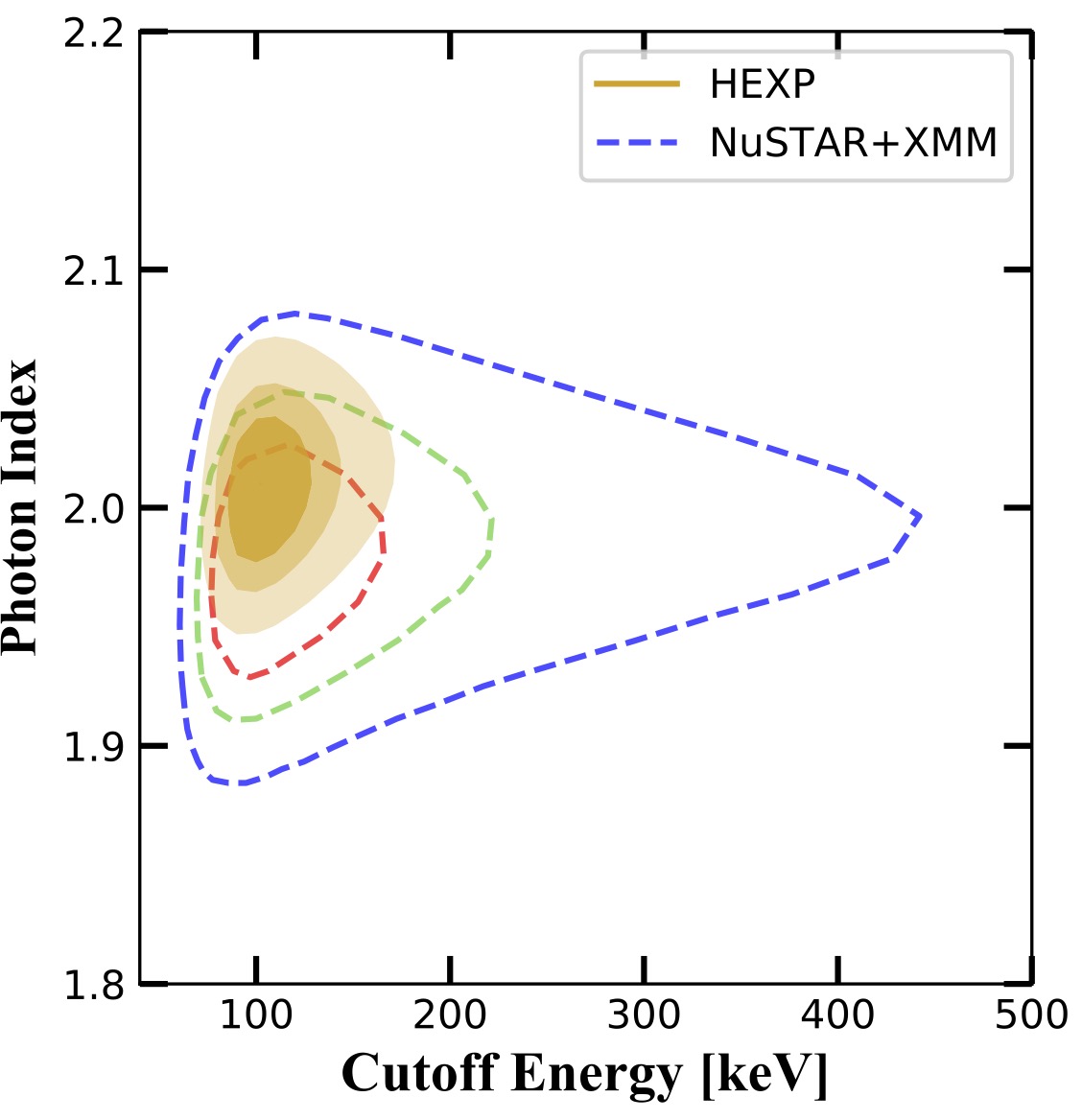}
\end{minipage}
\caption{{\it Left, upper panel:} 140~ks \nustar\ (blue) and 69~ks \xmm\ (red) spectra of 2MASS\,J1614346+470420 at $z=1.86$ \citep[adapted from ][]{Lanzuisi2019}. {\it Left, lower panel:} 140~ks \hexp\ simulation of the same quasar (HET in black, LET in red). {\it Right:} contours of \ecut\ and photon index of the \nustar+\xmm\ observations (red, green, and blue correspond to 68\%, 90\%, and 99\% confidence level, respectively) and \hexp\ simulations (dark yellow, yellow, and light yellow correspond to 68\%, 90\%, and 99\% confidence level, respectively).}\label{fig:simcontour}
\end{figure}  

\hexp, with larger effective area, lower background, and broad band coverage, is well suited to efficiently study the coronal properties of high-redshift, high-luminosity quasars. We compare the 140~ks \nustar + 69~ks \xmm\ observation of 2MASS\,J1614346+470420 ($z=1.86$) from \citet{Lanzuisi2019} to a simulated 140~ks \hexp\ observation in Figure\,\ref{fig:simcontour}.  Note that L1 orbit of \hexp\ implies that the \hexp\ observation would take half the total observatory time compared to the \nustar\ observation, since \nustar\ is in low-Earth orbit and does not re-point during Earth occultations. The 90\% confidence level constraints on \ecut\ are 106$_{-40}^{+100}$~keV from \nustar+\xmm\ and 103$_{-20}^{+30}$~keV from the simulated \hexp\ spectra, implying significantly improved constraints on \ecut\ by \hexp\ in the same exposure time and half the clock time. The high-energy cutoff and the photon index contours (at 68\%, 90\%, and 99\% confidence levels) derived from \nustar+\xmm\ and simulated \hexp\ spectra are plotted in the right panel of Figure\,\ref{fig:simcontour}. The contours show that \hexp\ can constrain \ecut\ at high confidence.

\begin{figure}[htbp]
\begin{center}
\includegraphics[width=1\linewidth]{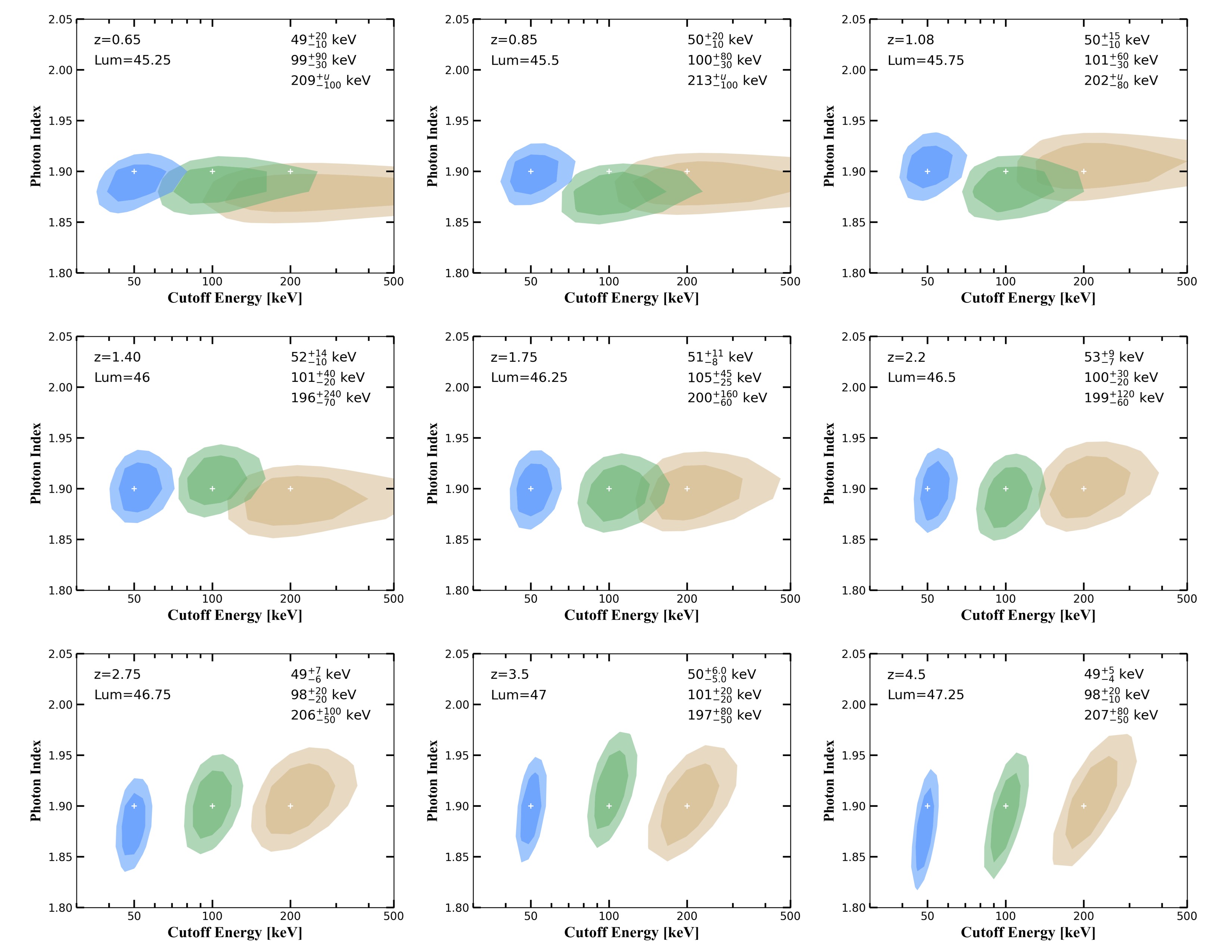}
\end{center}
\caption{$E_{\rm cut}-\Gamma$ contour plots (68\% and 90\% confidence levels) measured from the simulated spectra of the quasars with different redshifts ($0.65<z<4.5$), assuming an input 2--10\,keV flux of 1 $\times$ 10$^{-12}$ erg s$^{-1}$ cm$^{-2}$ and a \hexp\ exposure of 50~ks. We simulated three set of high-energy cutoff for each source, $E_{\rm cut}$ = 50, 100, and 200\,keV (blue, green, and yellow). The white crosses indicate the input values.The best-fit value of each spectrum with its 90\% confidence level uncertainty is indicated on the upper right corner of each panel.}\label{fig:high_z}
\end{figure}

\hexp\ will not only provide better constraints on AGN coronal properties but it will also efficiently probe AGN coron\ae\ in a large sample of high-redshift, high-luminosity quasars. We demonstrate the constraining power of \hexp\ for coronal cutoff energy measurements of intermediate-/high-redshift, high-luminosity quasars. We simulated nine \hexp\ observations of quasars with 50\,ks exposures assuming an input 2--10\,keV flux of $F_{\rm 2-10\,keV}$ = $10^{-12}$ erg s$^{-1}$ cm$^{-2}$. We note that cosmic X-ray background population synthesis models predict $\sim$ 200 quasars at $z>$1 with such flux level \citep[see e.g.,][]{Gilli07}. The input redshifts range goes from $z$ = 0.65 to $z$ = 4.5, so that their $2-10$\,keV intrinsic luminosity range over $\log \left( L_{\rm 2-10\,keV}/\rm erg\,s^{-1}\right) =$ 45.25--47.25. We simulated source spectra assuming a cutoff power law with a Galactic absorption of 10$^{20}$\,cm$^{-2}$ and photon index of $\Gamma$ = 1.90. We simulated three different high-energy cutoffs, $E_{\rm cut}$ = 50, 100, and 200\,keV, to explore the parameter space. Assuming the same supermassive black hole mass, log($M_{\rm BH}$/$M_{\odot}$) = 9.2 and a $2-10$\,keV luminosity to bolometric luminosity conversion factor of 20 \citep[e.g.,][]{Vasudevan2007, Lusso2012}, the Eddington ratios of the sample range over $\lambda_{\rm Edd}$ = 0.2--20. 

Figure~\ref{fig:high_z} shows the $E_{\rm cut}-\Gamma$ contours measured from the simulated spectra of the quasars with different redshifts and different high-energy cutoffs. The best-fit high-energy cutoffs of each simulated spectrum with its 90\% confidence level uncertainties are indicated on the upper right corner of each panel. As expected, $E_{\rm cut}$ is better constrained at higher redshift thanks to the cosmological redshifting, given the fact that we conserve the flux ($F_{\rm 2-10\,keV}$ = $10^{-12}$\,erg\,s$^{-1}$\,cm$^{-2}$) of each simulated source at different redshifts. Likewise, sources with $E_{\rm cut}$ = 50\,keV are better constrained than higher $E_{\rm cut}$ sources at all redshifts due to the band coverage of \hexp. We found that it is difficult to constrain the high-energy cutoff of intermediate redshifts ($z<$1.1) sources with $E_{\rm cut}$ = 200\,keV in a 50\,ks exposure. 

We tested the $E_{\rm cut}$ constrained by \hexp\ assuming different photon indices ($\Gamma$ = 1.60, 1.90, and 2.20) under a few redshifts ($z$ = 0.85, 1.40, 2.20, and 3.50) assuming the same flux as above. The results are reported in Table~\ref{Table:constraints}. We found that harder photon index provides better constraints on $E_{\rm cut}$ than softer photon index at all redshifts because more high energy photons are obtained for harder photon indices and a given $2-10\,\rm keV$ flux. The differences of the constraints on $E_{\rm cut}$ between softer and harder photon indices are much larger at lower redshifts than at higher redshifts. Nevertheless, \hexp\ could constrain the $E_{\rm cut}$ in most circumstances even for the sources with quite soft photon index, $\Gamma$ = 2.20.

\begingroup
\renewcommand*{\arraystretch}{1.25}
\begin{table*}
\small
\caption{Constraints on the high-energy cutoff with different photon indices under different redshifts.}
\centering
\label{Table:constraints}
\vspace{.3cm}
 \begin{tabular}{c|ccc|ccc|ccc}
    \hline
    \hline   
	$z$&$\Gamma$ = 1.6&$\Gamma$ = 1.9&$\Gamma$ = 2.2&$\Gamma$ = 1.6&$\Gamma$ = 1.9&$\Gamma$ = 2.2&$\Gamma$ = 1.6&$\Gamma$ = 1.9&$\Gamma$ = 2.2\\
	\hline
	&&50\,keV&&&100\,keV&&&200\,keV&\\
	\hline \\[-0.4cm]
	0.85&51$^{+12}_{-9}$&50$^{+17}_{-10}$&51$^{+42}_{-15}$&100$^{+42}_{-24}$&100$^{+78}_{-31}$&98$^{+104}_{-34}$&205$^{+237}_{-74}$&213$^{+u}_{-96}$&213$^{+u}_{-117}$\\[0.1cm]
	1.40&51$^{+9}_{-7}$&52$^{+14}_{-9}$&51$^{+23}_{-12}$&102$^{+30}_{-20}$&101$^{+40}_{-20}$&96$^{+70}_{-30}$&205$^{+140}_{-60}$&196$^{+240}_{-70}$&205$^{+u}_{-90}$\\[0.1cm]
	2.20&50$^{+6}_{-6}$&53$^{+9}_{-7}$&50$^{+11}_{-8}$&99$^{+23}_{-16}$&100$^{+30}_{-19}$&101$^{+39}_{-23}$&197$^{+80}_{-50}$&199$^{+120}_{-60}$&190$^{+175}_{-55}$\\[0.1cm]
	3.50&50$^{+6}_{-5}$&50$^{+6}_{-5}$&51$^{+7}_{-6}$&99$^{+20}_{-15}$&101$^{+20}_{-15}$&101$^{+23}_{-17}$&197$^{+80}_{-45}$&197$^{+80}_{-45}$&197$^{+80}_{-45}$\\[0.1cm]
	\hline
	\hline
\end{tabular}
\end{table*}
\endgroup

eROSITA \citep{Predehl2021} observations will provide an all-sky survey of X-ray sources, which will be the deepest survey in the soft X-rays ever performed. With the ongoing SDSS and DESI surveys, and future {\it Euclid}, 4MOST, Subaru Prime Focus Spectrograph, and {\it SPHEREx}  spectroscopic surveys, a large number of high-redshift, high-luminosity, radio-quiet AGN are expected to be discovered prior to the launch of \hexp. The combination of these large surveys should therefore provide a large sample of high-redshift quasars for \hexp\ coronal physics studies in the next decade.


\section{Physics of the corona in transient sources: Changing-look/changing-state AGN and Tidal Disruption Events}
\label{sec:CLAGN}

The growing number of time-domain surveys across the electromagnetic spectrum are revealing extreme variability in the accretion process onto supermassive black holes. These objects break from the standard stochastic variability seen in steadily accreting AGN, and thus provide a unique window into the formation, stability, and heating of the corona. 

Changing-look AGN (CLAGN; also often referred to as changing-state AGN) show rapid changes between optical spectral types through either the appearance or disappearance of broad emission lines on timescales of months to years \citep[for a recent review, see][]{Ricci2022}. These sources challenge the classic unified model for AGN, whereby the appearance of broad emission lines is solely a function of viewing angle. Instead, the change between optical spectral types is often coupled with an associated increase or decrease in the observed optical and X-ray flux, suggesting that the accretion rate is another key factor in AGN unification \citep[e.g.,][]{Elitzur2014}. The driving mechanism behind these changing-look events is still under debate, with theories including state transitions reminiscent of the behavior in black hole binaries \citep{Noda2018,Ruan2019}, radiation pressure disk instabilities \citep{Sniegowska2020}, propagating cooling fronts linked to changes in the magnetic torques in the innermost regions of the accretion flow \citep{Ross2018,Stern2018}, and transient events like tidal disruption events \citep[TDEs; e.g.,][]{Merloni2015,Ricci2020}.


Although CLAGN are still a relatively rare phenomenon with only about 100 such systems known to date, they provide unique insights into the dynamic behavior of the inner accretion flow and unveil physics that is impossible to probe in standard steady-state AGN. One extreme example is evidenced by the recent discovery of the disappearance and recreation of the X-ray corona in 1ES\,1927+654. In late 2017, the source underwent one of the fastest changing-look events to date, whereby broad emission lines were caught forming on timescales of months \citep{Trakhtenbrot2019}. Shortly after the start of the changing-look event, X-ray observations of 1ES\,1927+654 revealed negligible hard X-ray emission, indicating that the X-ray corona had vanished in this source \citep{Ricci2020,Ricci2021}. The corona began to reappear as the source brightened in the X-ray, although it remained extremely soft compared to standard AGN \citep[$\Gamma \approx 3-3.5$;][]{Ricci2021,Masterson2022}. Two competing theories have been suggested to explain the behavior in this kind of system: a TDE occurring in an AGN \citep{Ricci2020}, or an inversion in the magnetic flux polarity \citep{Scepi2021,Laha2022}. Both theories can effectively cut off the energy supply to the corona, leading to its destruction.  


\begin{figure}[htbp]
\begin{center}
\includegraphics[width=\textwidth]{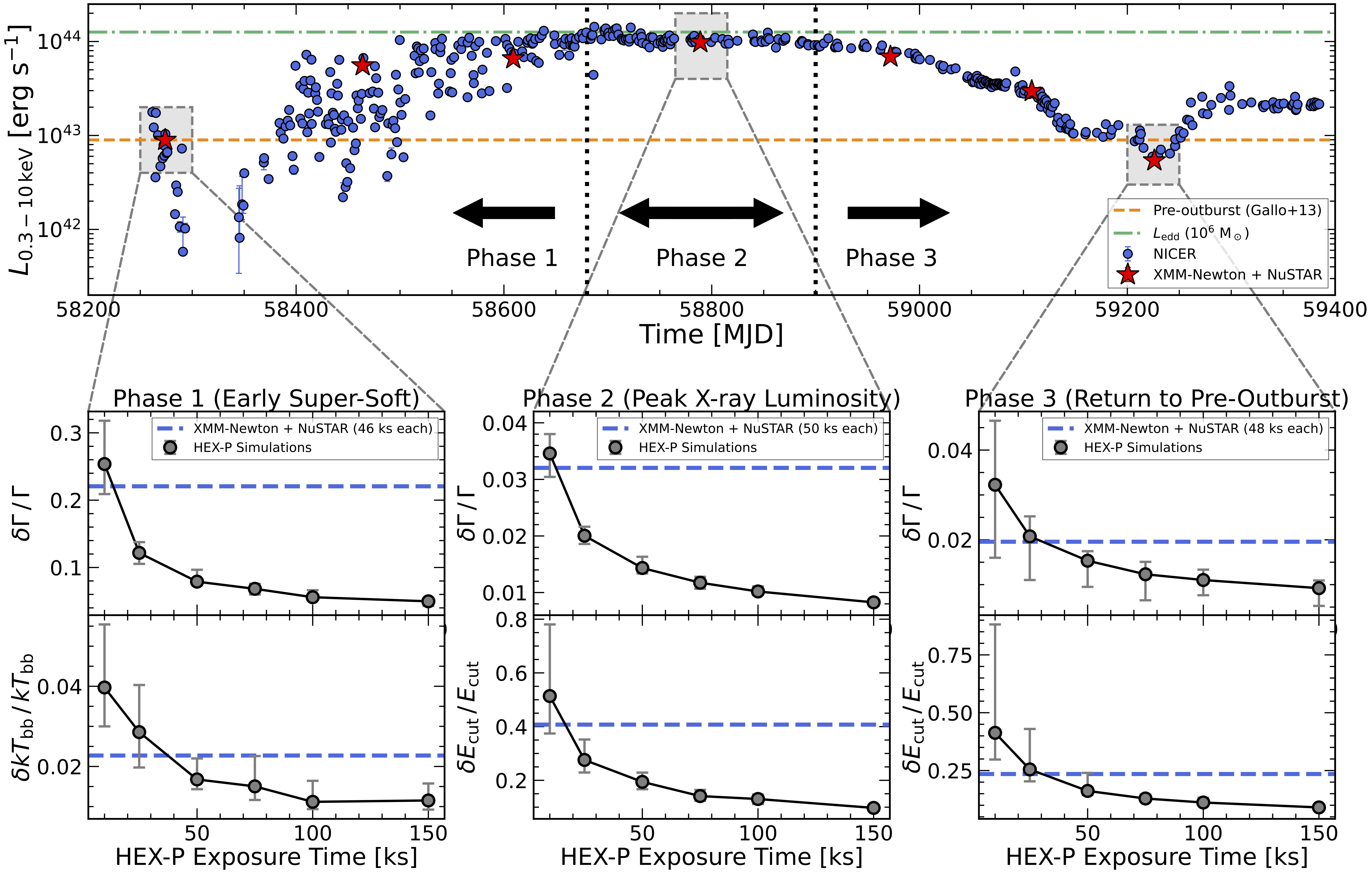}
\end{center}
\caption{\textit{Top:} X-ray light curve of 1ES\,1927+654 from \cite{Masterson2022}, including 7 simultaneous \xmm/\nustar\ observations and $\sim$ 500 NICER observations. The pre-outburst X-ray luminosity is shown as an orange dashed line, and the Eddington luminosity for a $10^6 \, M_\odot$ black hole is shown as a green dot-dashed line. The three evolutionary phases are indicated with vertical black dotted lines. The three \xmm/\nustar\ epochs simulated with \hexp\ are shown with grey boxes. \textit{Bottom:} Fractional uncertainty on $\Gamma$, $E_\mathrm{cut}$, and $kT_\mathrm{bb}$ for \hexp\ simulations of the changing-look AGN 1ES\,1927+654 in three different evolutionary stages during its 2018--2021 outburst. The left-most panel shows simulations during the early super-soft phase (June 2018/Epoch 1), the middle panel shows simulations during the peak X-ray luminosity phase (Nov. 2019/Epoch 4), and the right panel shows simulations during the return to the post-outburst phase (Jan. 2021/Epoch 7). For each evolutionary phase, we simulated 25 spectra for 6 different \hexp\ exposure times based on the best-fit \xmm/\nustar\ model from \cite{Masterson2022}. The blue dashed lines show the joint \xmm/\nustar\ constraints on $\Gamma$ and $E_\mathrm{cut}$ (or $kT_\mathrm{bb}$ for the early super-soft state, in which the cutoff energy cannot be constrained), which can be reached with \hexp\ with much shorter exposure times (10--25 ks). The spectral models and key coronal parameters used to simulate the \hexp\ spectra are given in Table \ref{tab:1ES}.}\label{fig:1ES}
\end{figure}

As a unique opportunity to witness, for the first time, the formation of the corona in an AGN, there have been extensive X-ray resources used to study this enigmatic outburst of 1ES\,1927+654. To date, \nicer\ has observed the target for more than 1 Ms, and there have been 8 simultaneous \xmm/\nustar\ monitoring periods, totalling more than 400 ks with each telescope. \nustar\ observations have been crucial for disentangling the soft photon index and low cutoff energy as the corona formed, which is not possible with either \xmm\ or \nicer\ alone due to their limited bandpasses. However, both \xmm\ and \nicer\ have been critical to unlocking information about the inner accretion flow with their soft X-ray spectra, including providing crucial information about rapid variability of the soft X-ray flux and the broad 1 keV line, that was recently linked to reflection from the inner disk \citep{Masterson2022}. To fully understand this system therefore requires a broad bandpass, which is provided by a single observatory in \hexp. 

To assess the contributions \hexp\ can make to our understanding of the corona in CLAGN, we simulated the three different phases in the outburst of 1ES\,1927+654 as would be seen with the LET and HET on \hexp. For each phase, we simulated \hexp\ spectra using the best-fitting \xmm/\nustar\ models from \cite{Masterson2022} with six exposure times ranging from 10 to 150 ks. The model used for each phase, along with the key coronal parameters used in the simulations, are given in Table \ref{tab:1ES}. Figure\,\ref{fig:1ES} shows the resulting constraints on key coronal parameters $\Gamma$ and $E_\mathrm{cut}$ for each evolutionary phase, as well as constraints on $kT_\mathrm{bb}$ of the dominant thermal component in the early super-soft phase (in which constraints on $E_\mathrm{cut}$ were not possible given the super-soft nature of the source). We compare these constraints to the ones achieved by 50-ks joint \xmm/\nustar\ observations, which vary with the changing spectral shape and flux of the source. We find that \hexp\ can reach similar constraints as from joint \xmm/\nustar\ observations in a fraction of the exposure time. These simulations show that \hexp\ could probe the evolution of the cutoff energy and photon index on roughly 10-25 ks timescales, meaning that either the source could be visited relatively frequently to track the evolution over weeks-months timescales, or that a single long stare could be broken into many 10-25 ks segments in which the evolution of $\Gamma$ and $E_\mathrm{cut}$ could be studied. The X-ray flux of 1ES\,1927+654 varied by an order of magnitude in timescales as short as 8\,hours while the corona was in the process of forming \citep{Ricci2020}, which would allow \hexp\ to provide unprecedented constraints on the evolution of the temperature and compactness of a newly forming corona for the first time. Since the discovery of 1ES\,1927+654, other sources with similarly dramatic X-ray variability have been discovered (e.g., a repeating TDE discovered by eROSITA, which shows a repetitive creation and collapse of the X-ray corona; \citealt{Liu2023}).

\begingroup
\renewcommand*{\arraystretch}{1.25}
\begin{table*}
\small
\caption{Spectral models and parameters used in simulations of 1ES\,1927+654 with \hexp}
\centering
\label{tab:1ES}
\vspace{.3cm}
 \begin{tabular}{c c c c c}
    \hline
    \hline   
	Phase & XSPEC Model & $\Gamma$ & $E_\mathrm{cut}$ (keV) & $kT_\mathrm{bb}$ (eV) \\
	\hline
	1 & tbabs $\times$ ztbabs  $\times$ (zbbody + zpower + gsmooth(xillverTDE)) & 3.4 & --$^\dagger$ & 88 \\
    2 & tbabs $\times$ ztbabs  $\times$ (zbbody + zcutoffpl + gsmooth(xillverTDE)) & 3.6 & 8.1 & 205 \\
    3 & tbabs $\times$ ztbabs  $\times$ (zcutoffpl + relxillCp) & 2.3 & 23.4 & -- \\
	\hline
\end{tabular}

{\raggedright $^\dagger$In this epoch, the source was too soft for proper determination of the cutoff energy, and hence, we used a simple power law to model the spectrum. Note that in the first joint \xmm/\nustar\ observation of 1ES 1927+654, the source was undetected above 3 keV with \xmm\ and was not detected with \nustar.\par}
\end{table*}
\endgroup

Moreover, non-jetted TDEs present another opportunity to witness the formation of the corona around SMBHs, and \hexp\ will play a major role in improving our understanding of various aspects of TDEs. Although detailed simulations go beyond the scope of the current manuscript, we highlight some of these science cases below. The majority of X-ray emitting, non-jetted TDEs show dominant thermal emission, presumably from the accretion disk, but a handful of sources have shown evidence for the formation of a corona in late-time ($\sim$ few years after ignition) X-ray observations \citep[e.g.,][]{Wevers2021,Yao2022}. As TDEs are believed to transition from super- to sub-Eddington when the mass fallback rate drops as stellar debris is accreted, they are another powerful probe of how the corona evolves with changes to the accretion state. A correlation between the Eddington ratio and X-ray spectral state has been already observed in a (small) sample study \citep{Wevers2020}, with a slope that is remarkably similar to that seen in AGN samples \citep[see e.g.,][]{Lusso2016}. \hexp\ studies will enable constraints on the formation and evolution of the corona while {\it simultaneously} constraining the high-energy cutoff energy over a wide range of accretion rates in individual SMBHs. This direct probe the disk-corona connection during different accretion states will allow for a comparison to state transitions in black hole binaries \citep{Remillard2006} and serve as a test of the scale-invariant nature of accretion.

X-ray observations of TDEs can also provide independent estimates of SMBH masses and spins, which are otherwise difficult to probe in dormant SMBHs. Detailed X-ray spectral modelling of the thermal continuum in TDEs can provide constraints on the mass and the spin of the central black hole \citep[e.g.,][]{Wen2020}. With its wide energy coverage, \hexp\ will be able to advance this technique by mitigating the inherent uncertainties and bias introduced by the presence of an X-ray corona. Furthermore, X-ray variability is also a powerful probe of SMBH properties. In particular, the SMBH spin can be estimated through the detection of X-ray quasi-periodic oscillations and/or disk precession \citep[e.g.,][]{Pasham2019}, and the SMBH mass can be measured with the break frequency of the power spectral distribution (PSD) \citep{McHardy2006}. In TDEs, the PSD behaviour has been shown to change with Eddington ratio \citep{Saxton2012, Wevers2021}, and hence can provide an independent estimate of the SMBH mass in TDEs. Additionally, if TDEs form a corona at late times when the accretion rate has dropped and the disk is geometrically thin, then the Fe\,K$\alpha$ line can be used to constrain the corona geometry and SMBH spin \citep[e.g.,][]{Yao2022}. Thanks to its broad energy coverage, high throughput, low background, and L1 orbit permitting long observations, \hexp\ will be able to improve our understanding of TDEs using all of these techniques. This will provide several independent mass and spin measurements of quiescent black holes, which would add invaluable information to constrain the demographics of quiescent low mass SMBHs ($<10^7 M_{\odot}$) that is not readily available through other means.


\section{Coronal geometry}
\label{sec:geometry}

One of the key advances in recent years has been the ability to measure the location, geometry and structure of the corona using X-ray reflection and reverberation from the inner regions of the accretion disk. Determining the structure of the corona in this way places important constraints on the mechanism by which the corona is formed and energized by the accretion flow, for example distinguishing between a corona that is formed in a jet or failed jet, from a corona that is formed over the surface of the inner accretion disk. 

In addition to understanding the physics of the corona itself, measurements of the location and geometry of the corona are important to validate assumptions that underpin the use of X-ray reflection spectroscopy to measure fundamental properties of the black hole, such as its spin \citep{Fabian1989, Dauser2013, Risaliti2013, Mallick2022}. One of the predominant techniques employed to measure black hole spin is based upon identifying the innermost radius of the accretion disk (assumed to coincide with the innermost stable circular orbit,  ISCO) from the extremal redshift detected in emission lines (namely the iron\,K line) within the reflection spectrum \citep[see][and Piotrowska et al., 2023]{Brenneman2006,Reynolds2021}. The intensity of the reflection and the line emission profile, however, depends upon the geometry of the corona, which provides the primary source of illumination. Typically, either a point source (or `lamppost') is assumed, or a phenomenological power law is used to model the emissivity profile of the disk (i.e., the intensity of the reflected flux as a function of radius). If the model corona or power law over-predicts reflection from the inner disk, the corona is at a greater height or is more extended than assumed in the model, and it is possible that the spin of the black hole is underestimated \citep{Fabian2014}. First-hand measurements of the location and geometry of the corona reduce this systematic uncertainty and the degeneracy that arises between extended coron\ae\ and low spins. 

\hexp\ will enable measurements of the location, geometry and structure of the corona via both broadband X-ray spectroscopy and timing. The former is based upon direct measurement of the emissivity profile of the disk from the reflection observed in time-averaged X-ray spectra. The latter technique utilizes the light travel time between the corona and disk as variations in the luminosity of the primary X-ray emission reverberate off of the accretion disk. In addition, further constrains on the geometry of the X-ray corona are available from X-ray polarisation measurements, which can distinguish between X-ray emission from a compact corona, a corona associated with a jet, or an extended slab-like corona extending over the inner accretion disk \citep[see e.g.,][]{Marinucci2022ixpe, Gianolli2023, Tagliacozzo2023, Ursini2023}.

\subsection{Measurements of the corona via X-ray spectroscopy}

The emissivity profile of the disk is encoded in the profile of the relativistically broadened emission lines. The emission line we observe is the line emission integrated over the entire disk, and is comprised of photons experiencing different Doppler shifts and gravitational redshifts, which vary as a function of position on the disk. This means that it is possible to fit the observed line profile as the sum of contributions from different radii on the disk, and the relative contribution of the line model from each radius provides a measurement of the emissivity profile \citep{Wilkins2011}. The measured profile can then be compared to theoretical predictions for the illumination of the disk by coron\ae\ with different geometries and at different locations \citep{Wilkins2012,Dauser2013}. Alternatively, a geometry can be assumed for the corona (usually a point source, or lamppost), and a model can be fit directly to full reflection spectrum to make a measurement of the coronal parameters; namely the height of the corona for the lamppost \citep{Dauser2016}. The emissivity profile of the disk is sensitive to the coronal height in the case of a lamppost geometry, or to the radial extent of the corona if it were extended over the surface of the disk \citep{Wilkins2012}.

A further measurement of the corona comes from the ratio of the reflected flux to the flux of the continuum that is observed directly from the corona, referred to as the reflection fraction \citep{Wilkins2015,Dauser2016}. In the case of an isotropically emitting point source above an infinite accretion disk in a flat spacetime, we expect a reflection fraction $R_{\rm f} = 1$, since exactly half of the emission from the corona is emitted downwards to illuminate the disk, and half is emitted upwards to escape to be observed as the continuum. Light bending around the black hole causes a greater fraction of the rays to be focused towards the inner accretion disk, enhancing the reflection fraction relative to the continuum when the corona is confined to a more compact region of space, closer to the black hole, providing another probe of the compactness of the corona \citep{Fabian2012,Parker2014, Walton2021}. On the other hand, the reflection fraction can be reduced below unity if it is outflowing at a mildly relativistic velocity (for example if the corona is part of a jet or failed jet). In this case, the coronal emission is relativistically beamed away from the accretion disk, enhancing the fraction of rays that are able to escape to be observed as part of the continuum \citep{Beloborodov+1999,Gonzalez2017}. The broad bandpass of \hexp\ is vital to obtain an accurate measurement of the total reflection fraction. In addition, by combining measurements of the reflection fraction with measurements of the relativistically broadened iron\,K line and the associated reverberation time lags, \hexp\ will be able to obtain measurements of the location, the geometry, and the motion of the X-ray emitting coron\ae\ around black holes.

\subsection{X-ray reverberation from the inner accretion disk}

The measurement of X-ray reverberation time lags adds a further dimension to the picture of the corona. The reflection and line emission from the accretion disk responds to short-timescale changes in luminosity of the primary X-ray emission from the corona. There is, however, a time delay between variations in the continuum and the correlated variations in the reflection due to the additional light travel time between the corona and disk \citep{Fabian2009,Uttley2014}. In the sample of AGN in which reverberation has been detected, the measured time delays are short and correspond to the light crossing time across distances of between approximately one and ten gravitational radii (this is approximately equal to the radial co-ordinate of the event horizon of a maximally spinning black hole). Such short time delays indicate that the corona must be compact and confined to a small region of space close to the black hole and innermost accretion flow \citep[e.g.,][]{DeMarco2013,Kara2016,Mallick2021}. The reverberation time scale is primarily sensitive to the scale height of the corona above the disk \citep{Wilkins2013,Cackett2014}. In recent years, there have also been advances in simultaneously modelling the X-ray spectrum and time lag measurements under the assumption of a point-like corona, using models such as \textsc{reltrans} \citep{Ingram2019}. With simultaneous fitting, it is possible to obtain simultaneous constraints on both the height of the corona and the mass of the black hole \citep[e.g.,][]{Mastroserio2020}.

\hexp\ measurements of X-ray reflection and reverberation were simulated for nearby AGN, typical of those in which reverberation from the inner disk has been detected \citep[e.g.,][]{Kara2015, Kara2016, Mallick2021}. These are predominantly narrow line Seyfert 1 (NLS1) AGN \citep{Gallo2006}, which show strong, relativistically broadened iron\,K lines from the inner accretion disk and a high degree of short-timescale variability in their X-ray light curves \citep[see e.g.,][for a review]{Gallo2018}. The observations were simulated from a full spectral-timing reverberation model, which begins with general relativistic ray tracing calculations between a point-like corona, the accretion disk, and the observer, using the \textsc{CUDAKerr} code of \citet{Wilkins2012} and \citet{Wilkins2016}. These ray tracing calculations self-consistently predict the illumination of the disk by the corona, and the energy shifts and time delays of the observed reflection and reverberation in the form of the impulse response function \citep{Reynolds1997,Cackett2014,Uttley2014}. The response function is then convolved with the \textsc{xillver} model for the reflection spectrum produced in the rest frame of the material in the disk \citep{Garcia2013} to predict the full spectral-timing response. The simulations generate the observed X-ray spectrum (including the instrumental background), and the light curves that would be observed in different energy bands, folded through the response functions of the \hexp\ telescopes and detectors, allowing measurements of the spectrum and time lags to be conducted as they would be for real observations.

\begin{figure} 
\begin{minipage}[b]{.56\textwidth}
\centering
\includegraphics[width=.9\textwidth]{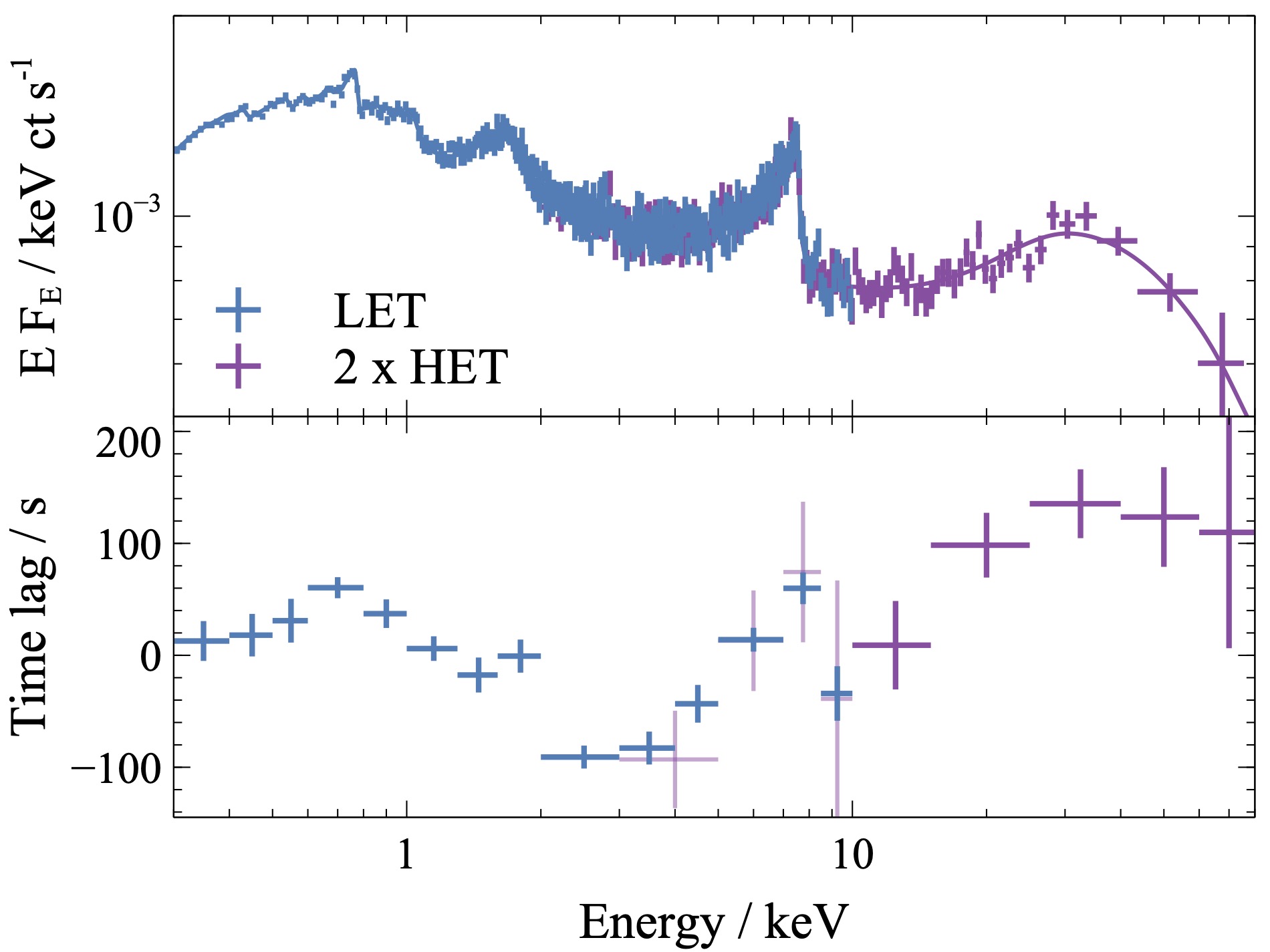}
\end{minipage}
\begin{minipage}[b]{.44\textwidth}
\centering
\includegraphics[width=.98\textwidth]{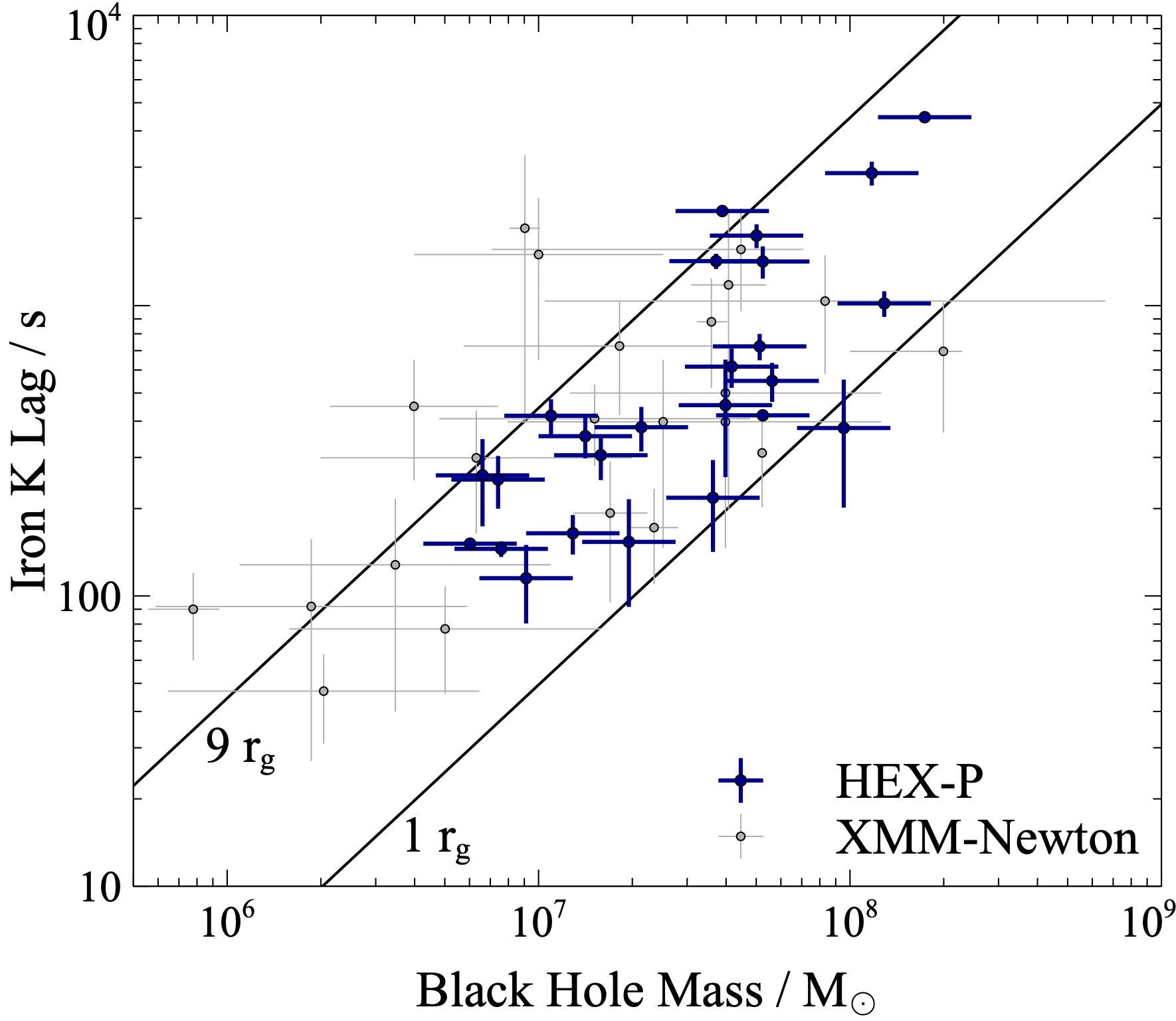}
\end{minipage}
\caption{\textit{Left:} Simulated 500\,ks \hexp\ observation of a typical nearby AGN, showing X-ray reflection and reverberation from the inner accretion disk. The top panel shows the simulated X-ray spectra obtained with the LET and HET. The relativistically broadened iron\,K line (around 6.4\,keV) and Compton hump (around 25\,kev) that are characteristic of reflection from the inner disk are clearly visible. The bottom panel shows the measured time lag \textit{vs.} energy spectrum, which shows the relative time at which different energy bands respond as variations in the continuum emission reverberate off the disk. The energy bands dominated by reflection from the disk (the soft X-ray excess, below 1\,keV, the broad iron line, and the Compton hump) lag behind the bands dominated by the primary continuum observed directly from the corona. This time delay arises from the additional light travel time between the corona and accretion disk, and is a sensitive probe of the location and geometry of the corona. \textit{Right:} Simulated iron\,K reverberation lag measurements as a function of black hole mass for a simulated sample of AGN based upon the BASS sample. For each simulated AGN, the mass, luminosity, and redshift are representative of the real sample, and the remaining parameters of the black hole, disk, and corona are drawn from random distributions. Each of the simulated AGN is observed for 300\,ks with \hexp. Error bars on the black hole mass values assume mass measurements can be made via optical reverberation, and correspond to a representative uncertainty of 0.15\,dex.}\label{fig:reverb_sim}
\end{figure}  

The left panel of Fig.\,\ref{fig:reverb_sim} shows the simulated broadband X-ray spectrum and time lag spectrum for a typical Seyfert-like AGN displaying X-ray reverberation for a deep 500\,ks observation with \hexp. Reverberation is simulated around a maximally spinning ($a=0.998$) black hole with a mass of $10^7\, M_\odot$ at a redshift of $z = 0.025$. The observed flux is equivalent to accretion at 0.3 times the Eddington limit, and 5 per cent of the total bolometric luminosity is produced by the corona. Parameters of the corona and accretion disk are selected to be representative of the typical reflection spectra observed in nearby AGN \citep[e.g.,][]{Mallick2018, Jiang2018}. The X-ray spectrum shows the characteristic features of X-ray reflection from the inner disk: a) the iron\,K line around 6.4\,keV, with an extended redshifted wing comprised of the line photons emitted from the innermost radii of the accretion disk; b) the Compton hump around 20\,keV; and c) a soft excess formed as soft X-ray lines are blended together by relativistic broadening. The ability to simultaneously measure the soft X-ray band, the broad iron\,K line, and the Compton hump will allow \hexp\ observations to unambiguously separate reflection from the accretion disk from other spectral components \citep[e.g.,][]{Walton2014}.

The characteristic form of the reflection spectrum is also evident in the time lag spectrum. The energy bands dominated by reprocessed emission from the disk, namely the soft excess, the iron\,K line, and the Compton hump, are seen to respond to variations at later times than the energy bands dominated by continuum emission observed directly from the corona (1.5--3\,keV and 8--10\,keV). The broad bandpass covered by the combination of the LET and HET on \hexp, combined with the available effective area, mean that \hexp\ is sensitive to reverberation time lags across the full range of the reflection spectrum. It will be possible to measure the differential time lag between the redshifted wing of the iron\,K line, which arises from the inner disk, closer to the black hole and corona, and the 6.4\,keV core of the iron\,K line from larger radii. The difference in time lag between the inner and outer disk is a sensitive probe of the structure of the corona \citep{Wilkins2016}. \hexp\ will be able to routinely measure the reverberation time delays associated with the Compton hump, only hints of which have so far appeared in \nustar\ observations \citep{Kara2015,Zoghbi2017}.

\subsection{Simulation of \hexp\ reverberation measurements}

To demonstrate the sensitivity of \hexp\ to X-ray reverberation and its effectiveness in measuring the structure of the corona across the supermassive black hole mass spectrum, iron\,K reverberation measurements were simulated for a sample representative of the 25 brightest nearby AGN in the BASS catalog, each observed for 300\,ks with \hexp. For each of these simulated AGN, the mass, bolometric luminosity, redshift and absorption were drawn from the BASS catalog \citep{Koss2022}. The remaining parameters of the black hole, accretion disk and corona were drawn randomly from distributions representing the reverberation sample (since detailed spectral analyses are not available for all of the BASS AGN\footnote{The probability distributions from which the parameter values are drawn were chosen to be consistent with the range of parameter values from the sub-set of the BASS AGN sample for which spectroscopic measurements are available, detailed at \url{www.bass-survey.com} and in \citet{Ricci2017}.}). The reverberation measurement was simulated as described above, predicting the time lag vs. Fourier frequency that would be measured by \hexp\ between the continuum-dominated 1.2--4\,keV band and the 4--7\,keV iron\,K band. We select the frequency bin with the highest signal-to-noise (i.e., the ratio between the lag and the lag uncertainty) as the representative iron\,K lag value.

The right panel of Figure~\ref{fig:reverb_sim} shows the simulated iron\,K reverberation lag measurements as a function of black hole mass for the BASS AGN sample, compared to the equivalent measurements made in a sample of nearby Seyfert galaxies using \xmm\ \citep{Kara2016}. \hexp\ will significantly detect reverberation in the iron\,K line and trace the relationship between the reverberation time lag and the mass of the black hole. From these simulations, we expect that \hexp\ will be able to measure iron\,K lags in a sample of AGN between redshifts of $z=0.005$ and $z=0.14$, with bolometric luminosities between $3\times 10^{43}$ and $10^{46}$\,erg\,s$^{-1}$ (assuming 5 per cent of this is emitted across the broad 0.1-100\,keV X-ray band). The simulations show that iron\,K lags will be measured to high precision, with a median uncertainty of 13 per cent across the sample from 300\,ks observations. Longer observations of around 500\,ks will enable an additional reverberation measurement to be made in the Compton hump to constrain the broadband time lag spectrum, to which more detailed reverberation models can be fit (as demonstrated in the left panel).\footnote{In Figure~\ref{fig:reverb_sim}, the error bars on mass measurements assume that the mass can be measured by optical reverberation. We assume a representative uncertainty of 0.15\,dex on the black hole mass, the median of those obtained by reverberation measurements  in the Black Hole Mass Database \citep{Bentz2015}.}

Notably, \hexp\ will also be able to detect reverberation around the most massive black holes in the sample ($\gtrsim 10^8$\msun), where the reverberation timescale is expected to be longer (if it scales simply with the gravitational radius around the black hole). The ability to measure long-timescale lags using common Fourier techniques \citep{Uttley2014} is limited by the duration of continuous light curves that can be obtained. The highly elliptical orbit of \xmm\ limits individual observations, and hence individual light curve segments, to $\sim 125$\,ks. At its L1 orbit, \hexp\ will be able to perform much longer continuous exposures (in principle up to 12 days), greatly lengthening the timescales that can be probed. Combined with the fact that the highest mass AGN are some of the brightest in the BASS sample, iron\,K reverberation is expected to be detected at high significance.

\section{Black hole X-ray binary coron\ae}
\label{sec:BHXRB}

This work has focused on coron\ae\ in the vicinity of extragalactic SMBHs, primarily observed through hard X-ray emission from AGN. However, Galactic black hole X-ray binaries (BHXRBs) can be viewed broadly as scaled-down versions of AGN, particularly from the point of view of the compact regions close to the BH \citep{Falcke2004,Koerding2006,McHardy2006, Walton2012}. Many BHXRBs -- those with low-mass companion stars-- are transient sources, undergoing outbursts lasting weeks to years, during which they display a full repertoire of accretion states and regularly exceed X-ray fluxes of 1 Crab (see, e.g., \citealt{Remillard2006,TetarenkoB2016}). Given the broad dynamical range one can probe via studies of BHXRBs, much can be learned about the long-term behaviour and evolution of AGN through observations possible on human timescales. In addition, the high fluxes of outbursting BHXRBs allow for characterization of the broadband X-ray continuum with exquisite detail, making them a primary target for \hexp. 

As briefly discussed in Section~\ref{sec:unobscured}, today's X-ray observatories already allow us to probe the coronal temperatures of BHXRBs as they decrease with increasing source flux \citep{Buisson2019,Cangemi2021,Zdziarski2021}, and these measurements have contributed greatly to our understanding of BH coron\ae. However, degeneracies remain in the broadband X-ray continuum emission and geometry of BHXRB coron\ae, as well as the AGN ones, and this has been a hindrance to reflection studies aimed at characterizing the inner accretion flow and measuring BH spin. With its broadband coverage, large collecting area, and high energy sensitivity, \hexp\ will: (i) break degeneracies in the geometry of the irradiating corona intrinsic to reflection models; and (ii) measure deviations from purely thermal Comptonization models, allowing explicit constraints on the non-thermal fraction of BHXRB coron\ae. The broadband spectral coverage of \hexp\ in the $0.2-80\,\rm keV$ range, combined with its high-energy sensitivity (owing to low X-ray background) will be especially powerful in constraining the weaker power law emission in BHXRB soft states, advancing coronal constraints spanning orders of magnitude in luminosity and across the spectral states. We refer the reader to Connors et al. (2023, in preparation) for a full description of \hexp\ simulations of BHXRB coron\ae. 

\section{Discussion and conclusions}
\label{sec:discussion}

In this paper, we show the advances that \hexp\ will make in understanding the physics of the X-ray corona in different types of AGN, ranging from local Seyferts to quasars at cosmological redshifts. The broad passband, large effective area, low background, and the ability to have continuous long exposures will enable addressing many open questions regarding the physics of the X-ray corona in AGN. \hexp\ will allow an accurate estimate of the continuum emission spectrum of the corona by measuring the photon index and the high-energy cutoff. Assuming a certain geometry, these parameters can then be mapped into optical depth and electron temperature within the framework of Comptonization \citep[see e.g.,][]{Middei19}. We demonstrate how \hexp\ will improve our understanding of the electron distribution in AGN coron\ae, measure coronal properties of obscured sources, and extend these analyses to $z \geq 1$. Furthermore, we show that \hexp\ will track variations of bright transient events such as CLAGN and TDEs on timescales of a few hours. \hexp\ will allow us to study the properties of the X-ray corona and probe the innermost regions of the accretion disk via reverberation mapping. This will be crucial for understanding how the corona interacts with the accretion disk, studying disk reflection and consequently measure the BH spin in AGN over a broad range of mass and luminosity, and probing AGN feedback (see Piotrowska et al., 2023, in preparation). Furthermore, \hexp\ will allow a detailed study of obscuration in AGN and identify the fraction of Compton-thick AGN in deep surveys (see Boorman et al., 2023, Civano et al., 2023, in preparation). All this will help achieving a more complete picture of how SMBHs grow and evolve across cosmic time, and the role they play in galaxy evolution.

\begin{figure}[htbp]
\begin{center}

\includegraphics[width=0.95\linewidth]{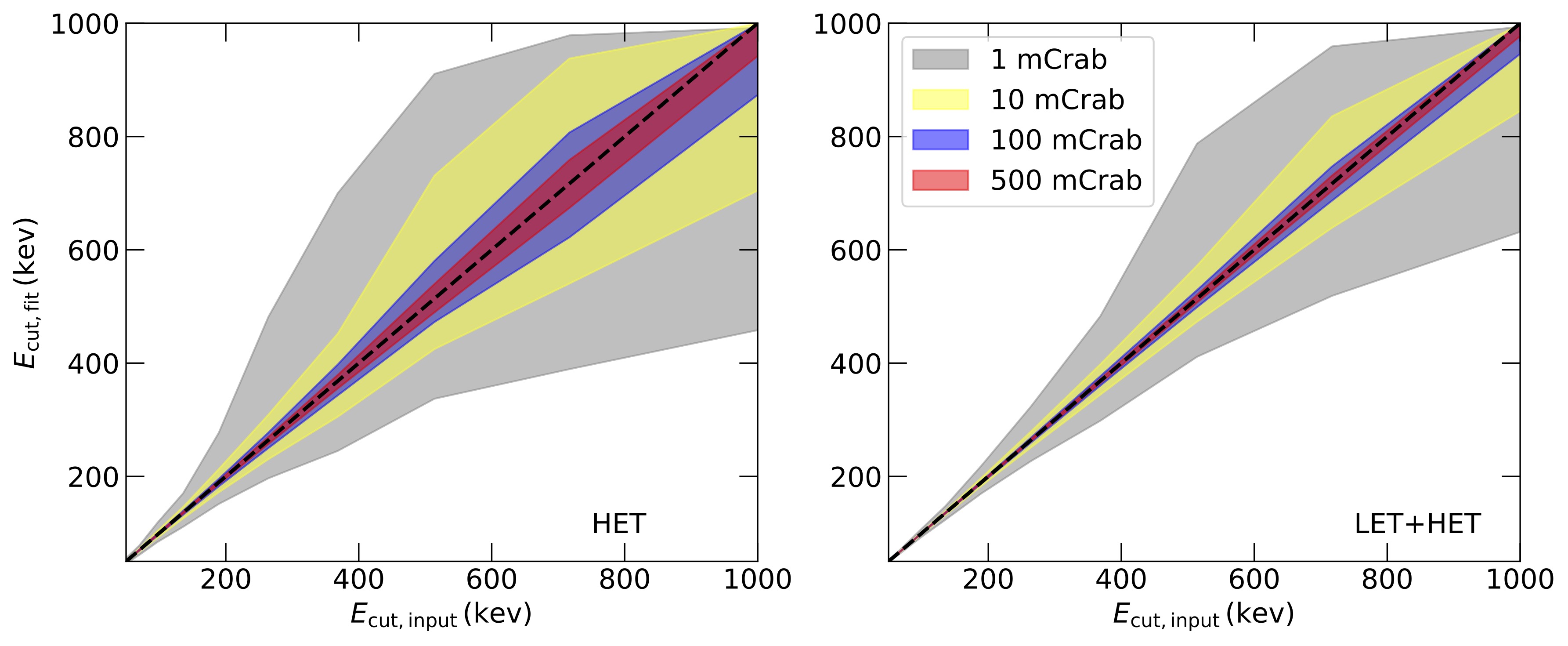}

\end{center}
\caption{Results of fitting \hexp\ simulations assuming relativistic reflection, by employing only HET (left) and by adding the soft X-rays LET+HET (right). The grey, yellow, blue, and red areas correspond to the 90\%\ confidence level on estimating \ecut\ for observed flux levels of 1\,mCrab, 10\,mCrab, 100\,mCrab, and 500\,mCrab, respectively.}\label{fig:Ecut_relxill}
\end{figure}

Simultaneous observations with LET and HET, covering the $0.2-80\,\rm keV$ range, will allow an accurate measurement of \ecut\ up to $\sim 1\,\rm MeV$ in bright sources where ionized disk reflection is important. This has been discussed in detail by \cite{Garcia2015} who showed that the soft part of the reflection spectrum, rich in fluorescent lines and atomic features, depends strongly on the properties of the corona, illuminating the accretion disk. \cite{Garcia2015} demonstrated this in the case of joint \nustar\ and \suzaku\ spectra. Here, we revisit the analysis presented by those authors for \hexp. We consider a simple spectrum comprised of neutral Galactic absorption \citep[{\tt TBabs};][]{Wilms2000} and a power law plus ionized reflection \citep[\texttt{relxill;}][]{Dauser2013, Garcia2014}. In {\tt XSPEC} parlance, the model is written as: ${\tt Model = TBabs \times relxill}$. We assume a Galactic absorption $N_{\rm H} = 5 \times 10^{20}\,\rm cm^{-2}$. As for the reflection component, we assume a photon index $\Gamma = 2$, a black hole spin $a^\ast = 0.998$, a disk ionization parameter $\log \left (\xi/\rm erg\,cm\,s^{-1} \right) = 1.5$, an inclination $i = 45^\circ$, a disk emissivity index $q = 3$, and an iron abundance equal to the solar value. \cite{Garcia2015} used  similar parameters except for the reflection fraction which they assumed to be $R_{\rm f} = 3$. This enhances the reflection features in the spectrum. We choose a more conservative value of $R_{\rm f} = 1$. We then simulate 100-ks observations of \hexp\ for \ecut\ ranging between 50\,keV and 1\,MeV for $2-10\,\rm keV$ flux levels of 1\,mCrab, 10\,mCrab, 100\,mCrab, and 500\,mCrab. These simulations are intended to represent a wide range of Galactic and extragalactic accreting black holes. We fit the simulated spectra using only the HET and by using both LET and HET simultaneously. Similarly to \cite{Garcia2015}, we find that \hexp\ will allow us to accurately estimate \ecut\ up to $\sim 1\,\rm MeV$ for bright sources (Fig.\,\ref{fig:Ecut_relxill}). In addition, we also find that the inclusion of LET significantly reduces the uncertainty on \ecut. We note that these results assume that the soft X-ray excess is in totality caused by the ionized relativistic reflection from the inner parts of the disk. In fact, the nature of this excess is still a matter of debate as it has been suggested that it could also be caused by the presence of a warm optically thick corona \citep[e.g.,][]{Petrucci2018, Petrucci2020}. Studying the effect of such model on the properties of the hot corona is beyond the scope of this paper.

\subsection{Synergies with other facilities}

We have demonstrated the power of \hexp\ as a stand-alone instrument to advance our understanding of the physics of X-ray coron\ae\ in accreting black holes. However, simultaneous broadband X-ray and multi-wavelength coverage will provide a more comprehensive physical picture of accretion onto these objects. \hexp\ will play an instrumental role in such studies thanks to its broad passband and high energy sensitivity as demonstrated throughout this paper. The ability to probe the emission from the X-ray corona for a variety of sources with a wide range of mass, luminosity, distance, and variability timescales will make \hexp\ an essential tool to provide a comprehensive understanding of the behavior and physics of AGN coron\ae, which is the dominant source of X-rays in the universe. 

For example, the recent advancements made by \ixpe, observing X-ray polarization in AGN for the first time, required a full characterization of the broadband X-ray spectrum. This has been ensured via simultaneous observations with \xmm, \nustar, and \chandra. \hexp\ will provide equivalent energy coverage to the combination of \xmm/\chandra\ and \nustar, with significant improvements to the high energy sensitivity. \hexp\ will be an essential complementary observatory to future X-ray polarization missions such as the {\it X-ray Polarization Probe} \citep[{\it XPP};][]{Krawczynski2019,Jahoda2019} and the {\it enhanced X-ray Timing and Polarimetry} mission \citep[{\it eXTP};][]{Zhang2019extp}. 

As mentioned earlier, a correlation between the hard X-ray emission and mm-band continuum has been observed in various samples of AGN \citep[see e.g.,][]{Behar2018, Kawamuro2022, Ricci2023}. \hexp\ will accurately measure the optical depth and temperature of the X-ray corona, as well as determine the fraction of non-thermal electrons producing the X-ray emission for a large sample of AGN. In coordination with ALMA and similar facilities, these parameters can then be used to model the continuum emission in the mm-band and probe the magnetic field powering the corona. \hexp\ will also complement observations with the Square Kilometer Array and the next generation of cosmic microwave background experiments \citep[e.g., CMB-S4;][]{cmbs4_2019,cmbs4_2021}.

Moreover, \hexp\ will play a crucial part in connecting soft X-ray and soft gamma-ray (i.e., up to a few MeV) observations of BHXRBs. In particular, a primary science goal of the {\em Compton Spectrometer and Imager} \citep[{\em COSI};][]{Tomsick2019}, scheduled for launch in 2027 as a NASA Small Explorer (SMEX) mission, is to characterize the Galactic $511\,\rm keV$ positron annihilation line and the gamma-ray polarization of compact objects. {\em COSI} will be able to detect a broadened and redshifted pair line from pair-dominated coron\ae. The broadband X-ray coverage of \hexp\ will help constrain the continuum coronal emission, which will be important for modeling the higher energy emission and polarization measurements provided by {\em COSI}. \hexp\ will also play a major role in constraining the X-ray emission from gamma-ray emitting AGN detected by {\em COSI} (see Marcotulli et al., 2023, in preparation), which will probe the connection between the accretion disk and jets \citep{Ajello2019}, shedding yet more light on the nature of the X-ray corona.

Recent systematic studies of the optical night sky, such as All Sky Automated Survey for Super Nov\ae\ \citep[ASASSN;][]{Shappee2014}, the Asteroid Terrestrial-impact Last Alert System \citep[ATLAS;][]{Tonry2018}, and the  Zwicky Transient Facility \citep[ZTF;][]{Bellm2014, Graham2019}, are providing large numbers of Galactic and extragalactic transients, which are often associated with accretion onto compact objects. The discovery rate of nuclear transients with dramatic X-ray variability, similar to 1ES\,1927+654, is expected to increase significantly in the next decade with the continued advance of large-scale, time-domain surveys in the optical (e.g., photometry from the Vera Rubin Observatory, spectroscopy from SDSS-V) and UV \citep[e.g., {\it ULTRASAT};][]{Sagiv2014}. Therefore, \hexp\ will be well-positioned to make significant discoveries and contributions to the X-ray evolution and nature of the corona in transient accretion systems (as discussed in Section\,\ref{sec:CLAGN}).

Finally, \hexp\ will play a major role in the era of the high-resolution X-ray spectroscopy, with the development of microcalorimeters achieving a resolving power ($E/\Delta E$) larger than 1000 at energies up to 7\,keV. The Resolve instrument on \xrism\ \citep{Tashiro2018, xrism2022} launched in August 2023, and the X-ray Integral Field Unit (X-IFU) on \athena\ \citep{Nandra2013, Barret2023}, expected to be launched in the late 2030s, will revolutionize our view of accretion onto compact objects. These instruments will probe the fine structure in absorption and emission from gas around accreting black holes. However, these gaseous structures are dictated by the hard ionizing X-rays, and the accurate measurements of such features will require a proper estimate of the underlying X-ray continuum out to tens of keV. This energy coverage can be provided by \nustar\ in the 2020s and \hexp\ will extend and improve this coverage in the decades to come. In particular, \hexp\ will be able to constrain the coronal emission and reflection spectrum at hard X-rays. This will be crucial for determining the illumination profile of the accretion disk by the X-ray corona, and accurately measuring BH spin for a broad range of mass and luminosity (see also Piotrowska et al., 2023, in preparation). Combining the resolving power of the future generation of microcalorimeters to the broad energy coverage of \hexp\ will allow us to address important questions about black hole growth and evolution, and to understand the role of these objects in shaping the Universe.

\subsection{Conclusions}

We have demonstrated that \hexp\ will provide an unprecedented look at the X-ray coron\ae\ of accreting black holes with a main focus on AGN, thanks to its high sensitivity and broad energy coverage. The main results of the paper can be summarised as follows:

\begin{itemize}
    \item \hexp\ will accurately estimate the coronal properties of unobscured AGN. In particular, it will study evolution of coronal temperature in bright, variable AGN. This will allow us to measure the non-thermal fraction of electrons in AGN coron\ae\ and test the runaway pair production hypothesis.

    \item  \hexp\ will also accurately estimate the coronal properties of obscured AGN up to $z \sim 0.5$. This will be a crucial test of the orientation-based unified model of AGN.

    \item \hexp\ will efficiently constrain the coronal properties of high-redshift quasars ($z>$1.40) even with high energy cutoff up to 200~keV. \hexp\ will also extend the estimates of the coronal properties of powerful quasars beyond cosmic noon up to $z \simeq 4.5 $. This will allow us to study the evolution of the X-ray properties of AGN across cosmic times.

    \item \hexp\ will play a major role in studying transient events such as CLAGN and TDEs. It will allow us to witness the destruction, formation, and  evolution of the X-ray corona in these extreme events.

    \item Finally, the high sensitivity of \hexp\ that extends to hard X-rays and the possibility of long continuous observations of bright AGN will enable accurate reverberation mapping of the innermost regions of the accretion disk. This will allow us to estimate the location of the corona and its illuminating profile over a broad mass range.
    
\end{itemize}


\section*{Funding}
EK acknowledges financial support from the Centre National d’Etudes Spatiales (CNES). MB acknowledges support from the YCAA Prize Postdoctoral Fellowship. LM is supported by the CITA National Fellowship. CR acknowledges support from Fondecyt Regular grant 1230345 and ANID BASAL project FB210003. The work of DS was carried out at the Jet Propulsion Laboratory, California Institute of Technology, under a contract with NASA. XZ acknowledges NASA support under contract number 80NSSC22K0012.

\section*{Author Contributions}

EK is responsible for the creation of the manuscript, contributed to the science cases and authored Sections\,\ref{sec:intro},\ref{sec:obscured}, \ref{sec:high-z},\ref{sec:discussion}, and made Figures\,\ref{fig:obscured_spectra}, \ref{fig:obscured_contours} and \ref{fig:Ecut_relxill}.
AL\ contributed to the science case and authored Section\,\ref{sec:unobscured}, and made Figure\,\ref{fig:ltheta}.
MM\ contributed to the science case and authored Section\,\ref{sec:CLAGN}, and made Figure\,\ref{fig:1ES}.
DW\ contributed to the science case and authored Section\,\ref{sec:geometry}, and made Figure\,\ref{fig:reverb_sim}.
XZ\ contributed to the science case and authored Section\,\ref{sec:high-z}, and made Figures\,\ref{fig:high_z_lum},\ref{fig:simcontour},\ref{fig:high_z}.
MB\ contributed to the science case and co-authored Section\,\ref{sec:unobscured}.
PGB\ contributed to the science case and co-authored Section\,\ref{sec:unobscured}.
RMTC\ contributed to the science case and authored Section\,\ref{sec:BHXRB}.
JAG\ contributed to the science case and authored Section\,\ref{sec:hexp}.
KKM\ contributed to the science case and authored Section\,\ref{sec:hexp}.
DS\ contributed to the science case and authored Section\,\ref{sec:hexp}.
JB\ contributed to the science case and co-authored Section\,\ref{sec:unobscured}. 
TW\ co-authored Section\,\ref{sec:CLAGN}. 
All authors contributed to the final editing of the manuscript.



\bibliographystyle{Frontiers-Harvard} 
\bibliography{HEX-P_corona}

\begin{thebibliography}{237}
\providecommand{\natexlab}[1]{#1}
\expandafter\ifx\csname urlstyle\endcsname\relax
  \providecommand{\doi}[1]{doi:\discretionary{}{}{}#1}\else
  \providecommand{\doi}{doi:\discretionary{}{}{}\begingroup \urlstyle{rm}\Url}\fi
\providecommand{\selectlanguage}[1]{\relax}
\providecommand{\bibAnnoteFile}[1]{%
  \IfFileExists{#1}{\begin{quotation}\noindent\textsc{Key:} #1\\
  \textsc{Annotation:}\ \input{#1}\end{quotation}}{}}
\providecommand{\bibAnnote}[2]{%
  \begin{quotation}\noindent\textsc{Key:} #1\\
  \textsc{Annotation:}\ #2\end{quotation}}

\bibitem[{{Abazajian} et~al.(2022){Abazajian}, {Abdulghafour}, {Addison}, {Adshead}, {Ahmed}, {Ajello} et~al.}]{cmbs4_2021}
{Abazajian}, K., {Abdulghafour}, A., {Addison}, G.~E., {Adshead}, P., {Ahmed}, Z., {Ajello}, M., et~al. (2022).
\newblock {Snowmass 2021 CMB-S4 White Paper}.
\newblock \emph{arXiv e-prints} , arXiv:2203.08024\doi{10.48550/arXiv.2203.08024}
\bibAnnoteFile{cmbs4_2021}

\bibitem[{{Agol} and {Krolik}(2000)}]{Agol2000}
{Agol}, E. and {Krolik}, J.~H. (2000).
\newblock {Magnetic Stress at the Marginally Stable Orbit: Altered Disk Structure, Radiation, and Black Hole Spin Evolution}.
\newblock \emph{\apj} 528, 161--170.
\newblock \doi{10.1086/308177}
\bibAnnoteFile{Agol2000}

\bibitem[{{Ajello} et~al.(2019){Ajello}, {Paliya}, {Marcotulli}, {Perkins}, {Prandini}, {D'Ammando} et~al.}]{Ajello2019}
{Ajello}, M., {Paliya}, V., {Marcotulli}, L., {Perkins}, J.~S., {Prandini}, E., {D'Ammando}, F., et~al. (2019).
\newblock {Supermassive black holes at high redshifts}.
\newblock \emph{\baas} 51, 289
\bibAnnoteFile{Ajello2019}

\bibitem[{{Akylas} and {Georgantopoulos}(2021)}]{Akylas2021}
{Akylas}, A. and {Georgantopoulos}, I. (2021).
\newblock {Distribution of the coronal temperature in Seyfert 1 galaxies}.
\newblock \emph{\aap} 655, A60.
\newblock \doi{10.1051/0004-6361/202141186}
\bibAnnoteFile{Akylas2021}

\bibitem[{{Antonucci}(1993)}]{Antonucci1993}
{Antonucci}, R. (1993).
\newblock {Unified models for active galactic nuclei and quasars.}
\newblock \emph{\araa} 31, 473--521.
\newblock \doi{10.1146/annurev.aa.31.090193.002353}
\bibAnnoteFile{Antonucci1993}

\bibitem[{{Aversa} et~al.(2015){Aversa}, {Lapi}, {de Zotti}, {Shankar}, and {Danese}}]{Aversa2015}
{Aversa}, R., {Lapi}, A., {de Zotti}, G., {Shankar}, F., and {Danese}, L. (2015).
\newblock {Black Hole and Galaxy Coevolution from Continuity Equation and Abundance Matching}.
\newblock \emph{\apj} 810, 74.
\newblock \doi{10.1088/0004-637X/810/1/74}
\bibAnnoteFile{Aversa2015}

\bibitem[{{Ballantyne} et~al.(2014){Ballantyne}, {Bollenbacher}, {Brenneman}, {Madsen}, {Balokovi{\'c}}, {Boggs} et~al.}]{Ballantyne2014}
{Ballantyne}, D.~R., {Bollenbacher}, J.~M., {Brenneman}, L.~W., {Madsen}, K.~K., {Balokovi{\'c}}, M., {Boggs}, S.~E., et~al. (2014).
\newblock {NuSTAR Reveals the Comptonizing Corona of the Broad-line Radio Galaxy 3C 382}.
\newblock \emph{\apj} 794, 62.
\newblock \doi{10.1088/0004-637X/794/1/62}
\bibAnnoteFile{Ballantyne2014}

\bibitem[{{Balokovi{\'c}} et~al.(2018){Balokovi{\'c}}, {Brightman}, {Harrison}, {Comastri}, {Ricci}, {Buchner} et~al.}]{Balokovic2018}
{Balokovi{\'c}}, M., {Brightman}, M., {Harrison}, F.~A., {Comastri}, A., {Ricci}, C., {Buchner}, J., et~al. (2018).
\newblock {New Spectral Model for Constraining Torus Covering Factors from Broadband X-Ray Spectra of Active Galactic Nuclei}.
\newblock \emph{\apj} 854, 42.
\newblock \doi{10.3847/1538-4357/aaa7eb}
\bibAnnoteFile{Balokovic2018}

\bibitem[{{Balokovi{\'c}} et~al.(2020){Balokovi{\'c}}, {Harrison}, {Madejski}, {Comastri}, {Ricci}, {Annuar} et~al.}]{Balokovic2020}
{Balokovi{\'c}}, M., {Harrison}, F.~A., {Madejski}, G., {Comastri}, A., {Ricci}, C., {Annuar}, A., et~al. (2020).
\newblock {NuSTAR Survey of Obscured Swift/BAT-selected Active Galactic Nuclei. II. Median High-energy Cutoff in Seyfert II Hard X-Ray Spectra}.
\newblock \emph{\apj} 905, 41.
\newblock \doi{10.3847/1538-4357/abc342}
\bibAnnoteFile{Balokovic2020}

\bibitem[{{Balokovi{\'c}} et~al.(2015){Balokovi{\'c}}, {Matt}, {Harrison}, {Zoghbi}, {Ballantyne}, {Boggs} et~al.}]{Balokovic2015}
{Balokovi{\'c}}, M., {Matt}, G., {Harrison}, F.~A., {Zoghbi}, A., {Ballantyne}, D.~R., {Boggs}, S.~E., et~al. (2015).
\newblock {Coronal Properties of the Seyfert 1.9 Galaxy MCG-05-23-016 Determined from Hard X-Ray Spectroscopy with NuSTAR}.
\newblock \emph{\apj} 800, 62.
\newblock \doi{10.1088/0004-637X/800/1/62}
\bibAnnoteFile{Balokovic2015}

\bibitem[{{Barret} et~al.(2023){Barret}, {Albouys}, {Herder}, {Piro}, {Cappi}, {Huovelin} et~al.}]{Barret2023}
{Barret}, D., {Albouys}, V., {Herder}, J.-W.~d., {Piro}, L., {Cappi}, M., {Huovelin}, J., et~al. (2023).
\newblock {The Athena X-ray Integral Field Unit: a consolidated design for the system requirement review of the preliminary definition phase}.
\newblock \emph{Experimental Astronomy} 55, 373--426.
\newblock \doi{10.1007/s10686-022-09880-7}
\bibAnnoteFile{Barret2023}

\bibitem[{{Barua} et~al.(2020){Barua}, {Jithesh}, {Misra}, {Dewangan}, {Sarma}, and {Pathak}}]{Barua2020}
{Barua}, S., {Jithesh}, V., {Misra}, R., {Dewangan}, G.~C., {Sarma}, R., and {Pathak}, A. (2020).
\newblock {NuSTAR observation of Ark 564 reveals the variation of coronal temperature with flux}.
\newblock \emph{\mnras} 492, 3041--3046.
\newblock \doi{10.1093/mnras/staa067}
\bibAnnoteFile{Barua2020}

\bibitem[{{Baumgartner} et~al.(2013){Baumgartner}, {Tueller}, {Markwardt}, {Skinner}, {Barthelmy}, {Mushotzky} et~al.}]{Baumgartner2013}
{Baumgartner}, W.~H., {Tueller}, J., {Markwardt}, C.~B., {Skinner}, G.~K., {Barthelmy}, S., {Mushotzky}, R.~F., et~al. (2013).
\newblock {The 70 Month Swift-BAT All-sky Hard X-Ray Survey}.
\newblock \emph{\apjs} 207, 19.
\newblock \doi{10.1088/0067-0049/207/2/19}
\bibAnnoteFile{Baumgartner2013}

\bibitem[{{Behar} et~al.(2015){Behar}, {Baldi}, {Laor}, {Horesh}, {Stevens}, and {Tzioumis}}]{Behar2015}
{Behar}, E., {Baldi}, R.~D., {Laor}, A., {Horesh}, A., {Stevens}, J., and {Tzioumis}, T. (2015).
\newblock {Discovery of millimetre-wave excess emission in radio-quiet active galactic nuclei}.
\newblock \emph{\mnras} 451, 517--526.
\newblock \doi{10.1093/mnras/stv988}
\bibAnnoteFile{Behar2015}

\bibitem[{{Behar} et~al.(2018){Behar}, {Vogel}, {Baldi}, {Smith}, and {Mushotzky}}]{Behar2018}
{Behar}, E., {Vogel}, S., {Baldi}, R.~D., {Smith}, K.~L., and {Mushotzky}, R.~F. (2018).
\newblock {The mm-wave compact component of an AGN}.
\newblock \emph{\mnras} 478, 399--406.
\newblock \doi{10.1093/mnras/sty850}
\bibAnnoteFile{Behar2018}

\bibitem[{{Beheshtipour} et~al.(2017){Beheshtipour}, {Krawczynski}, and {Malzac}}]{Beheshtipour2017}
{Beheshtipour}, B., {Krawczynski}, H., and {Malzac}, J. (2017).
\newblock {The X-Ray Polarization of the Accretion Disk Coronae of Active Galactic Nuclei}.
\newblock \emph{\apj} 850, 14.
\newblock \doi{10.3847/1538-4357/aa906a}
\bibAnnoteFile{Beheshtipour2017}

\bibitem[{{Bellm}(2014)}]{Bellm2014}
{Bellm}, E. (2014).
\newblock {The Zwicky Transient Facility}.
\newblock In \emph{The Third Hot-wiring the Transient Universe Workshop}, eds. P.~R. {Wozniak}, M.~J. {Graham}, A.~A. {Mahabal}, and R.~{Seaman}. 27--33.
\newblock \doi{10.48550/arXiv.1410.8185}
\bibAnnoteFile{Bellm2014}

\bibitem[{{Belmont} et~al.(2008){Belmont}, {Malzac}, and {Marcowith}}]{Belmont2008}
{Belmont}, R., {Malzac}, J., and {Marcowith}, A. (2008).
\newblock {Simulating radiation and kinetic processes in relativistic plasmas}.
\newblock \emph{\aap} 491, 617--631.
\newblock \doi{10.1051/0004-6361:200809982}
\bibAnnoteFile{Belmont2008}

\bibitem[{{Beloborodov}(1999)}]{Beloborodov+1999}
{Beloborodov}, A.~M. (1999).
\newblock {Plasma Ejection from Magnetic Flares and the X-Ray Spectrum of Cygnus X-1}.
\newblock \emph{\apjl} 510, L123--L126.
\newblock \doi{10.1086/311810}
\bibAnnoteFile{Beloborodov+1999}

\bibitem[{{Bentz} and {Katz}(2015)}]{Bentz2015}
{Bentz}, M.~C. and {Katz}, S. (2015).
\newblock {The AGN Black Hole Mass Database}.
\newblock \emph{\pasp} 127, 67.
\newblock \doi{10.1086/679601}
\bibAnnoteFile{Bentz2015}

\bibitem[{{Bertola} et~al.(2022){Bertola}, {Vignali}, {Lanzuisi}, {Dadina}, {Cappi}, {Gilli} et~al.}]{Bertola+22}
{Bertola}, E., {Vignali}, C., {Lanzuisi}, G., {Dadina}, M., {Cappi}, M., {Gilli}, R., et~al. (2022).
\newblock {The properties of the X-ray corona in the distant (z = 3.91) quasar APM 08279+5255}.
\newblock \emph{\aap} 662, A98.
\newblock \doi{10.1051/0004-6361/202142642}
\bibAnnoteFile{Bertola+22}

\bibitem[{{Brenneman} and {Reynolds}(2006)}]{Brenneman2006}
{Brenneman}, L.~W. and {Reynolds}, C.~S. (2006).
\newblock {Constraining Black Hole Spin via X-Ray Spectroscopy}.
\newblock \emph{\apj} 652, 1028--1043.
\newblock \doi{10.1086/508146}
\bibAnnoteFile{Brenneman2006}

\bibitem[{{Buchner} et~al.(2021){Buchner}, {Brightman}, {Balokovi{\'c}}, {Wada}, {Bauer}, and {Nandra}}]{Buchner2021physicalobscurermodels}
{Buchner}, J., {Brightman}, M., {Balokovi{\'c}}, M., {Wada}, K., {Bauer}, F.~E., and {Nandra}, K. (2021).
\newblock {Physically motivated X-ray obscurer models}.
\newblock \emph{\aap} 651, A58.
\newblock \doi{10.1051/0004-6361/201834963}
\bibAnnoteFile{Buchner2021physicalobscurermodels}

\bibitem[{{Buchner} et~al.(2019){Buchner}, {Brightman}, {Nandra}, {Nikutta}, and {Bauer}}]{Buchner2019uxclumpy}
{Buchner}, J., {Brightman}, M., {Nandra}, K., {Nikutta}, R., and {Bauer}, F.~E. (2019).
\newblock {X-ray spectral and eclipsing model of the clumpy obscurer in active galactic nuclei}.
\newblock \emph{\aap} 629, A16.
\newblock \doi{10.1051/0004-6361/201834771}
\bibAnnoteFile{Buchner2019uxclumpy}

\bibitem[{{Buisson} et~al.(2019){Buisson}, {Fabian}, {Barret}, {F{\"u}rst}, {Gandhi}, {Garc{\'\i}a} et~al.}]{Buisson2019}
{Buisson}, D.~J.~K., {Fabian}, A.~C., {Barret}, D., {F{\"u}rst}, F., {Gandhi}, P., {Garc{\'\i}a}, J.~A., et~al. (2019).
\newblock {MAXI J1820+070 with NuSTAR I. An increase in variability frequency but a stable reflection spectrum: coronal properties and implications for the inner disc in black hole binaries}.
\newblock \emph{\mnras} 490, 1350--1362.
\newblock \doi{10.1093/mnras/stz2681}
\bibAnnoteFile{Buisson2019}

\bibitem[{{Buisson} et~al.(2018){Buisson}, {Fabian}, and {Lohfink}}]{buisson+2018}
{Buisson}, D.~J.~K., {Fabian}, A.~C., and {Lohfink}, A.~M. (2018).
\newblock {Coronal temperatures of the AGN ESO 103-035 and IGR 2124.7+5058 from NuSTAR observations}.
\newblock \emph{\mnras} 481, 4419--4426.
\newblock \doi{10.1093/mnras/sty2609}
\bibAnnoteFile{buisson+2018}

\bibitem[{{Cackett} et~al.(2014){Cackett}, {Zoghbi}, {Reynolds}, {Fabian}, {Kara}, {Uttley} et~al.}]{Cackett2014}
{Cackett}, E.~M., {Zoghbi}, A., {Reynolds}, C., {Fabian}, A.~C., {Kara}, E., {Uttley}, P., et~al. (2014).
\newblock {Modelling the broad Fe K{\ensuremath{\alpha}} reverberation in the AGN NGC 4151}.
\newblock \emph{\mnras} 438, 2980--2994.
\newblock \doi{10.1093/mnras/stt2424}
\bibAnnoteFile{Cackett2014}

\bibitem[{{Cadolle Bel} et~al.(2006){Cadolle Bel}, {Sizun}, {Goldwurm}, {Rodriguez}, {Laurent}, {Zdziarski} et~al.}]{Cadolle2006}
{Cadolle Bel}, M., {Sizun}, P., {Goldwurm}, A., {Rodriguez}, J., {Laurent}, P., {Zdziarski}, A.~A., et~al. (2006).
\newblock {The broad-band spectrum of Cygnus X-1 measured by INTEGRAL}.
\newblock \emph{\aap} 446, 591--602.
\newblock \doi{10.1051/0004-6361:20053068}
\bibAnnoteFile{Cadolle2006}

\bibitem[{{Cangemi} et~al.(2021){Cangemi}, {Beuchert}, {Siegert}, {Rodriguez}, {Grinberg}, {Belmont} et~al.}]{Cangemi2021}
{Cangemi}, F., {Beuchert}, T., {Siegert}, T., {Rodriguez}, J., {Grinberg}, V., {Belmont}, R., et~al. (2021).
\newblock {Potential origin of the state-dependent high-energy tail in the black hole microquasar Cygnus X-1 as seen with INTEGRAL}.
\newblock \emph{\aap} 650, A93.
\newblock \doi{10.1051/0004-6361/202038604}
\bibAnnoteFile{Cangemi2021}

\bibitem[{{Carlstrom} et~al.(2019){Carlstrom}, {Abazajian}, {Addison}, {Adshead}, {Ahmed}, {Allen} et~al.}]{cmbs4_2019}
{Carlstrom}, J., {Abazajian}, K., {Addison}, G., {Adshead}, P., {Ahmed}, Z., {Allen}, S.~W., et~al. (2019).
\newblock {CMB-S4}.
\newblock In \emph{Bulletin of the American Astronomical Society}. vol.~51, 209.
\newblock \doi{10.48550/arXiv.1908.01062}
\bibAnnoteFile{cmbs4_2019}

\bibitem[{{Chartas} et~al.(2002){Chartas}, {Brandt}, {Gallagher}, and {Garmire}}]{Chartas2002}
{Chartas}, G., {Brandt}, W.~N., {Gallagher}, S.~C., and {Garmire}, G.~P. (2002).
\newblock {CHANDRA Detects Relativistic Broad Absorption Lines from APM 08279+5255}.
\newblock \emph{\apj} 579, 169--175.
\newblock \doi{10.1086/342744}
\bibAnnoteFile{Chartas2002}

\bibitem[{{Chartas} et~al.(2009){Chartas}, {Kochanek}, {Dai}, {Poindexter}, and {Garmire}}]{Chartas2009}
{Chartas}, G., {Kochanek}, C.~S., {Dai}, X., {Poindexter}, S., and {Garmire}, G. (2009).
\newblock {X-Ray Microlensing in RXJ1131-1231 and HE1104-1805}.
\newblock \emph{\apj} 693, 174--185.
\newblock \doi{10.1088/0004-637X/693/1/174}
\bibAnnoteFile{Chartas2009}

\bibitem[{{Chartas} et~al.(2016){Chartas}, {Rhea}, {Kochanek}, {Dai}, {Morgan}, {Blackburne} et~al.}]{Chartas2016}
{Chartas}, G., {Rhea}, C., {Kochanek}, C., {Dai}, X., {Morgan}, C., {Blackburne}, J., et~al. (2016).
\newblock {Gravitational lensing size scales for quasars}.
\newblock \emph{Astronomische Nachrichten} 337, 356.
\newblock \doi{10.1002/asna.201612313}
\bibAnnoteFile{Chartas2016}

\bibitem[{{Coppi}(1999)}]{Coppi1999}
{Coppi}, P.~S. (1999).
\newblock {The Physics of Hybrid Thermal/Non-Thermal Plasmas}.
\newblock In \emph{High Energy Processes in Accreting Black Holes}, eds. J.~{Poutanen} and R.~{Svensson}. vol. 161 of \emph{Astronomical Society of the Pacific Conference Series}, 375.
\newblock \doi{10.48550/arXiv.astro-ph/9903158}
\bibAnnoteFile{Coppi1999}

\bibitem[{{Dadina}(2007)}]{dadina-2007}
{Dadina}, M. (2007).
\newblock {BeppoSAX observations in the 2--100 keV band of the nearby Seyfert galaxies: an atlas of spectra}.
\newblock \emph{\aap} 461, 1209--1252.
\newblock \doi{10.1051/0004-6361:20065734}
\bibAnnoteFile{dadina-2007}

\bibitem[{{Dauser} et~al.(2020){Dauser}, {Garc{\'\i}a}, {Parker}, {Fabian}, {Wilms}, {Lohfink} et~al.}]{Dauser2020}
[Dataset] {Dauser}, T., {Garc{\'\i}a}, J., {Parker}, M.~L., {Fabian}, A.~C., {Wilms}, J., {Lohfink}, A., et~al. (2020).
\newblock {relxill: Reflection models of black hole accretion disks}.
\newblock Astrophysics Source Code Library, record ascl:2010.015
\bibAnnoteFile{Dauser2020}

\bibitem[{{Dauser} et~al.(2016){Dauser}, {Garc{\'\i}a}, {Walton}, {Eikmann}, {Kallman}, {McClintock} et~al.}]{Dauser2016}
{Dauser}, T., {Garc{\'\i}a}, J., {Walton}, D.~J., {Eikmann}, W., {Kallman}, T., {McClintock}, J., et~al. (2016).
\newblock {Normalizing a relativistic model of X-ray reflection. Definition of the reflection fraction and its implementation in relxill}.
\newblock \emph{\aap} 590, A76.
\newblock \doi{10.1051/0004-6361/201628135}
\bibAnnoteFile{Dauser2016}

\bibitem[{{Dauser} et~al.(2013){Dauser}, {Garcia}, {Wilms}, {B{\"o}ck}, {Brenneman}, {Falanga} et~al.}]{Dauser2013}
{Dauser}, T., {Garcia}, J., {Wilms}, J., {B{\"o}ck}, M., {Brenneman}, L.~W., {Falanga}, M., et~al. (2013).
\newblock {Irradiation of an accretion disc by a jet: general properties and implications for spin measurements of black holes}.
\newblock \emph{\mnras} 430, 1694--1708.
\newblock \doi{10.1093/mnras/sts710}
\bibAnnoteFile{Dauser2013}

\bibitem[{{De Marco} et~al.(2013){De Marco}, {Ponti}, {Cappi}, {Dadina}, {Uttley}, {Cackett} et~al.}]{DeMarco2013}
{De Marco}, B., {Ponti}, G., {Cappi}, M., {Dadina}, M., {Uttley}, P., {Cackett}, E.~M., et~al. (2013).
\newblock {Discovery of a relation between black hole mass and soft X-ray time lags in active galactic nuclei}.
\newblock \emph{\mnras} 431, 2441--2452.
\newblock \doi{10.1093/mnras/stt339}
\bibAnnoteFile{DeMarco2013}

\bibitem[{{de Rosa} et~al.(2012){de Rosa}, {Panessa}, {Bassani}, {Bazzano}, {Bird}, {Landi} et~al.}]{deRosa+2012}
{de Rosa}, A., {Panessa}, F., {Bassani}, L., {Bazzano}, A., {Bird}, A., {Landi}, R., et~al. (2012).
\newblock {Broad-band study of hard X-ray-selected absorbed active galactic nuclei}.
\newblock \emph{\mnras} 420, 2087--2101.
\newblock \doi{10.1111/j.1365-2966.2011.20167.x}
\bibAnnoteFile{deRosa+2012}

\bibitem[{{Di Matteo} et~al.(1997){Di Matteo}, {Celotti}, and {Fabian}}]{DiMatteo1997}
{Di Matteo}, T., {Celotti}, A., and {Fabian}, A.~C. (1997).
\newblock {Cyclo-synchrotron emission from magnetically dominated active regions above accretion discs}.
\newblock \emph{\mnras} 291, 805--810.
\newblock \doi{10.1093/mnras/291.4.805}
\bibAnnoteFile{DiMatteo1997}

\bibitem[{{Doi} and {Inoue}(2016)}]{Doi2016}
{Doi}, A. and {Inoue}, Y. (2016).
\newblock {High-frequency excess in the radio continuum spectrum of the type-1 Seyfert galaxy NGC 985}.
\newblock \emph{\pasj} 68, 56.
\newblock \doi{10.1093/pasj/psw052}
\bibAnnoteFile{Doi2016}

\bibitem[{{Done} et~al.(2013){Done}, {Jin}, {Middleton}, and {Ward}}]{Done2013}
{Done}, C., {Jin}, C., {Middleton}, M., and {Ward}, M. (2013).
\newblock {A new way to measure supermassive black hole spin in accretion disc-dominated active galaxies}.
\newblock \emph{\mnras} 434, 1955--1963.
\newblock \doi{10.1093/mnras/stt1138}
\bibAnnoteFile{Done2013}

\bibitem[{{Dov{\v{c}}iak} et~al.(2022){Dov{\v{c}}iak}, {Papadakis}, {Kammoun}, and {Zhang}}]{Dovciak2022}
{Dov{\v{c}}iak}, M., {Papadakis}, I.~E., {Kammoun}, E.~S., and {Zhang}, W. (2022).
\newblock {Physical model for the broadband energy spectrum of X-ray illuminated accretion discs: Fitting the spectral energy distribution of NGC 5548}.
\newblock \emph{\aap} 661, A135.
\newblock \doi{10.1051/0004-6361/202142358}
\bibAnnoteFile{Dovciak2022}

\bibitem[{{Elitzur} et~al.(2014){Elitzur}, {Ho}, and {Trump}}]{Elitzur2014}
{Elitzur}, M., {Ho}, L.~C., and {Trump}, J.~R. (2014).
\newblock {Evolution of broad-line emission from active galactic nuclei}.
\newblock \emph{\mnras} 438, 3340--3351.
\newblock \doi{10.1093/mnras/stt2445}
\bibAnnoteFile{Elitzur2014}

\bibitem[{{Emmanoulopoulos} et~al.(2014){Emmanoulopoulos}, {Papadakis}, {Dov{\v{c}}iak}, and {McHardy}}]{Emmanoulopoulos2014}
{Emmanoulopoulos}, D., {Papadakis}, I.~E., {Dov{\v{c}}iak}, M., and {McHardy}, I.~M. (2014).
\newblock {General relativistic modelling of the negative reverberation X-ray time delays in AGN}.
\newblock \emph{\mnras} 439, 3931--3950.
\newblock \doi{10.1093/mnras/stu249}
\bibAnnoteFile{Emmanoulopoulos2014}

\bibitem[{{Eraerds} et~al.(2021){Eraerds}, {Antonelli}, {Davis}, {Hall}, {Hetherington}, {Holland} et~al.}]{Eraerds2021}
{Eraerds}, T., {Antonelli}, V., {Davis}, C., {Hall}, D., {Hetherington}, O., {Holland}, A., et~al. (2021).
\newblock {Enhanced simulations on the Athena/Wide Field Imager instrumental background}.
\newblock \emph{Journal of Astronomical Telescopes, Instruments, and Systems} 7, 034001.
\newblock \doi{10.1117/1.JATIS.7.3.034001}
\bibAnnoteFile{Eraerds2021}

\bibitem[{{Fabian} and {Iwasawa}(1999)}]{Fabian1999}
{Fabian}, A.~C. and {Iwasawa}, K. (1999).
\newblock {The mass density in black holes inferred from the X-ray background}.
\newblock \emph{\mnras} 303, L34--L36.
\newblock \doi{10.1046/j.1365-8711.1999.02404.x}
\bibAnnoteFile{Fabian1999}

\bibitem[{{Fabian} et~al.(2017){Fabian}, {Lohfink}, {Belmont}, {Malzac}, and {Coppi}}]{Fabian2017}
{Fabian}, A.~C., {Lohfink}, A., {Belmont}, R., {Malzac}, J., and {Coppi}, P. (2017).
\newblock {Properties of AGN coronae in the NuSTAR era - II. Hybrid plasma}.
\newblock \emph{\mnras} 467, 2566--2570.
\newblock \doi{10.1093/mnras/stx221}
\bibAnnoteFile{Fabian2017}

\bibitem[{{Fabian} et~al.(2015){Fabian}, {Lohfink}, {Kara}, {Parker}, {Vasudevan}, and {Reynolds}}]{Fabian+15}
{Fabian}, A.~C., {Lohfink}, A., {Kara}, E., {Parker}, M.~L., {Vasudevan}, R., and {Reynolds}, C.~S. (2015).
\newblock {Properties of AGN coronae in the NuSTAR era}.
\newblock \emph{\mnras} 451, 4375--4383.
\newblock \doi{10.1093/mnras/stv1218}
\bibAnnoteFile{Fabian+15}

\bibitem[{{Fabian} et~al.(2014){Fabian}, {Parker}, {Wilkins}, {Miller}, {Kara}, {Reynolds} et~al.}]{Fabian2014}
{Fabian}, A.~C., {Parker}, M.~L., {Wilkins}, D.~R., {Miller}, J.~M., {Kara}, E., {Reynolds}, C.~S., et~al. (2014).
\newblock {On the determination of the spin and disc truncation of accreting black holes using X-ray reflection}.
\newblock \emph{\mnras} 439, 2307--2313.
\newblock \doi{10.1093/mnras/stu045}
\bibAnnoteFile{Fabian2014}

\bibitem[{{Fabian} et~al.(1989){Fabian}, {Rees}, {Stella}, and {White}}]{Fabian1989}
{Fabian}, A.~C., {Rees}, M.~J., {Stella}, L., and {White}, N.~E. (1989).
\newblock {X-ray fluorescence from the inner disc in Cygnus X-1.}
\newblock \emph{\mnras} 238, 729--736.
\newblock \doi{10.1093/mnras/238.3.729}
\bibAnnoteFile{Fabian1989}

\bibitem[{{Fabian} et~al.(2009){Fabian}, {Zoghbi}, {Ross}, {Uttley}, {Gallo}, {Brandt} et~al.}]{Fabian2009}
{Fabian}, A.~C., {Zoghbi}, A., {Ross}, R.~R., {Uttley}, P., {Gallo}, L.~C., {Brandt}, W.~N., et~al. (2009).
\newblock {Broad line emission from iron K- and L-shell transitions in the active galaxy 1H0707-495}.
\newblock \emph{\nat} 459, 540--542.
\newblock \doi{10.1038/nature08007}
\bibAnnoteFile{Fabian2009}

\bibitem[{{Fabian} et~al.(2012){Fabian}, {Zoghbi}, {Wilkins}, {Dwelly}, {Uttley}, {Schartel} et~al.}]{Fabian2012}
{Fabian}, A.~C., {Zoghbi}, A., {Wilkins}, D., {Dwelly}, T., {Uttley}, P., {Schartel}, N., et~al. (2012).
\newblock {1H 0707-495 in 2011: an X-ray source within a gravitational radius of the event horizon}.
\newblock \emph{\mnras} 419, 116--123.
\newblock \doi{10.1111/j.1365-2966.2011.19676.x}
\bibAnnoteFile{Fabian2012}

\bibitem[{{Falcke} et~al.(2004){Falcke}, {K{\"o}rding}, and {Markoff}}]{Falcke2004}
{Falcke}, H., {K{\"o}rding}, E., and {Markoff}, S. (2004).
\newblock {A scheme to unify low-power accreting black holes. Jet-dominated accretion flows and the radio/X-ray correlation}.
\newblock \emph{\aap} 414, 895--903.
\newblock \doi{10.1051/0004-6361:20031683}
\bibAnnoteFile{Falcke2004}

\bibitem[{{Galeev} et~al.(1979){Galeev}, {Rosner}, and {Vaiana}}]{Galeev1979}
{Galeev}, A.~A., {Rosner}, R., and {Vaiana}, G.~S. (1979).
\newblock {Structured coronae of accretion disks.}
\newblock \emph{\apj} 229, 318--326.
\newblock \doi{10.1086/156957}
\bibAnnoteFile{Galeev1979}

\bibitem[{{Gallo}(2018)}]{Gallo2018}
{Gallo}, L. (2018).
\newblock {X-ray perspective of Narrow-line Seyfert 1 galaxies}.
\newblock In \emph{Revisiting Narrow-Line Seyfert 1 Galaxies and their Place in the Universe}. 34.
\newblock \doi{10.22323/1.328.0034}
\bibAnnoteFile{Gallo2018}

\bibitem[{{Gallo}(2006)}]{Gallo2006}
{Gallo}, L.~C. (2006).
\newblock {Investigating the nature of narrow-line Seyfert 1 galaxies with high-energy spectral complexity}.
\newblock \emph{\mnras} 368, 479--486.
\newblock \doi{10.1111/j.1365-2966.2006.10137.x}
\bibAnnoteFile{Gallo2006}

\bibitem[{{Gallo} et~al.(2021){Gallo}, {Gonzalez}, and {Miller}}]{Gallo2021}
{Gallo}, L.~C., {Gonzalez}, A.~G., and {Miller}, J.~M. (2021).
\newblock {Eclipsing the X-Ray Emitting Region in the Active Galaxy NGC 6814}.
\newblock \emph{\apjl} 908, L33.
\newblock \doi{10.3847/2041-8213/abdcb5}
\bibAnnoteFile{Gallo2021}

\bibitem[{{Garc{\'\i}a} et~al.(2014){Garc{\'\i}a}, {Dauser}, {Lohfink}, {Kallman}, {Steiner}, {McClintock} et~al.}]{Garcia2014}
{Garc{\'\i}a}, J., {Dauser}, T., {Lohfink}, A., {Kallman}, T.~R., {Steiner}, J.~F., {McClintock}, J.~E., et~al. (2014).
\newblock {Improved Reflection Models of Black Hole Accretion Disks: Treating the Angular Distribution of X-Rays}.
\newblock \emph{\apj} 782, 76.
\newblock \doi{10.1088/0004-637X/782/2/76}
\bibAnnoteFile{Garcia2014}

\bibitem[{{Garc{\'\i}a} et~al.(2013){Garc{\'\i}a}, {Dauser}, {Reynolds}, {Kallman}, {McClintock}, {Wilms} et~al.}]{Garcia2013}
{Garc{\'\i}a}, J., {Dauser}, T., {Reynolds}, C.~S., {Kallman}, T.~R., {McClintock}, J.~E., {Wilms}, J., et~al. (2013).
\newblock {X-Ray Reflected Spectra from Accretion Disk Models. III. A Complete Grid of Ionized Reflection Calculations}.
\newblock \emph{\apj} 768, 146.
\newblock \doi{10.1088/0004-637X/768/2/146}
\bibAnnoteFile{Garcia2013}

\bibitem[{García et~al.(2015)García, Dauser, Steiner, McClintock, Keck, and Wilms}]{Garcia2015}
García, J.~A., Dauser, T., Steiner, J.~F., McClintock, J.~E., Keck, M.~L., and Wilms, J. (2015).
\newblock On estimating the high-energy cutoff in the x-ray spectra of black holes via reflection spectroscopy.
\newblock \emph{The Astrophysical Journal Letters} 808, L37.
\newblock \doi{10.1088/2041-8205/808/2/L37}
\bibAnnoteFile{Garcia2015}

\bibitem[{{Ghisellini} et~al.(1993){Ghisellini}, {Haardt}, and {Fabian}}]{Ghisellini1993}
{Ghisellini}, G., {Haardt}, F., and {Fabian}, A.~C. (1993).
\newblock {On re-acceleration, pairs and the high-energy spectrum of AGN and galactic black hole candidates.}
\newblock \emph{\mnras} 263, L9--L12.
\newblock \doi{10.1093/mnras/263.1.L9}
\bibAnnoteFile{Ghisellini1993}

\bibitem[{{Ghisellini} et~al.(1994){Ghisellini}, {Haardt}, and {Matt}}]{Ghisellini1994}
{Ghisellini}, G., {Haardt}, F., and {Matt}, G. (1994).
\newblock {The contribution of the obscuring torus to the X-ray spectrum of Seyfert galaxies: a test for the unification model.}
\newblock \emph{\mnras} 267, 743--754.
\newblock \doi{10.1093/mnras/267.3.743}
\bibAnnoteFile{Ghisellini1994}

\bibitem[{{Ghisellini} et~al.(2004){Ghisellini}, {Haardt}, and {Matt}}]{Ghisellini2004}
{Ghisellini}, G., {Haardt}, F., and {Matt}, G. (2004).
\newblock {Aborted jets and the X-ray emission of radio-quiet AGNs}.
\newblock \emph{\aap} 413, 535--545.
\newblock \doi{10.1051/0004-6361:20031562}
\bibAnnoteFile{Ghisellini2004}

\bibitem[{{Gianolli} et~al.(2023){Gianolli}, {Kim}, {Bianchi}, {Ag{\'\i}s-Gonz{\'a}lez}, {Madejski}, {Marin} et~al.}]{Gianolli2023}
{Gianolli}, V.~E., {Kim}, D.~E., {Bianchi}, S., {Ag{\'\i}s-Gonz{\'a}lez}, B., {Madejski}, G., {Marin}, F., et~al. (2023).
\newblock {Uncovering the geometry of the hot X-ray corona in the Seyfert galaxy NGC 4151 with IXPE}.
\newblock \emph{\mnras} 523, 4468--4476.
\newblock \doi{10.1093/mnras/stad1697}
\bibAnnoteFile{Gianolli2023}

\bibitem[{{Gilli} et~al.(2007){Gilli}, {Comastri}, and {Hasinger}}]{Gilli07}
{Gilli}, R., {Comastri}, A., and {Hasinger}, G. (2007).
\newblock {The synthesis of the cosmic X-ray background in the Chandra and XMM-Newton era}.
\newblock \emph{\aap} 463, 79--96.
\newblock \doi{10.1051/0004-6361:20066334}
\bibAnnoteFile{Gilli07}

\bibitem[{{Gonzalez} et~al.(2017){Gonzalez}, {Wilkins}, and {Gallo}}]{Gonzalez2017}
{Gonzalez}, A.~G., {Wilkins}, D.~R., and {Gallo}, L.~C. (2017).
\newblock {Probing the geometry and motion of AGN coronae through accretion disc emissivity profiles}.
\newblock \emph{\mnras} 472, 1932--1945.
\newblock \doi{10.1093/mnras/stx2080}
\bibAnnoteFile{Gonzalez2017}

\bibitem[{{Goosmann} and {Matt}(2011)}]{Goosmann2011}
{Goosmann}, R.~W. and {Matt}, G. (2011).
\newblock {Spotting the misaligned outflows in NGC 1068 using X-ray polarimetry}.
\newblock \emph{\mnras} 415, 3119--3128.
\newblock \doi{10.1111/j.1365-2966.2011.18923.x}
\bibAnnoteFile{Goosmann2011}

\bibitem[{{Graham} et~al.(2019){Graham}, {Kulkarni}, {Bellm}, {Adams}, {Barbarino}, {Blagorodnova} et~al.}]{Graham2019}
{Graham}, M.~J., {Kulkarni}, S.~R., {Bellm}, E.~C., {Adams}, S.~M., {Barbarino}, C., {Blagorodnova}, N., et~al. (2019).
\newblock {The Zwicky Transient Facility: Science Objectives}.
\newblock \emph{\pasp} 131, 078001.
\newblock \doi{10.1088/1538-3873/ab006c}
\bibAnnoteFile{Graham2019}

\bibitem[{{Guedel} and {Benz}(1993)}]{Guedel1993}
{Guedel}, M. and {Benz}, A.~O. (1993).
\newblock {X-Ray/Microwave Relation of Different Types of Active Stars}.
\newblock \emph{\apjl} 405, L63.
\newblock \doi{10.1086/186766}
\bibAnnoteFile{Guedel1993}

\bibitem[{{Guilbert} et~al.(1983){Guilbert}, {Fabian}, and {Rees}}]{Guilbert1983}
{Guilbert}, P.~W., {Fabian}, A.~C., and {Rees}, M.~J. (1983).
\newblock {Spectral and variability constraints on compact sources}.
\newblock \emph{\mnras} 205, 593--603.
\newblock \doi{10.1093/mnras/205.3.593}
\bibAnnoteFile{Guilbert1983}

\bibitem[{{Gupta} et~al.(2023){Gupta}, {Sridhar}, and {Sironi}}]{Gupta+23}
{Gupta}, S., {Sridhar}, N., and {Sironi}, L. (2023).
\newblock {Comptonization by Reconnection Plasmoids in Black Hole Coronae III: Dependence on the Guide Field in Pair Plasma}.
\newblock \emph{arXiv e-prints} , arXiv:2310.04233\doi{10.48550/arXiv.2310.04233}
\bibAnnoteFile{Gupta+23}

\bibitem[{{Haardt} and {Maraschi}(1991)}]{Haardt1991}
{Haardt}, F. and {Maraschi}, L. (1991).
\newblock {A Two-Phase Model for the X-Ray Emission from Seyfert Galaxies}.
\newblock \emph{\apjl} 380, L51.
\newblock \doi{10.1086/186171}
\bibAnnoteFile{Haardt1991}

\bibitem[{{Haardt} and {Maraschi}(1993)}]{Haardt1993}
{Haardt}, F. and {Maraschi}, L. (1993).
\newblock {X-Ray Spectra from Two-Phase Accretion Disks}.
\newblock \emph{\apj} 413, 507.
\newblock \doi{10.1086/173020}
\bibAnnoteFile{Haardt1993}

\bibitem[{{Harrison} et~al.(2013){Harrison}, {Craig}, {Christensen}, {Hailey}, {Zhang}, {Boggs} et~al.}]{Harrison2013}
{Harrison}, F.~A., {Craig}, W.~W., {Christensen}, F.~E., {Hailey}, C.~J., {Zhang}, W.~W., {Boggs}, S.~E., et~al. (2013).
\newblock {The Nuclear Spectroscopic Telescope Array (NuSTAR) High-energy X-Ray Mission}.
\newblock \emph{\apj} 770, 103.
\newblock \doi{10.1088/0004-637X/770/2/103}
\bibAnnoteFile{Harrison2013}

\bibitem[{{Henri} and {Pelletier}(1991)}]{Henri1991}
{Henri}, G. and {Pelletier}, G. (1991).
\newblock {Relativistic Electron-Positron Beam Formation in the Framework of the Two-Flow Model for Active Galactic Nuclei}.
\newblock \emph{\apjl} 383, L7.
\newblock \doi{10.1086/186228}
\bibAnnoteFile{Henri1991}

\bibitem[{{Henri} and {Petrucci}(1997)}]{Henri1997}
{Henri}, G. and {Petrucci}, P.~O. (1997).
\newblock {Anisotropic illumination of AGN's accretion disk by a non thermal source. I. General theory and application to the Newtonian geometry.}
\newblock \emph{\aap} 326, 87--98.
\newblock \doi{10.48550/arXiv.astro-ph/9705233}
\bibAnnoteFile{Henri1997}

\bibitem[{{Hickox} and {Alexander}(2018)}]{hickox+alexander-2018}
{Hickox}, R.~C. and {Alexander}, D.~M. (2018).
\newblock {Obscured Active Galactic Nuclei}.
\newblock \emph{\araa} 56, 625--671.
\newblock \doi{10.1146/annurev-astro-081817-051803}
\bibAnnoteFile{hickox+alexander-2018}

\bibitem[{{Ingram} et~al.(2023){Ingram}, {Ewing}, {Marinucci}, {Tagliacozzo}, {Rosario}, {Veledina} et~al.}]{Ingram2023}
{Ingram}, A., {Ewing}, M., {Marinucci}, A., {Tagliacozzo}, D., {Rosario}, D.~J., {Veledina}, A., et~al. (2023).
\newblock {The X-ray polarisation of the Seyfert 1 galaxy IC 4329A}.
\newblock \emph{arXiv e-prints} , arXiv:2305.13028\doi{10.48550/arXiv.2305.13028}
\bibAnnoteFile{Ingram2023}

\bibitem[{{Ingram} et~al.(2019){Ingram}, {Mastroserio}, {Dauser}, {Hovenkamp}, {van der Klis}, and {Garc{\'\i}a}}]{Ingram2019}
{Ingram}, A., {Mastroserio}, G., {Dauser}, T., {Hovenkamp}, P., {van der Klis}, M., and {Garc{\'\i}a}, J.~A. (2019).
\newblock {A public relativistic transfer function model for X-ray reverberation mapping of accreting black holes}.
\newblock \emph{\mnras} 488, 324--347.
\newblock \doi{10.1093/mnras/stz1720}
\bibAnnoteFile{Ingram2019}

\bibitem[{{Inoue} and {Doi}(2014)}]{Inoue2014}
{Inoue}, Y. and {Doi}, A. (2014).
\newblock {Unveiling the nature of coronae in active galactic nuclei through submillimeter observations}.
\newblock \emph{\pasj} 66, L8.
\newblock \doi{10.1093/pasj/psu079}
\bibAnnoteFile{Inoue2014}

\bibitem[{{Inoue} and {Doi}(2018)}]{Inoue2018}
{Inoue}, Y. and {Doi}, A. (2018).
\newblock {Detection of Coronal Magnetic Activity in nearby Active Supermassive Black Holes}.
\newblock \emph{\apj} 869, 114.
\newblock \doi{10.3847/1538-4357/aaeb95}
\bibAnnoteFile{Inoue2018}

\bibitem[{{Jahoda} et~al.(2019){Jahoda}, {Krawczynski}, {Kislat}, {Marshall}, {Okajima}, {Agudo} et~al.}]{Jahoda2019}
{Jahoda}, K., {Krawczynski}, H., {Kislat}, F., {Marshall}, H., {Okajima}, T., {Agudo}, I., et~al. (2019).
\newblock {The X-ray Polarization Probe mission concept}.
\newblock \emph{arXiv e-prints} , arXiv:1907.10190\doi{10.48550/arXiv.1907.10190}
\bibAnnoteFile{Jahoda2019}

\bibitem[{{Jansen} et~al.(2001){Jansen}, {Lumb}, {Altieri}, {Clavel}, {Ehle}, {Erd} et~al.}]{Jansen2001}
{Jansen}, F., {Lumb}, D., {Altieri}, B., {Clavel}, J., {Ehle}, M., {Erd}, C., et~al. (2001).
\newblock {XMM-Newton observatory. I. The spacecraft and operations}.
\newblock \emph{\aap} 365, L1--L6.
\newblock \doi{10.1051/0004-6361:20000036}
\bibAnnoteFile{Jansen2001}

\bibitem[{{Jiang} et~al.(2018){Jiang}, {Parker}, {Fabian}, {Alston}, {Buisson}, {Cackett} et~al.}]{Jiang2018}
{Jiang}, J., {Parker}, M.~L., {Fabian}, A.~C., {Alston}, W.~N., {Buisson}, D.~J.~K., {Cackett}, E.~M., et~al. (2018).
\newblock {The 1.5 Ms observing campaign on IRAS 13224-3809 - I. X-ray spectral analysis}.
\newblock \emph{\mnras} 477, 3711--3726.
\newblock \doi{10.1093/mnras/sty836}
\bibAnnoteFile{Jiang2018}

\bibitem[{{Kammoun} et~al.(2023){Kammoun}, {Igo}, {Miller}, {Fabian}, {Reynolds}, {Merloni} et~al.}]{Kammoun2023}
{Kammoun}, E.~S., {Igo}, Z., {Miller}, J.~M., {Fabian}, A.~C., {Reynolds}, M.~T., {Merloni}, A., et~al. (2023).
\newblock {The first X-ray look at SMSS J114447.77-430859.3: the most luminous quasar in the last 9 Gyr}.
\newblock \emph{\mnras} 522, 5217--5237.
\newblock \doi{10.1093/mnras/stad952}
\bibAnnoteFile{Kammoun2023}

\bibitem[{{Kammoun} et~al.(2020){Kammoun}, {Miller}, {Koss}, {Oh}, {Zoghbi}, {Mushotzky} et~al.}]{Kammoun2020}
{Kammoun}, E.~S., {Miller}, J.~M., {Koss}, M., {Oh}, K., {Zoghbi}, A., {Mushotzky}, R.~F., et~al. (2020).
\newblock {A Hard Look at Local, Optically Selected, Obscured Seyfert Galaxies}.
\newblock \emph{\apj} 901, 161.
\newblock \doi{10.3847/1538-4357/abb29f}
\bibAnnoteFile{Kammoun2020}

\bibitem[{{Kammoun} et~al.(2017){Kammoun}, {Risaliti}, {Stern}, {Jun}, {Graham}, {Celotti} et~al.}]{Kammoun2017}
{Kammoun}, E.~S., {Risaliti}, G., {Stern}, D., {Jun}, H.~D., {Graham}, M., {Celotti}, A., et~al. (2017).
\newblock {Coronal properties of the luminous radio-quiet quasar QSO B2202-209}.
\newblock \emph{\mnras} 465, 1665--1671.
\newblock \doi{10.1093/mnras/stw2897}
\bibAnnoteFile{Kammoun2017}

\bibitem[{{Kamraj} et~al.(2022){Kamraj}, {Brightman}, {Harrison}, {Stern}, {Garc{\'\i}a}, {Balokovi{\'c}} et~al.}]{Kamraj2022}
{Kamraj}, N., {Brightman}, M., {Harrison}, F.~A., {Stern}, D., {Garc{\'\i}a}, J.~A., {Balokovi{\'c}}, M., et~al. (2022).
\newblock {X-Ray Coronal Properties of Swift/BAT-selected Seyfert 1 Active Galactic Nuclei}.
\newblock \emph{\apj} 927, 42.
\newblock \doi{10.3847/1538-4357/ac45f6}
\bibAnnoteFile{Kamraj2022}

\bibitem[{{Kamraj} et~al.(2018){Kamraj}, {Harrison}, {Balokovi{\'c}}, {Lohfink}, and {Brightman}}]{Kamraj2018}
{Kamraj}, N., {Harrison}, F.~A., {Balokovi{\'c}}, M., {Lohfink}, A., and {Brightman}, M. (2018).
\newblock {Coronal Properties of Swift/BAT-selected Seyfert 1 AGNs Observed with NuSTAR}.
\newblock \emph{\apj} 866, 124.
\newblock \doi{10.3847/1538-4357/aadd0d}
\bibAnnoteFile{Kamraj2018}

\bibitem[{{Kara} et~al.(2016){Kara}, {Alston}, {Fabian}, {Cackett}, {Uttley}, {Reynolds} et~al.}]{Kara2016}
{Kara}, E., {Alston}, W.~N., {Fabian}, A.~C., {Cackett}, E.~M., {Uttley}, P., {Reynolds}, C.~S., et~al. (2016).
\newblock {A global look at X-ray time lags in Seyfert galaxies}.
\newblock \emph{\mnras} 462, 511--531.
\newblock \doi{10.1093/mnras/stw1695}
\bibAnnoteFile{Kara2016}

\bibitem[{{Kara} et~al.(2017){Kara}, {Garc{\'\i}a}, {Lohfink}, {Fabian}, {Reynolds}, {Tombesi} et~al.}]{Kara2017}
{Kara}, E., {Garc{\'\i}a}, J.~A., {Lohfink}, A., {Fabian}, A.~C., {Reynolds}, C.~S., {Tombesi}, F., et~al. (2017).
\newblock {The high-Eddington NLS1 Ark 564 has the coolest corona}.
\newblock \emph{\mnras} 468, 3489--3498.
\newblock \doi{10.1093/mnras/stx792}
\bibAnnoteFile{Kara2017}

\bibitem[{{Kara} et~al.(2015){Kara}, {Zoghbi}, {Marinucci}, {Walton}, {Fabian}, {Risaliti} et~al.}]{Kara2015}
{Kara}, E., {Zoghbi}, A., {Marinucci}, A., {Walton}, D.~J., {Fabian}, A.~C., {Risaliti}, G., et~al. (2015).
\newblock {Iron K and Compton hump reverberation in SWIFT J2127.4+5654 and NGC 1365 revealed by NuSTAR and XMM-Newton}.
\newblock \emph{\mnras} 446, 737--749.
\newblock \doi{10.1093/mnras/stu2136}
\bibAnnoteFile{Kara2015}

\bibitem[{{Kawamuro} et~al.(2022){Kawamuro}, {Ricci}, {Imanishi}, {Mushotzky}, {Izumi}, {Ricci} et~al.}]{Kawamuro2022}
{Kawamuro}, T., {Ricci}, C., {Imanishi}, M., {Mushotzky}, R.~F., {Izumi}, T., {Ricci}, F., et~al. (2022).
\newblock {BASS XXXII: Studying the Nuclear Millimeter-wave Continuum Emission of AGNs with ALMA at Scales {\ensuremath{\lesssim}}100-200 pc}.
\newblock \emph{\apj} 938, 87.
\newblock \doi{10.3847/1538-4357/ac8794}
\bibAnnoteFile{Kawamuro2022}

\bibitem[{{Kawamuro} et~al.(2023){Kawamuro}, {Ricci}, {Mushotzky}, {Imanishi}, {Bauer}, {Ricci} et~al.}]{Kawamuro2023}
{Kawamuro}, T., {Ricci}, C., {Mushotzky}, R.~F., {Imanishi}, M., {Bauer}, F.~E., {Ricci}, F., et~al. (2023).
\newblock {BASS XXXIV: A Catalog of the Nuclear Mm-wave Continuum Emission Properties of AGNs Constrained on Scales $\lesssim$ 100--200 pc}.
\newblock \emph{arXiv e-prints} , arXiv:2309.02776\doi{10.48550/arXiv.2309.02776}
\bibAnnoteFile{Kawamuro2023}

\bibitem[{{Kochanek} et~al.(2007{\natexlab{a}}){Kochanek}, {Dai}, {Morgan}, {Morgan}, and {Poindexter}}]{Kochanek2004}
{Kochanek}, C.~S., {Dai}, X., {Morgan}, C., {Morgan}, N., and {Poindexter}, G., S.~Chartas (2007{\natexlab{a}}).
\newblock {Turning AGN Microlensing from a Curiosity into a Tool}.
\newblock In \emph{Statistical Challenges in Modern Astronomy IV}, eds. G.~J. {Babu} and E.~D. {Feigelson}. vol. 371 of \emph{Astronomical Society of the Pacific Conference Series}, 43.
\newblock \doi{10.48550/arXiv.astro-ph/0609112}
\bibAnnoteFile{Kochanek2004}

\bibitem[{{Kochanek} et~al.(2007{\natexlab{b}}){Kochanek}, {Dai}, {Morgan}, {Morgan}, and {Poindexter}}]{Kochanek2007}
{Kochanek}, C.~S., {Dai}, X., {Morgan}, C., {Morgan}, N., and {Poindexter}, G., S.~Chartas (2007{\natexlab{b}}).
\newblock {Turning AGN Microlensing from a Curiosity into a Tool}.
\newblock In \emph{Statistical Challenges in Modern Astronomy IV}, eds. G.~J. {Babu} and E.~D. {Feigelson}. vol. 371 of \emph{Astronomical Society of the Pacific Conference Series}, 43.
\newblock \doi{10.48550/arXiv.astro-ph/0609112}
\bibAnnoteFile{Kochanek2007}

\bibitem[{{K{\"o}rding} et~al.(2006){K{\"o}rding}, {Jester}, and {Fender}}]{Koerding2006}
{K{\"o}rding}, E.~G., {Jester}, S., and {Fender}, R. (2006).
\newblock {Accretion states and radio loudness in active galactic nuclei: analogies with X-ray binaries}.
\newblock \emph{\mnras} 372, 1366--1378.
\newblock \doi{10.1111/j.1365-2966.2006.10954.x}
\bibAnnoteFile{Koerding2006}

\bibitem[{{Koss} et~al.(2022){Koss}, {Ricci}, {Trakhtenbrot}, {Oh}, {den Brok}, {Mej{\'\i}a-Restrepo} et~al.}]{Koss2022}
{Koss}, M.~J., {Ricci}, C., {Trakhtenbrot}, B., {Oh}, K., {den Brok}, J.~S., {Mej{\'\i}a-Restrepo}, J.~E., et~al. (2022).
\newblock {BASS. XXII. The BASS DR2 AGN Catalog and Data}.
\newblock \emph{\apjs} 261, 2.
\newblock \doi{10.3847/1538-4365/ac6c05}
\bibAnnoteFile{Koss2022}

\bibitem[{{Krawczynski} et~al.(2019){Krawczynski}, {Matt}, {Ingram}, {Taverna}, {Turolla}, {Kislat} et~al.}]{Krawczynski2019}
{Krawczynski}, H., {Matt}, G., {Ingram}, A.~R., {Taverna}, R., {Turolla}, R., {Kislat}, F., et~al. (2019).
\newblock {Astro2020 Science White Paper: Using X-Ray Polarimetry to Probe the Physics of Black Holes and Neutron Stars}.
\newblock \emph{arXiv e-prints} , arXiv:1904.09313\doi{10.48550/arXiv.1904.09313}
\bibAnnoteFile{Krawczynski2019}

\bibitem[{{Krawczynski} et~al.(2022){Krawczynski}, {Muleri}, {Dov{\v{c}}iak}, {Veledina}, {Rodriguez Cavero}, {Svoboda} et~al.}]{Krawczynski2022}
{Krawczynski}, H., {Muleri}, F., {Dov{\v{c}}iak}, M., {Veledina}, A., {Rodriguez Cavero}, N., {Svoboda}, J., et~al. (2022).
\newblock {Polarized x-rays constrain the disk-jet geometry in the black hole x-ray binary Cygnus X-1}.
\newblock \emph{Science} 378, 650--654.
\newblock \doi{10.1126/science.add5399}
\bibAnnoteFile{Krawczynski2022}

\bibitem[{{Laha} et~al.(2022){Laha}, {Meyer}, {Roychowdhury}, {Becerra Gonzalez}, {Acosta-Pulido}, {Thapa} et~al.}]{Laha2022}
{Laha}, S., {Meyer}, E., {Roychowdhury}, A., {Becerra Gonzalez}, J., {Acosta-Pulido}, J.~A., {Thapa}, A., et~al. (2022).
\newblock {A Radio, Optical, UV, and X-Ray View of the Enigmatic Changing-look Active Galactic Nucleus 1ES 1927+654 from Its Pre- to Postflare States}.
\newblock \emph{\apj} 931, 5.
\newblock \doi{10.3847/1538-4357/ac63aa}
\bibAnnoteFile{Laha2022}

\bibitem[{{Lanzuisi} et~al.(2019){Lanzuisi}, {Gilli}, {Cappi}, {Dadina}, {Bianchi}, {Brusa} et~al.}]{Lanzuisi2019}
{Lanzuisi}, G., {Gilli}, R., {Cappi}, M., {Dadina}, M., {Bianchi}, S., {Brusa}, M., et~al. (2019).
\newblock {NuSTAR Measurement of Coronal Temperature in Two Luminous, High-redshift Quasars}.
\newblock \emph{\apjl} 875, L20.
\newblock \doi{10.3847/2041-8213/ab15dc}
\bibAnnoteFile{Lanzuisi2019}

\bibitem[{{Laor} and {Behar}(2008)}]{Laor2008}
{Laor}, A. and {Behar}, E. (2008).
\newblock {On the origin of radio emission in radio-quiet quasars}.
\newblock \emph{\mnras} 390, 847--862.
\newblock \doi{10.1111/j.1365-2966.2008.13806.x}
\bibAnnoteFile{Laor2008}

\bibitem[{{Lewin} et~al.(2022){Lewin}, {Kara}, {Wilkins}, {Mastroserio}, {Garc{\'\i}a}, {Zhang} et~al.}]{Lewin2022}
{Lewin}, C., {Kara}, E., {Wilkins}, D., {Mastroserio}, G., {Garc{\'\i}a}, J.~A., {Zhang}, R.~C., et~al. (2022).
\newblock {X-Ray Reverberation Mapping of Ark 564 Using Gaussian Process Regression}.
\newblock \emph{\apj} 939, 109.
\newblock \doi{10.3847/1538-4357/ac978f}
\bibAnnoteFile{Lewin2022}

\bibitem[{{Liu} et~al.(2014){Liu}, {Wang}, {Yang}, {Zhu}, and {Zhou}}]{liu+2014}
{Liu}, T., {Wang}, J.-X., {Yang}, H., {Zhu}, F.-F., and {Zhou}, Y.-Y. (2014).
\newblock {Are X-Ray Emitting Coronae around Supermassive Black Holes Outflowing?}
\newblock \emph{\apj} 783, 106.
\newblock \doi{10.1088/0004-637X/783/2/106}
\bibAnnoteFile{liu+2014}

\bibitem[{{Liu} et~al.(2023){Liu}, {Malyali}, {Krumpe}, {Homan}, {Goodwin}, {Grotova} et~al.}]{Liu2023}
{Liu}, Z., {Malyali}, A., {Krumpe}, M., {Homan}, D., {Goodwin}, A.~J., {Grotova}, I., et~al. (2023).
\newblock {Deciphering the extreme X-ray variability of the nuclear transient eRASSt J045650.3{\ensuremath{-}}203750. A likely repeating partial tidal disruption event}.
\newblock \emph{\aap} 669, A75.
\newblock \doi{10.1051/0004-6361/202244805}
\bibAnnoteFile{Liu2023}

\bibitem[{{Lubi{\'n}ski} et~al.(2016){Lubi{\'n}ski}, {Beckmann}, {Gibaud}, {Paltani}, {Papadakis}, {Ricci} et~al.}]{Lubinski2016}
{Lubi{\'n}ski}, P., {Beckmann}, V., {Gibaud}, L., {Paltani}, S., {Papadakis}, I.~E., {Ricci}, C., et~al. (2016).
\newblock {A comprehensive analysis of the hard X-ray spectra of bright Seyfert galaxies}.
\newblock \emph{\mnras} 458, 2454--2475.
\newblock \doi{10.1093/mnras/stw454}
\bibAnnoteFile{Lubinski2016}

\bibitem[{{Lusso} et~al.(2012){Lusso}, {Comastri}, {Simmons}, {Mignoli}, {Zamorani}, {Vignali} et~al.}]{Lusso2012}
{Lusso}, E., {Comastri}, A., {Simmons}, B.~D., {Mignoli}, M., {Zamorani}, G., {Vignali}, C., et~al. (2012).
\newblock {Bolometric luminosities and Eddington ratios of X-ray selected active galactic nuclei in the XMM-COSMOS survey}.
\newblock \emph{\mnras} 425, 623--640.
\newblock \doi{10.1111/j.1365-2966.2012.21513.x}
\bibAnnoteFile{Lusso2012}

\bibitem[{{Lusso} and {Risaliti}(2016)}]{Lusso2016}
{Lusso}, E. and {Risaliti}, G. (2016).
\newblock {The Tight Relation between X-Ray and Ultraviolet Luminosity of Quasars}.
\newblock \emph{\apj} 819, 154.
\newblock \doi{10.3847/0004-637X/819/2/154}
\bibAnnoteFile{Lusso2016}

\bibitem[{{Malizia} et~al.(2014){Malizia}, {Molina}, {Bassani}, {Stephen}, {Bazzano}, {Ubertini} et~al.}]{Malizia2014}
{Malizia}, A., {Molina}, M., {Bassani}, L., {Stephen}, J.~B., {Bazzano}, A., {Ubertini}, P., et~al. (2014).
\newblock {The INTEGRAL High-energy Cut-off Distribution of Type 1 Active Galactic Nuclei}.
\newblock \emph{\apjl} 782, L25.
\newblock \doi{10.1088/2041-8205/782/2/L25}
\bibAnnoteFile{Malizia2014}

\bibitem[{{Mallick} et~al.(2018){Mallick}, {Alston}, {Parker}, {Fabian}, {Pinto}, {Dewangan} et~al.}]{Mallick2018}
{Mallick}, L., {Alston}, W.~N., {Parker}, M.~L., {Fabian}, A.~C., {Pinto}, C., {Dewangan}, G.~C., et~al. (2018).
\newblock {A high-density relativistic reflection origin for the soft and hard X-ray excess emission from Mrk 1044}.
\newblock \emph{\mnras} 479, 615--634.
\newblock \doi{10.1093/mnras/sty1487}
\bibAnnoteFile{Mallick2018}

\bibitem[{{Mallick} et~al.(2022){Mallick}, {Fabian}, {Garc{\'\i}a}, {Tomsick}, {Parker}, {Dauser} et~al.}]{Mallick2022}
{Mallick}, L., {Fabian}, A.~C., {Garc{\'\i}a}, J.~A., {Tomsick}, J.~A., {Parker}, M.~L., {Dauser}, T., et~al. (2022).
\newblock {High-density disc reflection spectroscopy of low-mass active galactic nuclei}.
\newblock \emph{\mnras} 513, 4361--4379.
\newblock \doi{10.1093/mnras/stac990}
\bibAnnoteFile{Mallick2022}

\bibitem[{{Mallick} et~al.(2021){Mallick}, {Wilkins}, {Alston}, {Markowitz}, {De Marco}, {Parker} et~al.}]{Mallick2021}
{Mallick}, L., {Wilkins}, D.~R., {Alston}, W.~N., {Markowitz}, A., {De Marco}, B., {Parker}, M.~L., et~al. (2021).
\newblock {Discovery of soft and hard X-ray time lags in low-mass AGNs}.
\newblock \emph{\mnras} 503, 3775--3783.
\newblock \doi{10.1093/mnras/stab627}
\bibAnnoteFile{Mallick2021}

\bibitem[{{Malzac} et~al.(2001){Malzac}, {Beloborodov}, and {Poutanen}}]{Malzac2001}
{Malzac}, J., {Beloborodov}, A.~M., and {Poutanen}, J. (2001).
\newblock {X-ray spectra of accretion discs with dynamic coronae}.
\newblock \emph{\mnras} 326, 417--427.
\newblock \doi{10.1046/j.1365-8711.2001.04450.x}
\bibAnnoteFile{Malzac2001}

\bibitem[{{Marconi} et~al.(2004){Marconi}, {Risaliti}, {Gilli}, {Hunt}, {Maiolino}, and {Salvati}}]{Marconi2004}
{Marconi}, A., {Risaliti}, G., {Gilli}, R., {Hunt}, L.~K., {Maiolino}, R., and {Salvati}, M. (2004).
\newblock {Local supermassive black holes, relics of active galactic nuclei and the X-ray background}.
\newblock \emph{\mnras} 351, 169--185.
\newblock \doi{10.1111/j.1365-2966.2004.07765.x}
\bibAnnoteFile{Marconi2004}

\bibitem[{{Marin} et~al.(2018){Marin}, {Dov{\v{c}}iak}, {Muleri}, {Kislat}, and {Krawczynski}}]{Marin2018}
{Marin}, F., {Dov{\v{c}}iak}, M., {Muleri}, F., {Kislat}, F.~F., and {Krawczynski}, H.~S. (2018).
\newblock {Predicting the X-ray polarization of type 2 Seyfert galaxies}.
\newblock \emph{\mnras} 473, 1286--1316.
\newblock \doi{10.1093/mnras/stx2382}
\bibAnnoteFile{Marin2018}

\bibitem[{{Marinucci} et~al.(2022{\natexlab{a}}){Marinucci}, {Muleri}, {Dovciak}, {Bianchi}, {Marin}, {Matt} et~al.}]{Marinucci2022ixpe}
{Marinucci}, A., {Muleri}, F., {Dovciak}, M., {Bianchi}, S., {Marin}, F., {Matt}, G., et~al. (2022{\natexlab{a}}).
\newblock {Polarization constraints on the X-ray corona in Seyfert Galaxies: MCG-05-23-16}.
\newblock \emph{\mnras} 516, 5907--5913.
\newblock \doi{10.1093/mnras/stac2634}
\bibAnnoteFile{Marinucci2022ixpe}

\bibitem[{{Marinucci} et~al.(2022{\natexlab{b}}){Marinucci}, {Vietri}, {Piconcelli}, {Bianchi}, {Guainazzi}, {Lanzuisi} et~al.}]{Marinucci2022}
{Marinucci}, A., {Vietri}, G., {Piconcelli}, E., {Bianchi}, S., {Guainazzi}, M., {Lanzuisi}, G., et~al. (2022{\natexlab{b}}).
\newblock {Breaking the rules at z = 0.45: The rebel case of RBS 1055}.
\newblock \emph{\aap} 666, A169.
\newblock \doi{10.1051/0004-6361/202244272}
\bibAnnoteFile{Marinucci2022}

\bibitem[{{Markoff} et~al.(2005){Markoff}, {Nowak}, and {Wilms}}]{Markoff2005}
{Markoff}, S., {Nowak}, M.~A., and {Wilms}, J. (2005).
\newblock {Going with the Flow: Can the Base of Jets Subsume the Role of Compact Accretion Disk Coronae?}
\newblock \emph{\apj} 635, 1203--1216.
\newblock \doi{10.1086/497628}
\bibAnnoteFile{Markoff2005}

\bibitem[{{Martocchia} and {Matt}(1996)}]{Martocchia1996}
{Martocchia}, A. and {Matt}, G. (1996).
\newblock {Iron Kalpha line intensity from accretion discs around rotating black holes}.
\newblock \emph{\mnras} 282, L53--L57.
\newblock \doi{10.1093/mnras/282.4.L53}
\bibAnnoteFile{Martocchia1996}

\bibitem[{{Masterson} et~al.(2022){Masterson}, {Kara}, {Ricci}, {Garc{\'\i}a}, {Fabian}, {Pinto} et~al.}]{Masterson2022}
{Masterson}, M., {Kara}, E., {Ricci}, C., {Garc{\'\i}a}, J.~A., {Fabian}, A.~C., {Pinto}, C., et~al. (2022).
\newblock {Evolution of a Relativistic Outflow and X-Ray Corona in the Extreme Changing-look AGN 1ES 1927+654}.
\newblock \emph{\apj} 934, 35.
\newblock \doi{10.3847/1538-4357/ac76c0}
\bibAnnoteFile{Masterson2022}

\bibitem[{{Mastroserio} et~al.(2020){Mastroserio}, {Ingram}, and {van der Klis}}]{Mastroserio2020}
{Mastroserio}, G., {Ingram}, A., and {van der Klis}, M. (2020).
\newblock {Multi-timescale reverberation mapping of Mrk 335}.
\newblock \emph{\mnras} 498, 4971--4982.
\newblock \doi{10.1093/mnras/staa2735}
\bibAnnoteFile{Mastroserio2020}

\bibitem[{{Mastroserio} et~al.(2021){Mastroserio}, {Ingram}, {Wang}, {Garc{\'\i}a}, {van der Klis}, {Cavecchi} et~al.}]{Mastroserio2021}
{Mastroserio}, G., {Ingram}, A., {Wang}, J., {Garc{\'\i}a}, J.~A., {van der Klis}, M., {Cavecchi}, Y., et~al. (2021).
\newblock {Modelling correlated variability in accreting black holes: the effect of high density and variable ionization on reverberation lags}.
\newblock \emph{\mnras} 507, 55--73.
\newblock \doi{10.1093/mnras/stab2056}
\bibAnnoteFile{Mastroserio2021}

\bibitem[{{Matt} et~al.(2014){Matt}, {Marinucci}, {Guainazzi}, {Brenneman}, {Elvis}, {Lohfink} et~al.}]{Matt2014}
{Matt}, G., {Marinucci}, A., {Guainazzi}, M., {Brenneman}, L.~W., {Elvis}, M., {Lohfink}, A., et~al. (2014).
\newblock {The soft-X-ray emission of Ark 120. XMM-Newton, NuSTAR, and the importance of taking the broad view}.
\newblock \emph{\mnras} 439, 3016--3021.
\newblock \doi{10.1093/mnras/stu159}
\bibAnnoteFile{Matt2014}

\bibitem[{{Matt} et~al.(1991){Matt}, {Perola}, and {Piro}}]{Matt1991}
{Matt}, G., {Perola}, G.~C., and {Piro}, L. (1991).
\newblock {The iron line and high energy bump as X-ray signatures of cold matter in Seyfert 1 galaxies.}
\newblock \emph{\aap} 247, 25
\bibAnnoteFile{Matt1991}

\bibitem[{{McHardy} et~al.(2006){McHardy}, {Koerding}, {Knigge}, {Uttley}, and {Fender}}]{McHardy2006}
{McHardy}, I.~M., {Koerding}, E., {Knigge}, C., {Uttley}, P., and {Fender}, R.~P. (2006).
\newblock {Active galactic nuclei as scaled-up Galactic black holes}.
\newblock \emph{\nat} 444, 730--732.
\newblock \doi{10.1038/nature05389}
\bibAnnoteFile{McHardy2006}

\bibitem[{{Meidinger} et~al.(2020){Meidinger}, {Albrecht}, {Beitler}, {Bonholzer}, {Emberger}, {Frank} et~al.}]{Meidinger2020}
{Meidinger}, N., {Albrecht}, S., {Beitler}, C., {Bonholzer}, M., {Emberger}, V., {Frank}, J., et~al. (2020).
\newblock {Development status of the wide field imager instrument for Athena}.
\newblock In \emph{Society of Photo-Optical Instrumentation Engineers (SPIE) Conference Series}. vol. 11444 of \emph{Society of Photo-Optical Instrumentation Engineers (SPIE) Conference Series}, 114440T.
\newblock \doi{10.1117/12.2560507}
\bibAnnoteFile{Meidinger2020}

\bibitem[{{Merloni} et~al.(2015){Merloni}, {Dwelly}, {Salvato}, {Georgakakis}, {Greiner}, {Krumpe} et~al.}]{Merloni2015}
{Merloni}, A., {Dwelly}, T., {Salvato}, M., {Georgakakis}, A., {Greiner}, J., {Krumpe}, M., et~al. (2015).
\newblock {A tidal disruption flare in a massive galaxy? Implications for the fuelling mechanisms of nuclear black holes}.
\newblock \emph{\mnras} 452, 69--87.
\newblock \doi{10.1093/mnras/stv1095}
\bibAnnoteFile{Merloni2015}

\bibitem[{{Merloni} and {Fabian}(2001{\natexlab{a}})}]{Merloni2001a}
{Merloni}, A. and {Fabian}, A.~C. (2001{\natexlab{a}}).
\newblock {Accretion disc coronae as magnetic reservoirs}.
\newblock \emph{\mnras} 321, 549--552.
\newblock \doi{10.1046/j.1365-8711.2001.04060.x}
\bibAnnoteFile{Merloni2001a}

\bibitem[{{Merloni} and {Fabian}(2001{\natexlab{b}})}]{Merloni2001b}
{Merloni}, A. and {Fabian}, A.~C. (2001{\natexlab{b}}).
\newblock {Thunderclouds and accretion discs: a model for the spectral and temporal variability of Seyfert 1 galaxies}.
\newblock \emph{\mnras} 328, 958--968.
\newblock \doi{10.1046/j.1365-8711.2001.04925.x}
\bibAnnoteFile{Merloni2001b}

\bibitem[{{Middei} et~al.(2019){Middei}, {Bianchi}, {Marinucci}, {Matt}, {Petrucci}, {Tamborra} et~al.}]{Middei19}
{Middei}, R., {Bianchi}, S., {Marinucci}, A., {Matt}, G., {Petrucci}, P.~O., {Tamborra}, F., et~al. (2019).
\newblock {Relations between phenomenological and physical parameters in the hot coronae of AGNs computed with the MoCA code}.
\newblock \emph{\aap} 630, A131.
\newblock \doi{10.1051/0004-6361/201935881}
\bibAnnoteFile{Middei19}

\bibitem[{{Miniutti} and {Fabian}(2004)}]{Miniutti2004}
{Miniutti}, G. and {Fabian}, A.~C. (2004).
\newblock {A light bending model for the X-ray temporal and spectral properties of accreting black holes}.
\newblock \emph{\mnras} 349, 1435--1448.
\newblock \doi{10.1111/j.1365-2966.2004.07611.x}
\bibAnnoteFile{Miniutti2004}

\bibitem[{{Molina} et~al.(2013){Molina}, {Bassani}, {Malizia}, {Stephen}, {Bird}, {Bazzano} et~al.}]{molina+2013}
{Molina}, M., {Bassani}, L., {Malizia}, A., {Stephen}, J.~B., {Bird}, A.~J., {Bazzano}, A., et~al. (2013).
\newblock {Hard-X-ray spectra of active galactic nuclei in the INTEGRAL complete sample}.
\newblock \emph{\mnras} 433, 1687--1700.
\newblock \doi{10.1093/mnras/stt844}
\bibAnnoteFile{molina+2013}

\bibitem[{{Mosquera} et~al.(2013){Mosquera}, {Kochanek}, {Chen}, {Dai}, {Blackburne}, and {Chartas}}]{Mosquera2013}
{Mosquera}, A.~M., {Kochanek}, C.~S., {Chen}, B., {Dai}, X., {Blackburne}, J.~A., and {Chartas}, G. (2013).
\newblock {The Structure of the X-Ray and Optical Emitting Regions of the Lensed Quasar Q 2237+0305}.
\newblock \emph{\apj} 769, 53.
\newblock \doi{10.1088/0004-637X/769/1/53}
\bibAnnoteFile{Mosquera2013}

\bibitem[{{Nandra} et~al.(2013){Nandra}, {Barret}, {Barcons}, {Fabian}, {den Herder}, {Piro} et~al.}]{Nandra2013}
{Nandra}, K., {Barret}, D., {Barcons}, X., {Fabian}, A., {den Herder}, J.-W., {Piro}, L., et~al. (2013).
\newblock {The Hot and Energetic Universe: A White Paper presenting the science theme motivating the Athena+ mission}.
\newblock \emph{arXiv e-prints} , arXiv:1306.2307\doi{10.48550/arXiv.1306.2307}
\bibAnnoteFile{Nandra2013}

\bibitem[{{Nicastro} et~al.(2000){Nicastro}, {Piro}, {De Rosa}, {Feroci}, {Grandi}, {Fiore} et~al.}]{Nicastro2000}
{Nicastro}, F., {Piro}, L., {De Rosa}, A., {Feroci}, M., {Grandi}, P., {Fiore}, F., et~al. (2000).
\newblock {A Long Observation of NGC 5548 by BeppoSAX: The High-Energy Cutoff, Intrinsic Spectral Variability, and a Truly Warm Absorber}.
\newblock \emph{\apj} 536, 718--728.
\newblock \doi{10.1086/308950}
\bibAnnoteFile{Nicastro2000}

\bibitem[{{Nied{\'z}wiecki} et~al.(2016){Nied{\'z}wiecki}, {Zdziarski}, and {Szanecki}}]{Niedzwiecki2016}
{Nied{\'z}wiecki}, A., {Zdziarski}, A.~A., and {Szanecki}, M. (2016).
\newblock {On the Lamppost Model of Accreting Black Holes}.
\newblock \emph{\apjl} 821, L1.
\newblock \doi{10.3847/2041-8205/821/1/L1}
\bibAnnoteFile{Niedzwiecki2016}

\bibitem[{{Noda} and {Done}(2018)}]{Noda2018}
{Noda}, H. and {Done}, C. (2018).
\newblock {Explaining changing-look AGN with state transition triggered by rapid mass accretion rate drop}.
\newblock \emph{\mnras} 480, 3898--3906.
\newblock \doi{10.1093/mnras/sty2032}
\bibAnnoteFile{Noda2018}

\bibitem[{{Oh} et~al.(2018){Oh}, {Koss}, {Markwardt}, {Schawinski}, {Baumgartner}, {Barthelmy} et~al.}]{Oh2018}
{Oh}, K., {Koss}, M., {Markwardt}, C.~B., {Schawinski}, K., {Baumgartner}, W.~H., {Barthelmy}, S.~D., et~al. (2018).
\newblock {The 105-Month Swift-BAT All-sky Hard X-Ray Survey}.
\newblock \emph{\apjs} 235, 4.
\newblock \doi{10.3847/1538-4365/aaa7fd}
\bibAnnoteFile{Oh2018}

\bibitem[{{Panessa} et~al.(2019){Panessa}, {Baldi}, {Laor}, {Padovani}, {Behar}, and {McHardy}}]{Panessa2019}
{Panessa}, F., {Baldi}, R.~D., {Laor}, A., {Padovani}, P., {Behar}, E., and {McHardy}, I. (2019).
\newblock {The origin of radio emission from radio-quiet active galactic nuclei}.
\newblock \emph{Nature Astronomy} 3, 387--396.
\newblock \doi{10.1038/s41550-019-0765-4}
\bibAnnoteFile{Panessa2019}

\bibitem[{{Parker} et~al.(2014){Parker}, {Wilkins}, {Fabian}, {Grupe}, {Dauser}, {Matt} et~al.}]{Parker2014}
{Parker}, M.~L., {Wilkins}, D.~R., {Fabian}, A.~C., {Grupe}, D., {Dauser}, T., {Matt}, G., et~al. (2014).
\newblock {The NuSTAR spectrum of Mrk 335: extreme relativistic effects within two gravitational radii of the event horizon?}
\newblock \emph{\mnras} 443, 1723--1732.
\newblock \doi{10.1093/mnras/stu1246}
\bibAnnoteFile{Parker2014}

\bibitem[{{Pasham} et~al.(2019){Pasham}, {Remillard}, {Fragile}, {Franchini}, {Stone}, {Lodato} et~al.}]{Pasham2019}
{Pasham}, D.~R., {Remillard}, R.~A., {Fragile}, P.~C., {Franchini}, A., {Stone}, N.~C., {Lodato}, G., et~al. (2019).
\newblock {A loud quasi-periodic oscillation after a star is disrupted by a massive black hole}.
\newblock \emph{Science} 363, 531--534.
\newblock \doi{10.1126/science.aar7480}
\bibAnnoteFile{Pasham2019}

\bibitem[{{Petrucci} et~al.(2020){Petrucci}, {Gronkiewicz}, {Rozanska}, {Belmont}, {Bianchi}, {Czerny} et~al.}]{Petrucci2020}
{Petrucci}, P.~O., {Gronkiewicz}, D., {Rozanska}, A., {Belmont}, R., {Bianchi}, S., {Czerny}, B., et~al. (2020).
\newblock {Radiation spectra of warm and optically thick coronae in AGNs}.
\newblock \emph{\aap} 634, A85.
\newblock \doi{10.1051/0004-6361/201937011}
\bibAnnoteFile{Petrucci2020}

\bibitem[{{Petrucci} et~al.(2001{\natexlab{a}}){Petrucci}, {Haardt}, {Maraschi}, {Grandi}, {Malzac}, {Matt} et~al.}]{Petrucci01a}
{Petrucci}, P.~O., {Haardt}, F., {Maraschi}, L., {Grandi}, P., {Malzac}, J., {Matt}, G., et~al. (2001{\natexlab{a}}).
\newblock {Testing Comptonization Models Using BeppoSAX Observations of Seyfert 1 Galaxies}.
\newblock \emph{\apj} 556, 716--726.
\newblock \doi{10.1086/321629}
\bibAnnoteFile{Petrucci01a}

\bibitem[{{Petrucci} et~al.(2004){Petrucci}, {Maraschi}, {Haardt}, and {Nandra}}]{Petrucci2004}
{Petrucci}, P.~O., {Maraschi}, L., {Haardt}, F., and {Nandra}, K. (2004).
\newblock {Physical interpretation of the NGC 7469 UV/X-ray variability}.
\newblock \emph{\aap} 413, 477--487.
\newblock \doi{10.1051/0004-6361:20031499}
\bibAnnoteFile{Petrucci2004}

\bibitem[{{Petrucci} et~al.(2001{\natexlab{b}}){Petrucci}, {Merloni}, {Fabian}, {Haardt}, and {Gallo}}]{Petrucci01b}
{Petrucci}, P.~O., {Merloni}, A., {Fabian}, A., {Haardt}, F., and {Gallo}, E. (2001{\natexlab{b}}).
\newblock {The effects of a Comptonizing corona on the appearance of the reflection components in accreting black hole spectra}.
\newblock \emph{\mnras} 328, 501--510.
\newblock \doi{10.1046/j.1365-8711.2001.04897.x}
\bibAnnoteFile{Petrucci01b}

\bibitem[{{Petrucci} et~al.(2013){Petrucci}, {Paltani}, {Malzac}, {Kaastra}, {Cappi}, {Ponti} et~al.}]{Petrucci2013}
{Petrucci}, P.~O., {Paltani}, S., {Malzac}, J., {Kaastra}, J.~S., {Cappi}, M., {Ponti}, G., et~al. (2013).
\newblock {Multiwavelength campaign on Mrk 509. XII. Broad band spectral analysis}.
\newblock \emph{\aap} 549, A73.
\newblock \doi{10.1051/0004-6361/201219956}
\bibAnnoteFile{Petrucci2013}

\bibitem[{{Petrucci} et~al.(2023){Petrucci}, {Pi{\'e}tu}, {Behar}, {Clavel}, {Bianchi}, {Henri} et~al.}]{Petrucci2023}
{Petrucci}, P.~O., {Pi{\'e}tu}, V., {Behar}, E., {Clavel}, M., {Bianchi}, S., {Henri}, G., et~al. (2023).
\newblock {Simultaneous millimetric and X-ray intraday variability in the radio-quiet AGN MCG+08-11-11}.
\newblock \emph{\aap} 678, L4.
\newblock \doi{10.1051/0004-6361/202347495}
\bibAnnoteFile{Petrucci2023}

\bibitem[{{Petrucci} et~al.(2018){Petrucci}, {Ursini}, {De Rosa}, {Bianchi}, {Cappi}, {Matt} et~al.}]{Petrucci2018}
{Petrucci}, P.~O., {Ursini}, F., {De Rosa}, A., {Bianchi}, S., {Cappi}, M., {Matt}, G., et~al. (2018).
\newblock {Testing warm Comptonization models for the origin of the soft X-ray excess in AGNs}.
\newblock \emph{\aap} 611, A59.
\newblock \doi{10.1051/0004-6361/201731580}
\bibAnnoteFile{Petrucci2018}

\bibitem[{{Porquet} et~al.(2019){Porquet}, {Done}, {Reeves}, {Grosso}, {Marinucci}, {Matt} et~al.}]{Porquet2019}
{Porquet}, D., {Done}, C., {Reeves}, J.~N., {Grosso}, N., {Marinucci}, A., {Matt}, G., et~al. (2019).
\newblock {A deep X-ray view of the bare AGN Ark 120. V. Spin determination from disc-Comptonisation efficiency method}.
\newblock \emph{\aap} 623, A11.
\newblock \doi{10.1051/0004-6361/201834448}
\bibAnnoteFile{Porquet2019}

\bibitem[{{Poutanen} and {Svensson}(1996)}]{Poutanen1996}
{Poutanen}, J. and {Svensson}, R. (1996).
\newblock {The Two-Phase Pair Corona Model for Active Galactic Nuclei and X-Ray Binaries: How to Obtain Exact Solutions}.
\newblock \emph{\apj} 470, 249.
\newblock \doi{10.1086/177865}
\bibAnnoteFile{Poutanen1996}

\bibitem[{{Predehl} et~al.(2021){Predehl}, {Andritschke}, {Arefiev}, {Babyshkin}, {Batanov}, {Becker} et~al.}]{Predehl2021}
{Predehl}, P., {Andritschke}, R., {Arefiev}, V., {Babyshkin}, V., {Batanov}, O., {Becker}, W., et~al. (2021).
\newblock {The eROSITA X-ray telescope on SRG}.
\newblock \emph{\aap} 647, A1.
\newblock \doi{10.1051/0004-6361/202039313}
\bibAnnoteFile{Predehl2021}

\bibitem[{{Raginski} and {Laor}(2016)}]{Raginski2016}
{Raginski}, I. and {Laor}, A. (2016).
\newblock {AGN coronal emission models - I. The predicted radio emission}.
\newblock \emph{\mnras} 459, 2082--2096.
\newblock \doi{10.1093/mnras/stw772}
\bibAnnoteFile{Raginski2016}

\bibitem[{{Raimundo} and {Fabian}(2009)}]{Raimundo2009}
{Raimundo}, S.~I. and {Fabian}, A.~C. (2009).
\newblock {Eddington ratio and accretion efficiency in active galactic nuclei evolution}.
\newblock \emph{\mnras} 396, 1217--1221.
\newblock \doi{10.1111/j.1365-2966.2009.14796.x}
\bibAnnoteFile{Raimundo2009}

\bibitem[{{Raimundo} et~al.(2012){Raimundo}, {Fabian}, {Vasudevan}, {Gandhi}, and {Wu}}]{Raimundo2012}
{Raimundo}, S.~I., {Fabian}, A.~C., {Vasudevan}, R.~V., {Gandhi}, P., and {Wu}, J. (2012).
\newblock {Can we measure the accretion efficiency of active galactic nuclei?}
\newblock \emph{\mnras} 419, 2529--2544.
\newblock \doi{10.1111/j.1365-2966.2011.19904.x}
\bibAnnoteFile{Raimundo2012}

\bibitem[{{Ratheesh} et~al.(2021){Ratheesh}, {Matt}, {Tombesi}, {Soffitta}, {Pesce-Rollins}, and {Di Marco}}]{Ratheesh2021}
{Ratheesh}, A., {Matt}, G., {Tombesi}, F., {Soffitta}, P., {Pesce-Rollins}, M., and {Di Marco}, A. (2021).
\newblock {Exploring the accretion-ejection geometry of GRS 1915+105 in the obscured state with future X-ray spectro-polarimetry}.
\newblock \emph{\aap} 655, A96.
\newblock \doi{10.1051/0004-6361/202140701}
\bibAnnoteFile{Ratheesh2021}

\bibitem[{{Reeves} et~al.(2021){Reeves}, {Braito}, {Porquet}, {Lobban}, {Matzeu}, and {Nardini}}]{Reeves2017}
{Reeves}, J.~N., {Braito}, V., {Porquet}, D., {Lobban}, A.~P., {Matzeu}, G.~A., and {Nardini}, E. (2021).
\newblock {The flaring X-ray corona in the quasar PDS 456}.
\newblock \emph{\mnras} 500, 1974--1991.
\newblock \doi{10.1093/mnras/staa3377}
\bibAnnoteFile{Reeves2017}

\bibitem[{{Reis} and {Miller}(2013)}]{Reis2013}
{Reis}, R.~C. and {Miller}, J.~M. (2013).
\newblock {On the Size and Location of the X-Ray Emitting Coronae around Black Holes}.
\newblock \emph{\apjl} 769, L7.
\newblock \doi{10.1088/2041-8205/769/1/L7}
\bibAnnoteFile{Reis2013}

\bibitem[{{Remillard} and {McClintock}(2006)}]{Remillard2006}
{Remillard}, R.~A. and {McClintock}, J.~E. (2006).
\newblock {X-Ray Properties of Black-Hole Binaries}.
\newblock \emph{\araa} 44, 49--92.
\newblock \doi{10.1146/annurev.astro.44.051905.092532}
\bibAnnoteFile{Remillard2006}

\bibitem[{{Reynolds}(2021)}]{Reynolds2021}
{Reynolds}, C.~S. (2021).
\newblock {Observational Constraints on Black Hole Spin}.
\newblock \emph{\araa} 59, 117--154.
\newblock \doi{10.1146/annurev-astro-112420-035022}
\bibAnnoteFile{Reynolds2021}

\bibitem[{{Reynolds} and {Begelman}(1997)}]{Reynolds1997}
{Reynolds}, C.~S. and {Begelman}, M.~C. (1997).
\newblock {Iron Fluorescence from within the Innermost Stable Orbit of Black Hole Accretion Disks}.
\newblock \emph{\apj} 488, 109--118.
\newblock \doi{10.1086/304703}
\bibAnnoteFile{Reynolds1997}

\bibitem[{{Ricci} et~al.(2023){Ricci}, {Chang}, {Kawamuro}, {Privon}, {Mushotzky}, {Trakhtenbrot} et~al.}]{Ricci2023}
{Ricci}, C., {Chang}, C.-S., {Kawamuro}, T., {Privon}, G., {Mushotzky}, R., {Trakhtenbrot}, B., et~al. (2023).
\newblock {A Tight Correlation Between Millimeter and X-ray Emission in Accreting Massive Black Holes from $<$100 Milliarcsecond-resolution ALMA Observations}.
\newblock \emph{arXiv e-prints} , arXiv:2306.04679\doi{10.48550/arXiv.2306.04679}
\bibAnnoteFile{Ricci2023}

\bibitem[{{Ricci} et~al.(2018){Ricci}, {Ho}, {Fabian}, {Trakhtenbrot}, {Koss}, {Ueda} et~al.}]{Ricci2018}
{Ricci}, C., {Ho}, L.~C., {Fabian}, A.~C., {Trakhtenbrot}, B., {Koss}, M.~J., {Ueda}, Y., et~al. (2018).
\newblock {BAT AGN Spectroscopic Survey - XII. The relation between coronal properties of active galactic nuclei and the Eddington ratio}.
\newblock \emph{\mnras} 480, 1819--1830.
\newblock \doi{10.1093/mnras/sty1879}
\bibAnnoteFile{Ricci2018}

\bibitem[{{Ricci} et~al.(2020){Ricci}, {Kara}, {Loewenstein}, {Trakhtenbrot}, {Arcavi}, {Remillard} et~al.}]{Ricci2020}
{Ricci}, C., {Kara}, E., {Loewenstein}, M., {Trakhtenbrot}, B., {Arcavi}, I., {Remillard}, R., et~al. (2020).
\newblock {The Destruction and Recreation of the X-Ray Corona in a Changing-look Active Galactic Nucleus}.
\newblock \emph{\apjl} 898, L1.
\newblock \doi{10.3847/2041-8213/ab91a1}
\bibAnnoteFile{Ricci2020}

\bibitem[{{Ricci} et~al.(2021){Ricci}, {Loewenstein}, {Kara}, {Remillard}, {Trakhtenbrot}, {Arcavi} et~al.}]{Ricci2021}
{Ricci}, C., {Loewenstein}, M., {Kara}, E., {Remillard}, R., {Trakhtenbrot}, B., {Arcavi}, I., et~al. (2021).
\newblock {The 450 Day X-Ray Monitoring of the Changing-look AGN 1ES 1927+654}.
\newblock \emph{\apjs} 255, 7.
\newblock \doi{10.3847/1538-4365/abe94b}
\bibAnnoteFile{Ricci2021}

\bibitem[{{Ricci} and {Trakhtenbrot}(2022)}]{Ricci2022}
{Ricci}, C. and {Trakhtenbrot}, B. (2022).
\newblock {Changing-look Active Galactic Nuclei}.
\newblock \emph{arXiv e-prints} , arXiv:2211.05132\doi{10.48550/arXiv.2211.05132}
\bibAnnoteFile{Ricci2022}

\bibitem[{{Ricci} et~al.(2017){Ricci}, {Trakhtenbrot}, {Koss}, {Ueda}, {Del Vecchio}, {Treister} et~al.}]{Ricci2017}
{Ricci}, C., {Trakhtenbrot}, B., {Koss}, M.~J., {Ueda}, Y., {Del Vecchio}, I., {Treister}, E., et~al. (2017).
\newblock {BAT AGN Spectroscopic Survey. V. X-Ray Properties of the Swift/BAT 70-month AGN Catalog}.
\newblock \emph{\apjs} 233, 17.
\newblock \doi{10.3847/1538-4365/aa96ad}
\bibAnnoteFile{Ricci2017}

\bibitem[{{Risaliti} et~al.(2013){Risaliti}, {Harrison}, {Madsen}, {Walton}, {Boggs}, {Christensen} et~al.}]{Risaliti2013}
{Risaliti}, G., {Harrison}, F.~A., {Madsen}, K.~K., {Walton}, D.~J., {Boggs}, S.~E., {Christensen}, F.~E., et~al. (2013).
\newblock {A rapidly spinning supermassive black hole at the centre of NGC 1365}.
\newblock \emph{\nat} 494, 449--451.
\newblock \doi{10.1038/nature11938}
\bibAnnoteFile{Risaliti2013}

\bibitem[{{Risaliti} et~al.(2011){Risaliti}, {Nardini}, {Salvati}, {Elvis}, {Fabbiano}, {Maiolino} et~al.}]{Risaliti2011}
{Risaliti}, G., {Nardini}, E., {Salvati}, M., {Elvis}, M., {Fabbiano}, G., {Maiolino}, R., et~al. (2011).
\newblock {X-ray absorption by broad-line region clouds in Mrk 766}.
\newblock \emph{\mnras} 410, 1027--1035.
\newblock \doi{10.1111/j.1365-2966.2010.17503.x}
\bibAnnoteFile{Risaliti2011}

\bibitem[{{Rivers} et~al.(2013){Rivers}, {Markowitz}, and {Rothschild}}]{Rivers2013}
{Rivers}, E., {Markowitz}, A., and {Rothschild}, R. (2013).
\newblock {Full Spectral Survey of Active Galactic Nuclei in the Rossi X-ray Timing Explorer Archive}.
\newblock \emph{\apj} 772, 114.
\newblock \doi{10.1088/0004-637X/772/2/114}
\bibAnnoteFile{Rivers2013}

\bibitem[{{Ross} et~al.(2018){Ross}, {Ford}, {Graham}, {McKernan}, {Stern}, {Meisner} et~al.}]{Ross2018}
{Ross}, N.~P., {Ford}, K.~E.~S., {Graham}, M., {McKernan}, B., {Stern}, D., {Meisner}, A.~M., et~al. (2018).
\newblock {A new physical interpretation of optical and infrared variability in quasars}.
\newblock \emph{\mnras} 480, 4468--4479.
\newblock \doi{10.1093/mnras/sty2002}
\bibAnnoteFile{Ross2018}

\bibitem[{{Ruan} et~al.(2019){Ruan}, {Anderson}, {Eracleous}, {Green}, {Haggard}, {MacLeod} et~al.}]{Ruan2019}
{Ruan}, J.~J., {Anderson}, S.~F., {Eracleous}, M., {Green}, P.~J., {Haggard}, D., {MacLeod}, C.~L., et~al. (2019).
\newblock {The Analogous Structure of Accretion Flows in Supermassive and Stellar Mass Black Holes: New Insights from Faded Changing-look Quasars}.
\newblock \emph{\apj} 883, 76.
\newblock \doi{10.3847/1538-4357/ab3c1a}
\bibAnnoteFile{Ruan2019}

\bibitem[{{Rybicki} and {Lightman}(1979)}]{Rybicki1979}
{Rybicki}, G.~B. and {Lightman}, A.~P. (1979).
\newblock \emph{{Radiative processes in astrophysics}} ({John Wiley \& Sons, Ltd})
\bibAnnoteFile{Rybicki1979}

\bibitem[{{Sagiv} et~al.(2014){Sagiv}, {Gal-Yam}, {Ofek}, {Waxman}, {Aharonson}, {Kulkarni} et~al.}]{Sagiv2014}
{Sagiv}, I., {Gal-Yam}, A., {Ofek}, E.~O., {Waxman}, E., {Aharonson}, O., {Kulkarni}, S.~R., et~al. (2014).
\newblock {Science with a Wide-field UV Transient Explorer}.
\newblock \emph{\aj} 147, 79.
\newblock \doi{10.1088/0004-6256/147/4/79}
\bibAnnoteFile{Sagiv2014}

\bibitem[{{Sanfrutos} et~al.(2013){Sanfrutos}, {Miniutti}, {Ag{\'\i}s-Gonz{\'a}lez}, {Fabian}, {Miller}, {Panessa} et~al.}]{Sanfrutos2013}
{Sanfrutos}, M., {Miniutti}, G., {Ag{\'\i}s-Gonz{\'a}lez}, B., {Fabian}, A.~C., {Miller}, J.~M., {Panessa}, F., et~al. (2013).
\newblock {The size of the X-ray emitting region in SWIFT J2127.4+5654 via a broad line region cloud X-ray eclipse}.
\newblock \emph{\mnras} 436, 1588--1594.
\newblock \doi{10.1093/mnras/stt1675}
\bibAnnoteFile{Sanfrutos2013}

\bibitem[{{Saxton} et~al.(2012){Saxton}, {Read}, {Esquej}, {Komossa}, {Dougherty}, {Rodriguez-Pascual} et~al.}]{Saxton2012}
{Saxton}, R.~D., {Read}, A.~M., {Esquej}, P., {Komossa}, S., {Dougherty}, S., {Rodriguez-Pascual}, P., et~al. (2012).
\newblock {A tidal disruption-like X-ray flare from the quiescent galaxy SDSS J120136.02+300305.5}.
\newblock \emph{\aap} 541, A106.
\newblock \doi{10.1051/0004-6361/201118367}
\bibAnnoteFile{Saxton2012}

\bibitem[{{Scepi} et~al.(2021){Scepi}, {Begelman}, and {Dexter}}]{Scepi2021}
{Scepi}, N., {Begelman}, M.~C., and {Dexter}, J. (2021).
\newblock {Magnetic flux inversion in a peculiar changing look AGN}.
\newblock \emph{\mnras} 502, L50--L54.
\newblock \doi{10.1093/mnrasl/slab002}
\bibAnnoteFile{Scepi2021}

\bibitem[{{Schnittman} and {Krolik}(2010)}]{Schnittman2010}
{Schnittman}, J.~D. and {Krolik}, J.~H. (2010).
\newblock {X-ray Polarization from Accreting Black Holes: Coronal Emission}.
\newblock \emph{\apj} 712, 908--924.
\newblock \doi{10.1088/0004-637X/712/2/908}
\bibAnnoteFile{Schnittman2010}

\bibitem[{{Shankar} et~al.(2004){Shankar}, {Salucci}, {Granato}, {De Zotti}, and {Danese}}]{Shankar2004}
{Shankar}, F., {Salucci}, P., {Granato}, G.~L., {De Zotti}, G., and {Danese}, L. (2004).
\newblock {Supermassive black hole demography: the match between the local and accreted mass functions}.
\newblock \emph{\mnras} 354, 1020--1030.
\newblock \doi{10.1111/j.1365-2966.2004.08261.x}
\bibAnnoteFile{Shankar2004}

\bibitem[{{Shankar} et~al.(2020){Shankar}, {Weinberg}, {Marsden}, {Grylls}, {Bernardi}, {Yang} et~al.}]{Shankar2020}
{Shankar}, F., {Weinberg}, D.~H., {Marsden}, C., {Grylls}, P.~J., {Bernardi}, M., {Yang}, G., et~al. (2020).
\newblock {Probing black hole accretion tracks, scaling relations, and radiative efficiencies from stacked X-ray active galactic nuclei}.
\newblock \emph{\mnras} 493, 1500--1511.
\newblock \doi{10.1093/mnras/stz3522}
\bibAnnoteFile{Shankar2020}

\bibitem[{{Shankar} et~al.(2009){Shankar}, {Weinberg}, and {Miralda-Escud{\'e}}}]{Shankar2009}
{Shankar}, F., {Weinberg}, D.~H., and {Miralda-Escud{\'e}}, J. (2009).
\newblock {Self-Consistent Models of the AGN and Black Hole Populations: Duty Cycles, Accretion Rates, and the Mean Radiative Efficiency}.
\newblock \emph{\apj} 690, 20--41.
\newblock \doi{10.1088/0004-637X/690/1/20}
\bibAnnoteFile{Shankar2009}

\bibitem[{{Shappee} et~al.(2014){Shappee}, {Prieto}, {Grupe}, {Kochanek}, {Stanek}, {De Rosa} et~al.}]{Shappee2014}
{Shappee}, B.~J., {Prieto}, J.~L., {Grupe}, D., {Kochanek}, C.~S., {Stanek}, K.~Z., {De Rosa}, G., et~al. (2014).
\newblock {The Man behind the Curtain: X-Rays Drive the UV through NIR Variability in the 2013 Active Galactic Nucleus Outburst in NGC 2617}.
\newblock \emph{\apj} 788, 48.
\newblock \doi{10.1088/0004-637X/788/1/48}
\bibAnnoteFile{Shappee2014}

\bibitem[{{Sironi} and {Beloborodov}(2020)}]{Sironi&Beloborodov20}
{Sironi}, L. and {Beloborodov}, A.~M. (2020).
\newblock {Kinetic Simulations of Radiative Magnetic Reconnection in the Coronae of Accreting Black Holes}.
\newblock \emph{\apj} 899, 52.
\newblock \doi{10.3847/1538-4357/aba622}
\bibAnnoteFile{Sironi&Beloborodov20}

\bibitem[{{Smith} et~al.(2016){Smith}, {Mushotzky}, {Vogel}, {Shimizu}, and {Miller}}]{Smith2016}
{Smith}, K.~L., {Mushotzky}, R.~F., {Vogel}, S., {Shimizu}, T.~T., and {Miller}, N. (2016).
\newblock {Radio Properties of the BAT AGNs: the FIR-radio Relation, the Fundamental Plane, and the Main Sequence of Star Formation}.
\newblock \emph{\apj} 832, 163.
\newblock \doi{10.3847/0004-637X/832/2/163}
\bibAnnoteFile{Smith2016}

\bibitem[{{Sniegowska} et~al.(2020){Sniegowska}, {Czerny}, {Bon}, and {Bon}}]{Sniegowska2020}
{Sniegowska}, M., {Czerny}, B., {Bon}, E., and {Bon}, N. (2020).
\newblock {Possible mechanism for multiple changing-look phenomena in active galactic nuclei}.
\newblock \emph{\aap} 641, A167.
\newblock \doi{10.1051/0004-6361/202038575}
\bibAnnoteFile{Sniegowska2020}

\bibitem[{{Soltan}(1982)}]{Soltan1982}
{Soltan}, A. (1982).
\newblock {Masses of quasars.}
\newblock \emph{\mnras} 200, 115--122.
\newblock \doi{10.1093/mnras/200.1.115}
\bibAnnoteFile{Soltan1982}

\bibitem[{{Sridhar} et~al.(2021){Sridhar}, {Sironi}, and {Beloborodov}}]{Sridhar+21}
{Sridhar}, N., {Sironi}, L., and {Beloborodov}, A.~M. (2021).
\newblock {Comptonization by reconnection plasmoids in black hole coronae I: Magnetically dominated pair plasma}.
\newblock \emph{\mnras} 507, 5625--5640.
\newblock \doi{10.1093/mnras/stab2534}
\bibAnnoteFile{Sridhar+21}

\bibitem[{{Sridhar} et~al.(2023){Sridhar}, {Sironi}, and {Beloborodov}}]{Sridhar+23}
{Sridhar}, N., {Sironi}, L., and {Beloborodov}, A.~M. (2023).
\newblock {Comptonization by reconnection plasmoids in black hole coronae II: Electron-ion plasma}.
\newblock \emph{\mnras} 518, 1301--1315.
\newblock \doi{10.1093/mnras/stac2730}
\bibAnnoteFile{Sridhar+23}

\bibitem[{{Stern} et~al.(1995){Stern}, {Poutanen}, {Svensson}, {Sikora}, and {Begelman}}]{Stern1995}
{Stern}, B.~E., {Poutanen}, J., {Svensson}, R., {Sikora}, M., and {Begelman}, M.~C. (1995).
\newblock {On the Geometry of the X-Ray--Emitting Region in Seyfert Galaxies}.
\newblock \emph{\apjl} 449, L13.
\newblock \doi{10.1086/309617}
\bibAnnoteFile{Stern1995}

\bibitem[{{Stern} et~al.(2018){Stern}, {McKernan}, {Graham}, {Ford}, {Ross}, {Meisner} et~al.}]{Stern2018}
{Stern}, D., {McKernan}, B., {Graham}, M.~J., {Ford}, K.~E.~S., {Ross}, N.~P., {Meisner}, A.~M., et~al. (2018).
\newblock {A Mid-IR Selected Changing-look Quasar and Physical Scenarios for Abrupt AGN Fading}.
\newblock \emph{\apj} 864, 27.
\newblock \doi{10.3847/1538-4357/aac726}
\bibAnnoteFile{Stern2018}

\bibitem[{{Svensson}(1984)}]{Svensson1984}
{Svensson}, R. (1984).
\newblock {Steady mildly relativistic thermal plasmas - Processes and properties}.
\newblock \emph{\mnras} 209, 175--208.
\newblock \doi{10.1093/mnras/209.2.175}
\bibAnnoteFile{Svensson1984}

\bibitem[{{Tagliacozzo} et~al.(2023){Tagliacozzo}, {Marinucci}, {Ursini}, {Matt}, {Bianchi}, {Baldini} et~al.}]{Tagliacozzo2023}
{Tagliacozzo}, D., {Marinucci}, A., {Ursini}, F., {Matt}, G., {Bianchi}, S., {Baldini}, L., et~al. (2023).
\newblock {The geometry of the hot corona in MCG-05-23-16 constrained by X-ray polarimetry}.
\newblock \emph{arXiv e-prints} , arXiv:2305.10213\doi{10.48550/arXiv.2305.10213}
\bibAnnoteFile{Tagliacozzo2023}

\bibitem[{{Tamborra} et~al.(2018){Tamborra}, {Matt}, {Bianchi}, and {Dov{\v{c}}iak}}]{Tamborra2018}
{Tamborra}, F., {Matt}, G., {Bianchi}, S., and {Dov{\v{c}}iak}, M. (2018).
\newblock {MoCA: A Monte Carlo code for Comptonisation in Astrophysics. I. Description of the code and first results}.
\newblock \emph{\aap} 619, A105.
\newblock \doi{10.1051/0004-6361/201732023}
\bibAnnoteFile{Tamborra2018}

\bibitem[{{Tashiro} et~al.(2018){Tashiro}, {Maejima}, {Toda}, {Kelley}, {Reichenthal}, {Lobell} et~al.}]{Tashiro2018}
{Tashiro}, M., {Maejima}, H., {Toda}, K., {Kelley}, R., {Reichenthal}, L., {Lobell}, J., et~al. (2018).
\newblock {Concept of the X-ray Astronomy Recovery Mission}.
\newblock In \emph{Space Telescopes and Instrumentation 2018: Ultraviolet to Gamma Ray}, eds. J.-W.~A. {den Herder}, S.~{Nikzad}, and K.~{Nakazawa}. vol. 10699 of \emph{Society of Photo-Optical Instrumentation Engineers (SPIE) Conference Series}, 1069922.
\newblock \doi{10.1117/12.2309455}
\bibAnnoteFile{Tashiro2018}

\bibitem[{{Tazaki} et~al.(2011){Tazaki}, {Ueda}, {Terashima}, and {Mushotzky}}]{tazaki+2011}
{Tazaki}, F., {Ueda}, Y., {Terashima}, Y., and {Mushotzky}, R.~F. (2011).
\newblock {Suzaku View of the Swift/BAT Active Galactic Nuclei. IV. Nature of Two Narrow-line Radio Galaxies (3C 403 and IC 5063)}.
\newblock \emph{\apj} 738, 70.
\newblock \doi{10.1088/0004-637X/738/1/70}
\bibAnnoteFile{tazaki+2011}

\bibitem[{{Tetarenko} et~al.(2016){Tetarenko}, {Sivakoff}, {Heinke}, and {Gladstone}}]{TetarenkoB2016}
{Tetarenko}, B.~E., {Sivakoff}, G.~R., {Heinke}, C.~O., and {Gladstone}, J.~C. (2016).
\newblock {WATCHDOG: A Comprehensive All-sky Database of Galactic Black Hole X-ray Binaries}.
\newblock \emph{\apjs} 222, 15.
\newblock \doi{10.3847/0067-0049/222/2/15}
\bibAnnoteFile{TetarenkoB2016}

\bibitem[{{Titarchuk}(1994)}]{Titarchuk1994}
{Titarchuk}, L. (1994).
\newblock {Generalized Comptonization Models and Application to the Recent High-Energy Observations}.
\newblock \emph{\apj} 434, 570.
\newblock \doi{10.1086/174760}
\bibAnnoteFile{Titarchuk1994}

\bibitem[{{Titarchuk} and {Lyubarskij}(1995)}]{Titarchuk95}
{Titarchuk}, L. and {Lyubarskij}, Y. (1995).
\newblock {Power-Law Spectra as a Result of Comptonization of the Soft Radiation in a Plasma Cloud}.
\newblock \emph{\apj} 450, 876.
\newblock \doi{10.1086/176191}
\bibAnnoteFile{Titarchuk95}

\bibitem[{{Tomsick} et~al.(2019){Tomsick}, {Zoglauer}, {Sleator}, {Lazar}, {Beechert}, {Boggs} et~al.}]{Tomsick2019}
{Tomsick}, J., {Zoglauer}, A., {Sleator}, C., {Lazar}, H., {Beechert}, J., {Boggs}, S., et~al. (2019).
\newblock {The Compton Spectrometer and Imager}.
\newblock In \emph{Bulletin of the American Astronomical Society}. vol.~51, 98.
\newblock \doi{10.48550/arXiv.1908.04334}
\bibAnnoteFile{Tomsick2019}

\bibitem[{{Tonry} et~al.(2018){Tonry}, {Denneau}, {Heinze}, {Stalder}, {Smith}, {Smartt} et~al.}]{Tonry2018}
{Tonry}, J.~L., {Denneau}, L., {Heinze}, A.~N., {Stalder}, B., {Smith}, K.~W., {Smartt}, S.~J., et~al. (2018).
\newblock {ATLAS: A High-cadence All-sky Survey System}.
\newblock \emph{\pasp} 130, 064505.
\newblock \doi{10.1088/1538-3873/aabadf}
\bibAnnoteFile{Tonry2018}

\bibitem[{{Tortosa} et~al.(2018){Tortosa}, {Bianchi}, {Marinucci}, {Matt}, and {Petrucci}}]{Tortosa2018}
{Tortosa}, A., {Bianchi}, S., {Marinucci}, A., {Matt}, G., and {Petrucci}, P.~O. (2018).
\newblock {A NuSTAR census of coronal parameters in Seyfert galaxies}.
\newblock \emph{\aap} 614, A37.
\newblock \doi{10.1051/0004-6361/201732382}
\bibAnnoteFile{Tortosa2018}

\bibitem[{{Tortosa} et~al.(2023){Tortosa}, {Ricci}, {Ho}, {Tombesi}, {Du}, {Inayoshi} et~al.}]{Tortosa2023}
{Tortosa}, A., {Ricci}, C., {Ho}, L.~C., {Tombesi}, F., {Du}, P., {Inayoshi}, K., et~al. (2023).
\newblock {Systematic broad-band X-ray study of super-Eddington accretion on to supermassive black holes - I. X-ray continuum}.
\newblock \emph{\mnras} 519, 6267--6283.
\newblock \doi{10.1093/mnras/stac3590}
\bibAnnoteFile{Tortosa2023}

\bibitem[{{Trakhtenbrot} et~al.(2019){Trakhtenbrot}, {Arcavi}, {MacLeod}, {Ricci}, {Kara}, {Graham} et~al.}]{Trakhtenbrot2019}
{Trakhtenbrot}, B., {Arcavi}, I., {MacLeod}, C.~L., {Ricci}, C., {Kara}, E., {Graham}, M.~L., et~al. (2019).
\newblock {1ES 1927+654: An AGN Caught Changing Look on a Timescale of Months}.
\newblock \emph{\apj} 883, 94.
\newblock \doi{10.3847/1538-4357/ab39e4}
\bibAnnoteFile{Trakhtenbrot2019}

\bibitem[{{Urry} and {Padovani}(1995)}]{Urry1995}
{Urry}, C.~M. and {Padovani}, P. (1995).
\newblock {Unified Schemes for Radio-Loud Active Galactic Nuclei}.
\newblock \emph{\pasp} 107, 803.
\newblock \doi{10.1086/133630}
\bibAnnoteFile{Urry1995}

\bibitem[{{Ursini} et~al.(2019){Ursini}, {Bassani}, {Malizia}, {Bazzano}, {Bird}, {Stephen} et~al.}]{Ursini2019}
{Ursini}, F., {Bassani}, L., {Malizia}, A., {Bazzano}, A., {Bird}, A.~J., {Stephen}, J.~B., et~al. (2019).
\newblock {The coronal temperature of NGC 4388 and NGC 2110 measured with INTEGRAL}.
\newblock \emph{\aap} 629, A54.
\newblock \doi{10.1051/0004-6361/201936273}
\bibAnnoteFile{Ursini2019}

\bibitem[{{Ursini} et~al.(2023){Ursini}, {Marinucci}, {Matt}, {Bianchi}, {Marin}, {Marshall} et~al.}]{Ursini2023}
{Ursini}, F., {Marinucci}, A., {Matt}, G., {Bianchi}, S., {Marin}, F., {Marshall}, H.~L., et~al. (2023).
\newblock {Mapping the circumnuclear regions of the Circinus galaxy with the Imaging X-ray Polarimetry Explorer}.
\newblock \emph{\mnras} 519, 50--58.
\newblock \doi{10.1093/mnras/stac3189}
\bibAnnoteFile{Ursini2023}

\bibitem[{{Ursini} et~al.(2016){Ursini}, {Petrucci}, {Matt}, {Bianchi}, {Cappi}, {De Marco} et~al.}]{Ursini2016}
{Ursini}, F., {Petrucci}, P.~O., {Matt}, G., {Bianchi}, S., {Cappi}, M., {De Marco}, B., et~al. (2016).
\newblock {High-energy monitoring of NGC 4593 with XMM-Newton and NuSTAR. X-ray spectral analysis}.
\newblock \emph{\mnras} 463, 382--392.
\newblock \doi{10.1093/mnras/stw2022}
\bibAnnoteFile{Ursini2016}

\bibitem[{{Uttley} et~al.(2014){Uttley}, {Cackett}, {Fabian}, {Kara}, and {Wilkins}}]{Uttley2014}
{Uttley}, P., {Cackett}, E.~M., {Fabian}, A.~C., {Kara}, E., and {Wilkins}, D.~R. (2014).
\newblock {X-ray reverberation around accreting black holes}.
\newblock \emph{\aapr} 22, 72.
\newblock \doi{10.1007/s00159-014-0072-0}
\bibAnnoteFile{Uttley2014}

\bibitem[{{Vaiana} and {Rosner}(1978)}]{Vaiana1978}
{Vaiana}, G.~S. and {Rosner}, R. (1978).
\newblock {Recent advances in coronal physics.}
\newblock \emph{\araa} 16, 393--428.
\newblock \doi{10.1146/annurev.aa.16.090178.002141}
\bibAnnoteFile{Vaiana1978}

\bibitem[{{Vasudevan} and {Fabian}(2007)}]{Vasudevan2007}
{Vasudevan}, R.~V. and {Fabian}, A.~C. (2007).
\newblock {Piecing together the X-ray background: bolometric corrections for active galactic nuclei}.
\newblock \emph{\mnras} 381, 1235--1251.
\newblock \doi{10.1111/j.1365-2966.2007.12328.x}
\bibAnnoteFile{Vasudevan2007}

\bibitem[{{Walton} et~al.(2021){Walton}, {Balokovi{\'c}}, {Fabian}, {Gallo}, {Koss}, {Nardini} et~al.}]{Walton2021}
{Walton}, D.~J., {Balokovi{\'c}}, M., {Fabian}, A.~C., {Gallo}, L.~C., {Koss}, M., {Nardini}, E., et~al. (2021).
\newblock {Extreme relativistic reflection in the active galaxy ESO 033-G002}.
\newblock \emph{\mnras} 506, 1557--1572.
\newblock \doi{10.1093/mnras/stab1290}
\bibAnnoteFile{Walton2021}

\bibitem[{{Walton} et~al.(2012){Walton}, {Reis}, {Cackett}, {Fabian}, and {Miller}}]{Walton2012}
{Walton}, D.~J., {Reis}, R.~C., {Cackett}, E.~M., {Fabian}, A.~C., and {Miller}, J.~M. (2012).
\newblock {The similarity of broad iron lines in X-ray binaries and active galactic nuclei}.
\newblock \emph{\mnras} 422, 2510--2531.
\newblock \doi{10.1111/j.1365-2966.2012.20809.x}
\bibAnnoteFile{Walton2012}

\bibitem[{{Walton} et~al.(2014){Walton}, {Risaliti}, {Harrison}, {Fabian}, {Miller}, {Arevalo} et~al.}]{Walton2014}
{Walton}, D.~J., {Risaliti}, G., {Harrison}, F.~A., {Fabian}, A.~C., {Miller}, J.~M., {Arevalo}, P., et~al. (2014).
\newblock {NuSTAR and XMM-NEWTON Observations of NGC 1365: Extreme Absorption Variability and a Constant Inner Accretion Disk}.
\newblock \emph{\apj} 788, 76.
\newblock \doi{10.1088/0004-637X/788/1/76}
\bibAnnoteFile{Walton2014}

\bibitem[{{Wambsganss}(2006)}]{Wambsganss2006}
{Wambsganss}, J. (2006).
\newblock {Part 4: Gravitational microlensing}.
\newblock In \emph{Saas-Fee Advanced Course 33: Gravitational Lensing: Strong, Weak and Micro}, eds. G.~{Meylan}, P.~{Jetzer}, P.~{North}, P.~{Schneider}, C.~S. {Kochanek}, and J.~{Wambsganss}. 453--540
\bibAnnoteFile{Wambsganss2006}

\bibitem[{{Weisskopf} et~al.(2022){Weisskopf}, {Soffitta}, {Baldini}, {Ramsey}, {O'Dell}, {Romani} et~al.}]{Weisskopf2022}
{Weisskopf}, M.~C., {Soffitta}, P., {Baldini}, L., {Ramsey}, B.~D., {O'Dell}, S.~L., {Romani}, R.~W., et~al. (2022).
\newblock {The Imaging X-Ray Polarimetry Explorer (IXPE): Pre-Launch}.
\newblock \emph{Journal of Astronomical Telescopes, Instruments, and Systems} 8, 026002.
\newblock \doi{10.1117/1.JATIS.8.2.026002}
\bibAnnoteFile{Weisskopf2022}

\bibitem[{{Wen} et~al.(2020){Wen}, {Jonker}, {Stone}, {Zabludoff}, and {Psaltis}}]{Wen2020}
{Wen}, S., {Jonker}, P.~G., {Stone}, N.~C., {Zabludoff}, A.~I., and {Psaltis}, D. (2020).
\newblock {Continuum-fitting the X-Ray Spectra of Tidal Disruption Events}.
\newblock \emph{\apj} 897, 80.
\newblock \doi{10.3847/1538-4357/ab9817}
\bibAnnoteFile{Wen2020}

\bibitem[{{Wevers}(2020)}]{Wevers2020}
{Wevers}, T. (2020).
\newblock {Fainter harder brighter softer: a correlation between {\ensuremath{\alpha}}$_{ox}$, X-ray spectral state, and Eddington ratio in tidal disruption events}.
\newblock \emph{\mnras} 497, L1--L6.
\newblock \doi{10.1093/mnrasl/slaa097}
\bibAnnoteFile{Wevers2020}

\bibitem[{{Wevers} et~al.(2021){Wevers}, {Pasham}, {van Velzen}, {Miller-Jones}, {Uttley}, {Gendreau} et~al.}]{Wevers2021}
{Wevers}, T., {Pasham}, D.~R., {van Velzen}, S., {Miller-Jones}, J.~C.~A., {Uttley}, P., {Gendreau}, K.~C., et~al. (2021).
\newblock {Rapid Accretion State Transitions following the Tidal Disruption Event AT2018fyk}.
\newblock \emph{\apj} 912, 151.
\newblock \doi{10.3847/1538-4357/abf5e2}
\bibAnnoteFile{Wevers2021}

\bibitem[{{Wilkins} et~al.(2016){Wilkins}, {Cackett}, {Fabian}, and {Reynolds}}]{Wilkins2016}
{Wilkins}, D.~R., {Cackett}, E.~M., {Fabian}, A.~C., and {Reynolds}, C.~S. (2016).
\newblock {Towards modelling X-ray reverberation in AGN: piecing together the extended corona}.
\newblock \emph{\mnras} 458, 200--225.
\newblock \doi{10.1093/mnras/stw276}
\bibAnnoteFile{Wilkins2016}

\bibitem[{{Wilkins} and {Fabian}(2011)}]{Wilkins2011}
{Wilkins}, D.~R. and {Fabian}, A.~C. (2011).
\newblock {Determination of the X-ray reflection emissivity profile of 1H 0707-495}.
\newblock \emph{\mnras} 414, 1269--1277.
\newblock \doi{10.1111/j.1365-2966.2011.18458.x}
\bibAnnoteFile{Wilkins2011}

\bibitem[{{Wilkins} and {Fabian}(2012)}]{Wilkins2012}
{Wilkins}, D.~R. and {Fabian}, A.~C. (2012).
\newblock {Understanding X-ray reflection emissivity profiles in AGN: locating the X-ray source}.
\newblock \emph{\mnras} 424, 1284--1296.
\newblock \doi{10.1111/j.1365-2966.2012.21308.x}
\bibAnnoteFile{Wilkins2012}

\bibitem[{{Wilkins} and {Fabian}(2013)}]{Wilkins2013}
{Wilkins}, D.~R. and {Fabian}, A.~C. (2013).
\newblock {The origin of the lag spectra observed in AGN: Reverberation and the propagation of X-ray source fluctuations}.
\newblock \emph{\mnras} 430, 247--258.
\newblock \doi{10.1093/mnras/sts591}
\bibAnnoteFile{Wilkins2013}

\bibitem[{{Wilkins} and {Gallo}(2015)}]{Wilkins2015}
{Wilkins}, D.~R. and {Gallo}, L.~C. (2015).
\newblock {Driving extreme variability: the evolving corona and evidence for jet launching in Markarian 335 }.
\newblock \emph{\mnras} 449, 129--146.
\newblock \doi{10.1093/mnras/stv162}
\bibAnnoteFile{Wilkins2015}

\bibitem[{{Wilms} et~al.(2000){Wilms}, {Allen}, and {McCray}}]{Wilms2000}
{Wilms}, J., {Allen}, A., and {McCray}, R. (2000).
\newblock {On the Absorption of X-Rays in the Interstellar Medium}.
\newblock \emph{\apj} 542, 914--924.
\newblock \doi{10.1086/317016}
\bibAnnoteFile{Wilms2000}

\bibitem[{{XRISM Science Team}(2022)}]{xrism2022}
{XRISM Science Team} (2022).
\newblock {XRISM Quick Reference}.
\newblock \emph{arXiv e-prints} , arXiv:2202.05399\doi{10.48550/arXiv.2202.05399}
\bibAnnoteFile{xrism2022}

\bibitem[{{Yao} et~al.(2022){Yao}, {Lu}, {Guolo}, {Pasham}, {Gezari}, {Gilfanov} et~al.}]{Yao2022}
{Yao}, Y., {Lu}, W., {Guolo}, M., {Pasham}, D.~R., {Gezari}, S., {Gilfanov}, M., et~al. (2022).
\newblock {The Tidal Disruption Event AT2021ehb: Evidence of Relativistic Disk Reflection, and Rapid Evolution of the Disk-Corona System}.
\newblock \emph{\apj} 937, 8.
\newblock \doi{10.3847/1538-4357/ac898a}
\bibAnnoteFile{Yao2022}

\bibitem[{{Yu} and {Tremaine}(2002)}]{Yu2002}
{Yu}, Q. and {Tremaine}, S. (2002).
\newblock {Observational constraints on growth of massive black holes}.
\newblock \emph{\mnras} 335, 965--976.
\newblock \doi{10.1046/j.1365-8711.2002.05532.x}
\bibAnnoteFile{Yu2002}

\bibitem[{{Zdziarski} et~al.(1996){Zdziarski}, {Johnson}, and {Magdziarz}}]{Zdziarski96}
{Zdziarski}, A.~A., {Johnson}, W.~N., and {Magdziarz}, P. (1996).
\newblock {Broad-band {\ensuremath{\gamma}}-ray and X-ray spectra of NGC 4151 and their implications for physical processes and geometry.}
\newblock \emph{\mnras} 283, 193--206.
\newblock \doi{10.1093/mnras/283.1.193}
\bibAnnoteFile{Zdziarski96}

\bibitem[{{Zdziarski} et~al.(2021){Zdziarski}, {Jourdain}, {Lubi{\'n}ski}, {Szanecki}, {Nied{\'z}wiecki}, {Veledina} et~al.}]{Zdziarski2021}
{Zdziarski}, A.~A., {Jourdain}, E., {Lubi{\'n}ski}, P., {Szanecki}, M., {Nied{\'z}wiecki}, A., {Veledina}, A., et~al. (2021).
\newblock {Hybrid Comptonization and Electron-Positron Pair Production in the Black-hole X-Ray Binary MAXI J1820+070}.
\newblock \emph{\apjl} 914, L5.
\newblock \doi{10.3847/2041-8213/ac0147}
\bibAnnoteFile{Zdziarski2021}

\bibitem[{{Zdziarski} and {Lightman}(1985)}]{Zdziarski1985}
{Zdziarski}, A.~A. and {Lightman}, A.~P. (1985).
\newblock {Nonthermal electron-positron pair production and the 'universal' X-ray spectrum of active galactic nuclei}.
\newblock \emph{\apjl} 294, L79--L83.
\newblock \doi{10.1086/184513}
\bibAnnoteFile{Zdziarski1985}

\bibitem[{{Zdziarski} et~al.(1993){Zdziarski}, {Lightman}, and {Maciolek-Niedzwiecki}}]{Zdziarski1993}
{Zdziarski}, A.~A., {Lightman}, A.~P., and {Maciolek-Niedzwiecki}, A. (1993).
\newblock {Acceleration Efficiency in Nonthermal Sources and the Soft Gamma Rays from NGC 4151 Observed by OSSE and SIGMA}.
\newblock \emph{\apjl} 414, L93.
\newblock \doi{10.1086/187004}
\bibAnnoteFile{Zdziarski1993}

\bibitem[{{Zhang} et~al.(2019{\natexlab{a}}){Zhang}, {Santangelo}, {Feroci}, {Xu}, {Lu}, {Chen} et~al.}]{Zhang2019extp}
{Zhang}, S., {Santangelo}, A., {Feroci}, M., {Xu}, Y., {Lu}, F., {Chen}, Y., et~al. (2019{\natexlab{a}}).
\newblock {The enhanced X-ray Timing and Polarimetry mission{\textemdash}eXTP}.
\newblock \emph{Science China Physics, Mechanics, and Astronomy} 62, 29502.
\newblock \doi{10.1007/s11433-018-9309-2}
\bibAnnoteFile{Zhang2019extp}

\bibitem[{{Zhang} et~al.(2019{\natexlab{b}}){Zhang}, {Dov{\v{c}}iak}, and {Bursa}}]{Zhang2019}
{Zhang}, W., {Dov{\v{c}}iak}, M., and {Bursa}, M. (2019{\natexlab{b}}).
\newblock {Constraining the Size of the Corona with Fully Relativistic Calculations of Spectra of Extended Coronae. I. The Monte Carlo Radiative Transfer Code}.
\newblock \emph{\apj} 875, 148.
\newblock \doi{10.3847/1538-4357/ab1261}
\bibAnnoteFile{Zhang2019}

\bibitem[{{Zhang} and {Lu}(2017)}]{Zhang2017}
{Zhang}, X. and {Lu}, Y. (2017).
\newblock {On the mean radiative efficiency of accreting massive black holes in AGNs and QSOs}.
\newblock \emph{Science China Physics, Mechanics, and Astronomy} 60, 109511.
\newblock \doi{10.1007/s11433-017-9062-1}
\bibAnnoteFile{Zhang2017}

\bibitem[{{Zoghbi} et~al.(2017){Zoghbi}, {Matt}, {Miller}, {Lohfink}, {Walton}, {Ballantyne} et~al.}]{Zoghbi2017}
{Zoghbi}, A., {Matt}, G., {Miller}, J.~M., {Lohfink}, A.~M., {Walton}, D.~J., {Ballantyne}, D.~R., et~al. (2017).
\newblock {A Long Look at MCG-5-23-16 with NuSTAR. I. Relativistic Reflection and Coronal Properties}.
\newblock \emph{\apj} 836, 2.
\newblock \doi{10.3847/1538-4357/aa582c}
\bibAnnoteFile{Zoghbi2017}

\end{thebibliography}


\end{document}